\documentclass[aps,prc,bibnotes,superscriptaddress,nofootinbib,floatfix,twocolumn]{revtex4}
\usepackage{graphicx} 
\usepackage{float} 
\usepackage{amsmath}
\usepackage{url}
\usepackage[colorlinks,linkcolor=blue,citecolor=blue,filecolor=black,urlcolor=blue]{hyperref}
\usepackage{MnSymbol}
\usepackage{multirow}
\usepackage{bm}

\newcommand{\lr}[1]{\left\langle #1\right\rangle}

\newcommand{\pT} {\ensuremath{p_{\mathrm{T}}}}

\newcommand{\nch}{\mbox{$N_{\mathrm{ch}}$}}
\newcommand{\nchrec}{\mbox{$N_{\mathrm{ch}}^{\mathrm{rec}}$}}
\newcommand{\snn}{\mbox{$\sqrt{s_{\mathrm{NN}}}$}}

\usepackage{color}
\usepackage{harpoon}
\graphicspath{{figs/}{./}}
\definecolor{my}{rgb}{1, 0, 0}

\begin{document} 
\title{Imaging nuclear shape through anisotropic and radial flow in high-energy heavy-ion collisions}
\author{The STAR Collaboration}

\begin{abstract}
Most atomic nuclei exhibit ellipsoidal shapes characterized by quadrupole deformation $\beta_2$ and triaxiality $\gamma$, and sometimes even a pear-like octupole deformation $\beta_3$. The STAR experiment introduced a new ``imaging-by-smashing'' technique~\cite{STAR:2024eky,Jia:2025wey} to image the nuclear global shape by colliding nuclei at ultra-relativistic speeds and analyzing outgoing debris. Features of nuclear shape manifest in collective observables like anisotropic flow $v_n$ and radial flow via mean transverse momentum $[p_{\mathrm{T}}]$. We present new measurements of the variances of $v_n$ ($n=2$, 3, and 4) and $[p_{\mathrm{T}}]$, and the covariance of $v_n^2$ with $[p_{\mathrm{T}}]$, in collisions of highly deformed $^{238}$U and nearly spherical $^{197}$Au. Ratios of these observables between the two systems effectively suppress common final-state effects, isolating the strong impact of uranium's deformation. By comparing results with state-of-the-art hydrodynamic model calculations, we extract $\beta_{2\mathrm{U}}$ and $\gamma_{\mathrm{U}}$ values consistent with those deduced from low-energy nuclear structure measurements. Measurements of $v_3$ and its correlation with $[p_{\mathrm{T}}]$ also provide the first experimental suggestion of a possible octupole deformation for $^{238}$U. These findings provide significant support for using high-energy collisions to explore nuclear shapes on femtosecond timescales, with implications for both nuclear structure and quark-gluon plasma studies.
\end{abstract}
\maketitle
\tableofcontents
\section{Introduction}\label{sec:1}
High-energy heavy-ion collisions, conducted at the Relativistic Heavy Ion Collider (RHIC) and the Large Hadron Collider (LHC), have been used to produce and study the quark-gluon plasma (QGP). This deconfined state of matter undergoes a spacetime evolution that is well-described by relativistic hydrodynamics~\cite{Busza:2018rrf}. The hydrodynamic expansion translates initial spatial anisotropies and size variations within the QGP into measurable patterns in the azimuthal and radial distributions of outgoing particles in the momentum space, observed as anisotropic~\cite{Ollitrault:1992bk,STAR:2000ekf} and radial flow~\cite{Schnedermann:1993ws,STAR:2008med}, respectively. While these observables contain vital information about the transport properties of the QGP, their interpretation is complicated by inherent uncertainties in the initial conditions -- the spatial distribution of energy density before the hydrodynamic evolution. The initial conditions cannot be calculated directly from first principles; they must be inferred alongside QGP properties using hydrodynamic models, which introduces significant uncertainties in both~\cite{Bernhard:2016tnd,JETSCAPE:2020mzn,Nijs:2020ors}. Consequently, it is important to develop methods to disentangle the influence of initial conditions from final-state effects.

{{\bf  Initial condition and nuclear shape.}} The connection between the initial conditions and observed final-state flow is relatively well-established~\cite{Busza:2018rrf}. The initial geometry of the collision, characterized by eccentricities $\varepsilon_n$ ($n=2$, 3, 4, $\dots$), directly drives the azimuthal anisotropies in the particle distribution, described by \mbox{$dN/d\phi~\propto1+2\sum v_n\cos(n\phi)$}, where $v_2$, $v_3$, and $v_4$ represent the dominant elliptic, triangular, and quadrangular flow components~\cite{Luzum:2013yya,Jia:2014jca}, respectively. Similarly, the compactness of the initial conditions, quantified by the inverse area $d_{\perp}$ of the overlap region~\cite{Schenke:2020uqq}, influences radial flow, as reflected in the event-by-event average transverse momentum ($[\pT]$). Hydrodynamic models predict approximately linear relationships: $v_n \propto \varepsilon_n$ for $n=2$ and 3 (and for $n=4$ in central collisions)~\cite{Niemi:2012aj}, and $\delta \pT \propto \delta d_{\perp}$~\cite{Bozek:2012fw}, where $\delta \pT = [\pT]-\lr{[\pT]}$ and $\delta d_{\perp} = d_{\perp}-\lr{d_{\perp}}$ represent deviations from their mean values. However, accurately describing the initial conditions requires a good understanding of both the nucleon spatial distribution within colliding nuclei and the mechanisms of energy deposition~\cite{Bally:2022vgo}. 

A major challenge in this endeavor arises from the fact that most atomic nuclei are deformed. Collisions of randomly oriented deformed nuclei significantly enhance event-by-event variations in both the initial size and shape of the QGP~\cite{Jia:2021tzt}. This, in turn, increases fluctuations in anisotropic and radial flow, an effect clearly demonstrated in central $^{238}$U+$^{238}$U collisions by our previous measurement~\cite{STAR:2024eky}. 

One powerful approach to constrain these initial conditions is to compare collision systems with similar sizes but distinct nuclear structures, such as isobar collisions~\cite{STAR:2021mii} or, as in this study, isobar-like collisions of $^{238}$U+$^{238}$U and $^{197}$Au+$^{197}$Au. Uranium is known to be strongly prolate deformed, while gold exhibits mild oblate deformation. Despite these shape differences, the resulting QGPs are of comparable sizes, and the 20\% difference in mass number ($A$) has a small impact on QGP properties~\cite{Mantysaari:2024uwn}. Consequently, ratios of bulk observables between these two systems largely cancel common final-state effects, thereby isolating variations in the initial conditions that stem from the structural differences between $^{238}$U and $^{197}$Au~\cite{Giacalone:2021uhj,Zhang:2022fou}. Analyzing these ratios as a function of nuclear shape also provides valuable insights into the energy deposition mechanisms~\cite{Jia:2021oyt}.

Nuclear shape, including the dominant quadrupole component $\beta_2$, and smaller axial octupole $\beta_3$ and axial hexadecapole $\beta_4$ components, can be described by a Woods-Saxon density profile,
\small{
\begin{align}\nonumber
&\rho(r,\theta,\phi)\propto (1+\exp\left[r-R(\theta,\phi)/a\right])^{-1},\\\label{eq:1}
R(\theta,\phi)\! &=\! R_0\!\left(\!1+\!\beta_2 [\cos \gamma Y_{2,0}\!+\! \sin\gamma Y_{2,2}]\!+\!\beta_3 Y_{3,0}\!+\!\beta_4 Y_{4,0}\right),
\vspace*{-0.3cm}
\end{align}}\normalsize
where $R_0= 1.2A^{1/3}$ is the nuclear radius, $a$ is the surface or skin depth, and $Y_{l,m}(\theta,\phi)$ are real spherical harmonics in the intrinsic frame. The triaxiality parameter $\gamma$ (ranging from $0^{\circ}$ to $60^{\circ}$) determines the relative ordering of the three principal radii ($r_a,r_b,r_c$). Specifically, $\gamma=0^\circ$ corresponds to a prolate shape ($r_a=r_b<r_c)$, $\gamma=60^{\circ}$ to an oblate shape ($r_a<r_b=r_c$), and intermediate values to a triaxial shape ($r_a<r_b<r_c$).

{{\bf  Manifestation of nuclear shapes from low to high energies.}} Traditionally, nuclear shapes are inferred from spectroscopic or scattering experiments conducted at sub-hundred MeV/nucleon beam energies~\cite{Verney:2025efj,Cline:1986ik,Yang:2022wbl}, which probe collective nuclear shapes on timescales longer than $100$~fm/$c$. For instance, the $^{238}$U nucleus, with an even number of protons and neutrons, exhibits a rotational spectrum consistent with a prolate rigid rotor. Estimates based on electric quadrupole transition data and rigid rotor assumption yield $\beta_{2\mathrm{U}}=0.287\pm0.007$~\cite{Pritychenko:2013gwa} and $\gamma_{\mathrm{U}}\lesssim8^{\circ}$~\cite{PhysRevC.54.2356}. Direct measurements of $\beta_{3\mathrm{U}}$ and $\beta_{4\mathrm{U}}$ are lacking, but model-dependent estimates suggest values around $\beta_{3\mathrm{U}}\sim0.08$ and $\beta_{4\mathrm{U}}\sim0.1$~\cite{Moss:1971zz,Moller:2015fba}. The shape of the odd-mass $^{197}$Au nucleus is less certain, with low-energy model calculations suggesting $\beta_{2\mathrm{Au}}=$0.1--0.14 and $\gamma_{\mathrm{Au}}\sim 45^{\circ}$~\cite{Ryssens:2023fkv,Bally:2023dxi}. 

The quantum mechanical nature of nuclear ground states deserves careful consideration in this context. It is understood that the wavefunction of a ground state nucleus is a superposition of all possible orientations, and its shape is not directly observable if nucleus stays in its ground state~\cite{Verney:2025efj}. However, high-energy heavy-ion collisions are strongly interacting, dynamical systems, for which the stationary solution of an individual nucleus is not an eigenstate of the combined two-nuclei system. Instead, the extremely fast smashing process ($\lesssim 0.1$~fm/$c$, corresponding to the crossing time of colliding nuclei) acts as a quantum measurement that projects nucleons into their position space, from which a shape and orientation can be defined.

Furthermore, high-energy heavy-ion collisions are sensitive not only to these global nuclear shapes relevant at low-energy, but also to fluctuations of the nuclear wavefunction on much shorter timescales. This makes high-energy collisions a novel tool for investigating nuclear structure and its evolution with energy~\cite{Jia:2025wey}.
\begin{figure}
\begin{center}
\includegraphics[width=1\linewidth]{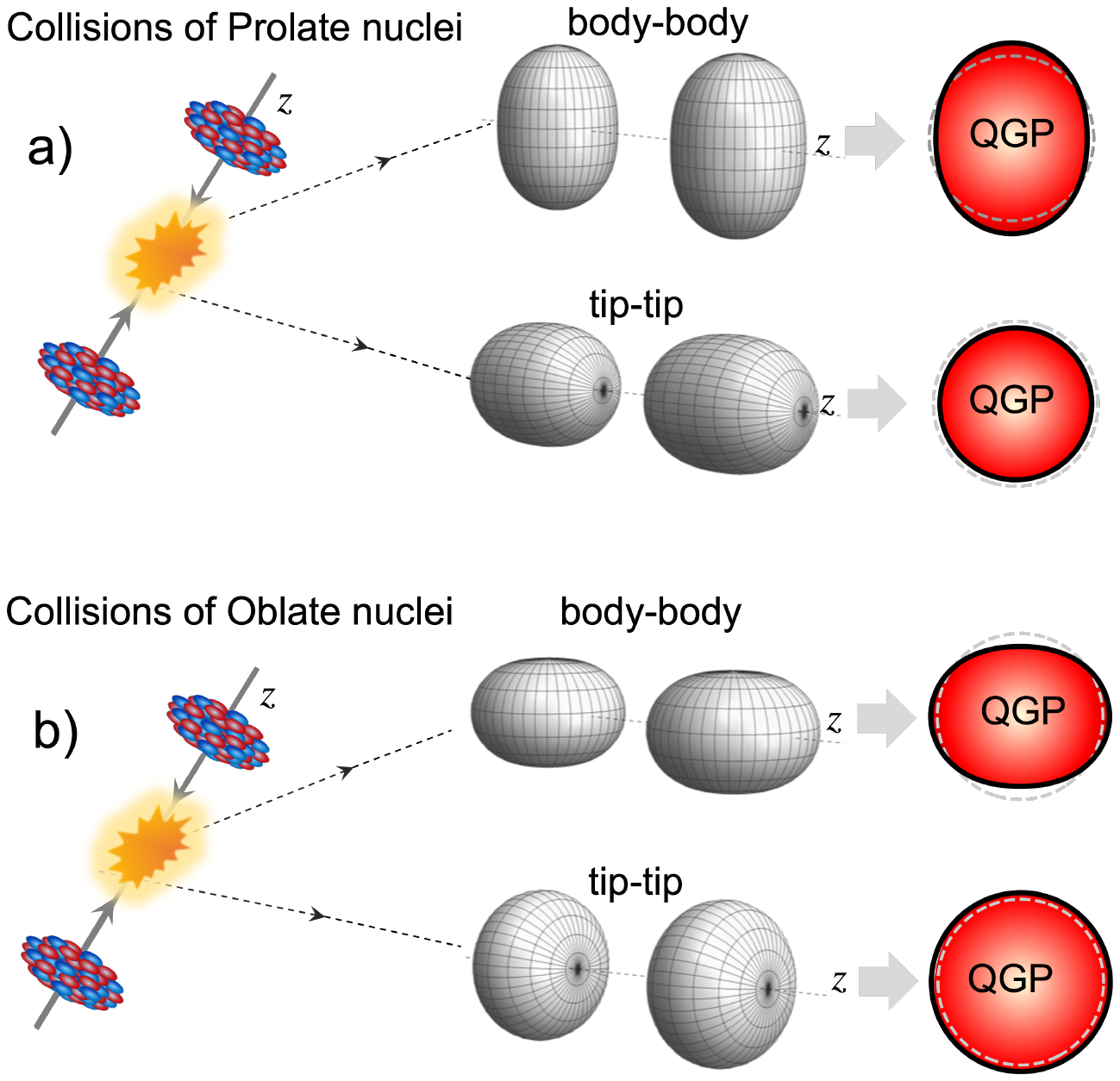}
\end{center}
\vspace*{-0.5cm}\caption{\label{fig:1} Illustration of how collision geometry and collective expansion are impacted by quadrupole deformation in collisions of prolate (\textbf{a}) and oblate (\textbf{b}) nuclei. The configurations vary between body-body and tip-tip, which are selected based on the momentum distribution of final state particles. The negative (positive) correlation between ellipticity and inverse size in (\textbf{a}) ((\textbf{b})) drives the negative (positive) correlation between elliptic flow and radial flow in the final state, highlighting the potential of constraining the triaxiality. The dashed grey circles represent the baseline shape of the QGP for spherical nuclei.}\vspace*{-0.2cm}
\end{figure}

Nuclear deformation influences high-energy collisions by modifying the nucleon distribution in the overlap region and, consequently, the initial conditions of the QGP. In collisions of prolate nuclei (Fig.~\ref{fig:1}a), the initial configuration varies between body-body (high $\varepsilon_2$, low $d_{\perp}$) and tip-tip (low $\varepsilon_2$, high $d_{\perp}$) collisions. This leads to enhanced, anti-correlated event-by-event fluctuations in $\varepsilon_2$ and $d_{\perp}$~\cite{Giacalone:2019pca}. Collisions of oblate nuclei also exhibit enhanced fluctuations, but with a positive correlation between $\varepsilon_2$ and $d_{\perp}$ (Fig.~\ref{fig:1}b). These effects are quantitatively captured by the moments of the $\varepsilon_2$ and $d_{\perp}$ distributions, including the variances $\lr{\varepsilon_2^2}$ and $\lr{(\delta d_{\perp})^2}$, and the skewness $\lr{\varepsilon_2^2\delta d_{\perp}}$. Here, the angle brackets ``$\lr{\cdot}$'' denote averages over events within a given centrality class. Similarly, the presence of modest higher-order deformations $\beta_3$ and $\beta_4$ is expected to slightly enhance the fluctuations of $\varepsilon_3$ and $\varepsilon_4$, respectively~\cite{Jia:2021tzt}.

{{\bf  Extraction of nuclear shape in high-energy collisions.}} Our analysis focuses on the moments of the final-state flow observables: $\lr{v_2^2}$, $\lr{(\delta \pT)^2}$, and $\lr{v_2^2\delta \pT}$~\cite{Bozek:2016yoj}. These observables are sensitive to quadrupole deformation due to their approximately linear relationships with $\lr{\varepsilon_2^2}$, $\lr{(\delta d_{\perp})^2}$, and $\lr{\varepsilon_2^2\delta d_{\perp}}$. For small deformations and using a Taylor expansion, one can show that these quantities follow simple parametric dependencies on the shape parameters~\cite{Jia:2021qyu}:
\begin{align}\nonumber
\lr{v_2^2}&= a_1+b_1\beta_2^2\;,\\\nonumber
\lr{(\delta \pT)^2}&= a_2+b_2\beta_2^2\;,\\\label{eq:2}
\lr{v_2^2\delta \pT} &= a_3-b_3\beta_2^3\cos(3\gamma)\;,
\vspace*{-0.3cm}
\end{align}
where the positive coefficients $a_n$ and $b_n$ depend on the collision geometry and QGP properties. The $b_n$ values are nearly independent of centrality, while the $a_n$ values are minimized in central collisions, making central collisions ideal for constraining nuclear shape. The $\cos(3\gamma)$ term distinguishes between prolate and oblate nuclei, as expected from Fig.~\ref{fig:1}. We also measure correlators involving high-order flow coefficients: $\lr{v_3^2}$, $\lr{v_4^2}$, $\lr{v_3^2\delta \pT}$ and $\lr{v_4^2\delta \pT}$. These observables are expected to be minimally sensitive to $\beta_2$ and $\gamma$~\cite{Jia:2021tzt,Jia:2021wbq}, thus serving as crucial control measurements. Conversely, they are sensitive to high-order deformations $\beta_3$ and $\beta_4$, potentially enabling searches for the presence of these deformations in $^{238}$U~\cite{Zhang:2025hvi}. For example, observables involving $v_3$ are sensitive to octupole deformation:
\begin{align}\nonumber
\lr{v_3^2}&= a_3+b_3\beta_2^2\;,\\\label{eq:2b}
\lr{v_3^2\delta \pT}&= a_4-b_4\beta_2\beta_3^2\;,
\vspace*{-0.3cm}
\end{align}

We employ two complementary approaches to probe the initial conditions and quantify the effects of nuclear deformation. The primary approach utilizes ratios of flow observables between U+U and Au+Au collisions:
\begin{align}\label{eq:3}
R_{\mathcal{O}}=\frac{\lr{\mathcal{O}}_{\mathrm{U}}}{\lr{\mathcal{O}}_{\mathrm{Au}}},\; \lr{\mathcal{O}} = \lr{v_n^2}, \lr{(\delta \pT)^2},\mbox{or}\; \lr{v_n^2\delta \pT}
\vspace*{-0.3cm}
\end{align}
The deformation of uranium is expected to cause these ratios to deviate significantly from unity, particularly in central collisions. The second approach constructs normalized quantities involving the three observables within a single collision system:
\begin{align}\label{eq:4}
\rho_{n}^{\mathrm{pcc}}&\equiv\frac{\lr{v_n^2\delta \pT}}{\sqrt{\mathrm{var}(v_n^2)}\sqrt{\lr{(\delta \pT)^2}}}\;,\\\label{eq:5}
\rho_{n}&\equiv \frac{\lr{v_n^2\delta \pT}}{\lr{v_n^2}\sqrt{\lr{(\delta \pT)^2}}}\;.
\vspace*{-0.3cm}
\end{align}
Here, var($v_n^2$)$=\lr{v_n^4} - \lr{v_n^2}^2$ is the variance of $v_n^2$. The observable $\rho_{n}^{\mathrm{pcc}}$ is the Pearson correlation coefficient~\cite{Bozek:2016yoj}, while $\rho_{n}$, introduced in Ref.~\cite{Jia:2021qyu}, better captures the spherical baseline. Both observables are expected to be strongly modified in U+U ultra-central collisions (UCC) for $n=2$, but not for $n=3$ and 4. These normalized observables are also sensitive to other aspects of the initial conditions, including nucleon width~\cite{Giacalone:2021clp}, initial momentum anisotropy~\cite{Giacalone:2020byk}, and sub-leading radial structure~\cite{Schenke:2020uqq}.

The influence of $\beta_2$ on $v_2$ was recognized in early studies~\cite{Li:1999bea,Shuryak:1999by,Filip:2009zz,Shou:2014eya,Goldschmidt:2015kpa}, and confirmed by experiments at RHIC~\cite{STAR:2015mki} and the LHC~\cite{Acharya:2018ihu,Sirunyan:2019wqp,Aad:2019xmh}. Recently, the STAR Collaboration observed enhanced $\lr{(\delta \pT)^2}$ and a clear negative $\lr{v_2^2\delta \pT}$ signal in central $^{238}$U+$^{238}$U collisions~\cite{STAR:2024eky}. The strength of $\lr{v_2^2\delta \pT}$ is much stronger than the correlation between $v_2$ and total multiplicity previously explored~\cite{STAR:2015mki}. By analyzing ratios (Eq.~\eqref{eq:3}) between $^{238}$U+$^{238}$U and $^{197}$Au+$^{197}$Au collisions and comparing them with hydrodynamic model calculations, STAR have extracted $\beta_{2\mathrm{U}}$ and $\gamma_{\mathrm{U}}$ values consistent with low-energy measurements. A similar extraction was also accomplished using a different hydrodynamic model framework~\cite{Fortier:2023xxy,Fortier:2024yxs}. However, no experimental attempt has yet been made to extract $\beta_3$ and $\beta_4$ deformations. 

{\bf Scope of this paper.} Building on these findings, this paper presents a comprehensive study of anisotropic flow and its correlations with $[\pT]$ in $^{238}$U+$^{238}$U collisions at $\snn=193$ GeV and $^{197}$Au+$^{197}$Au collisions at $\snn=200$ GeV. This study extends previous results in several key aspects:
\begin{itemize}
\item \underline{{\it Normalized observables:}} We compute Pearson correlation coefficients $\rho_2^{\mathrm{pcc}}$ and $\rho_n$ (Eqs.~\eqref{eq:4}--\eqref{eq:5}), which aim to isolate deformation-driven signals by partially canceling final-state effects in a single collision system.
\vspace*{-0.2cm}

\item  \underline{{\it Searching for higher-order deformations:}} Measurements of $\langle v_3^2 \rangle$, $\langle v_4^2 \rangle$, $\langle v_3^2 \delta \pT \rangle$, and $\langle v_4^2 \delta \pT \rangle$ provide critical baselines, as these observables are largely insensitive to $\beta_2$ and $\gamma$. These observables could potentially provide evidence for the presence of modest $\beta_3$ and $\beta_4$ in $^{238}$U.
\vspace*{-0.3cm}

\item  \underline{{\it Differential measurements:}} By varying $\pT$ ranges and pseudorapidity ($\eta$) selections, we assess the robustness of deformation signatures against non-flow correlations and sub-leading fluctuations in the initial geometry.
\vspace*{-0.2cm}

\item  \underline{{\it New ratio observables:}} We combine $\langle v_2^2 \rangle$, $\lr{(\delta \pT)^2}$, and $\lr{v_2^2\delta \pT}$ after taking the difference between the two collision systems. This approach is designed to better eliminate spherical baseline terms ($a_n$ in Eq.~\eqref{eq:2}), thereby enhancing sensitivity to nuclear deformation.
\vspace*{-0.2cm}

\item  \underline{{\it Model comparison:}} We perform detailed hydrodynamic model and Glauber model calculations to extract $\beta_{2\mathrm{U}}$ and $\gamma_{\mathrm{U}}$, validating the ``imaging-by-smashing'' approach against low-energy nuclear structure data.
\vspace*{-0.2cm}
\end{itemize}

This study demonstrates high-energy collisions as a promising imaging tool for nuclear structure research, while also advancing our understanding of the initial conditions of the QGP.

\section{Analysis}\label{sec:2}
\subsection{Event and track selections}\label{sec:21}
This analysis utilizes Au+Au collisions at $\snn=200$ GeV (2010 and 2011) and U+U collisions at $\snn=193$ GeV (2012), recorded by the STAR detector. Charged particles were reconstructed within $|\eta|<1$ using the STAR time projection chamber (TPC)~\cite{Anderson:2003ur}. Minimum-bias events were selected by requiring coincident signals in the two vertex position detectors in the forward regions ($4.4 < |\eta| < 4.9$)~\cite{Llope:2003ti}. To enhance statistics for central collisions, special triggers requiring large TPC multiplicities and minimal activity in the zero-degree calorimeters (ZDC)~\cite{Bieser:2002ah} were employed. 

Offline event selection required a primary vertex $z_{\mathrm{vtx}}$ within 30 cm of the TPC center along the beam axis and within 2 cm of the beam spot in the transverse plane. Pileup events (those containing more than one collision) and background events were reduced by selecting on the correlation between the number of TPC tracks and the number of tracks matched to the time of flight (TOF) detector~\cite{Llope:2012zz,Chen:2024aom}. One type of pileup event consists of a primary event and another central event that reduces TPC reconstruction efficiency. These low-efficiency events were rejected based on the number of TPC tracks and those matched to the primary collision vertex. After these selections, the dataset comprises approximately 528 million minimum-bias (370 million from 2011) and 120 million triggered central Au+Au events, as well as approximately 300 million minimum-bias and 5 million triggered central U+U events.

Charged particle tracks included in this analysis are required to be within $0.2 < \pT < 3.0 $ GeV/$c$ and $|\eta|<1$. Standard STAR track quality criteria were applied~\cite{STAR:2024eky}. To minimize contributions from secondary decays, the Distance of Closest Approach (DCA) of each track to the primary vertex was required to be less than 3 cm. Tracking efficiencies in the TPC, $\epsilon_{\mathrm{TPC}}$, were evaluated using the standard STAR embedding technique~\cite{STAR:2009sxc}. For $\pT > 0.5$ GeV/$c$, efficiency is nearly $\pT$-independent, ranging from 0.72 (0.69) in the most central Au+Au (U+U) collisions to 0.92 in the most peripheral collisions. The largest $\pT$-dependent variation ($\sim$10\% of the plateau value) occurs at $0.2 < \pT < 0.5$ GeV/$c$.

\begin{figure}[!b]
\begin{center}
\includegraphics[width=1\linewidth]{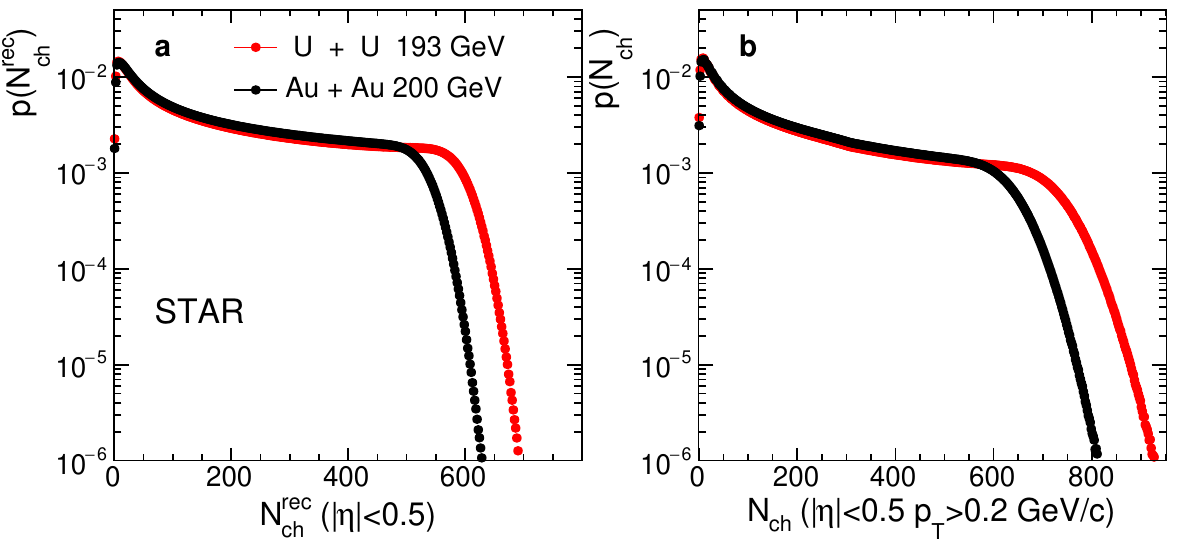}
\end{center}
\vspace*{-0.2cm}
\caption{\label{fig:2} Distributions of reconstructed TPC track multiplicity ($\nchrec$) in $|\eta|<0.5$ and $\pT> 0.15$ GeV/$c$ (left), and efficiency-corrected TPC track multiplicity ($\nch$) for $|\eta|<0.5$ and $\pT> 0.2$ GeV/$c$ (right), in U+U and Au+Au collisions.}
\end{figure}

\begin{figure}[!h]
\begin{center}
\includegraphics[width=1\linewidth]{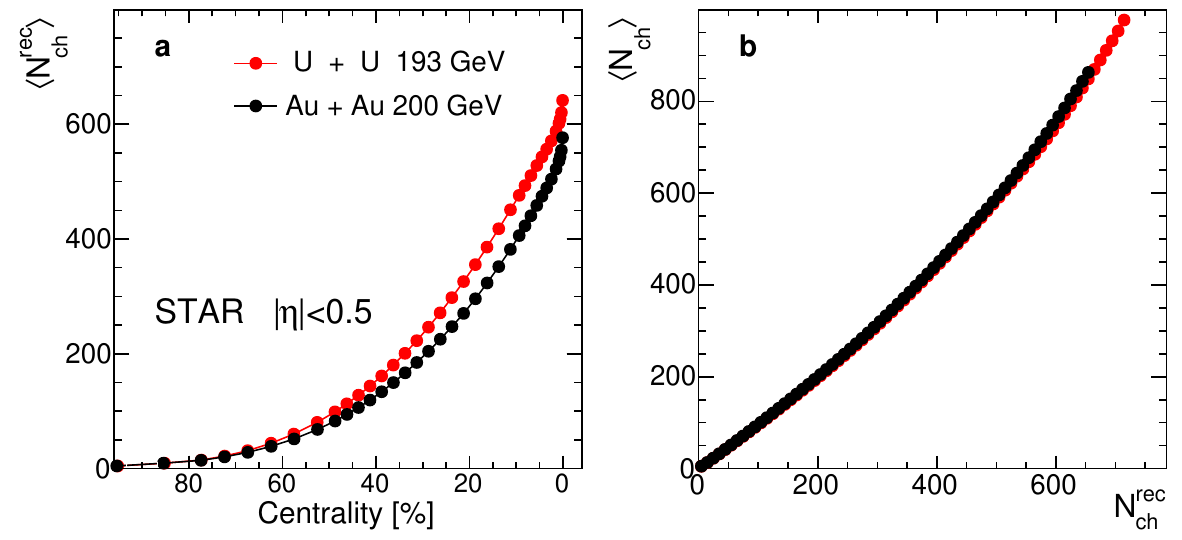}
\end{center}
\vspace*{-0.2cm}
\caption{\label{fig:3} Relationship between centrality and $\nchrec$ (left), and mapping between $\nchrec$ and $\nch$ (right) in U+U and Au+Au collisions.}
\end{figure}

Collision centrality was determined using $\nchrec$, the number of reconstructed tracks in $|\eta|<0.5$ with $\pT> 0.15$ GeV/$c$. The $\nchrec$ distribution, corrected for vertex position and luminosity dependence, was compared to a Monte Carlo (MC) Glauber calculation to define centrality intervals as percentages of the total nucleus-nucleus inelastic cross-section~\cite{STAR:2012och}. An efficiency-corrected multiplicity, $\nch$, was also computed by summing track weights for tracks with $\pT>0.2$ GeV/$c$ and $|\eta|<0.5$ (detailed in Section~\ref{sec:22}). 

Figure~\ref{fig:2} shows the distributions of $\nchrec$ and $\nch$ for both systems. The distribution for U+U spans a wider range than Au+Au due to uranium's larger mass and deformation. Figure~\ref{fig:3} summarizes the mapping between $\nchrec$ and centrality, and between $\nchrec$ and $\nch$. These mappings allow us to present the results in terms of centrality, $\nchrec$, or $\nch$. Most results are presented as a function of centrality in 0--60\% centrality range, with $\nch$ dependence results included in the Appendix.


\subsection{Calculation of observables}\label{sec:22}
The observables $v_n$ ($n=2$, 3, and 4), $\delta \pT= [\pT] -\lr{[\pT]}$, $\lr{v_n^2\delta \pT}$, and other related observables are calculated using charged tracks as:
\small{\begin{align}\nonumber
[\pT] &= \frac{\sum_{i}w_ip_{\mathrm{T},i}}{\sum_{i}w_i},\lr{\pT} \equiv \lr{[\pT]}_{\mathrm{evt}} \\\nonumber
\lr{(\delta \pT)^2}&=\lr{\frac{\sum_{i\neq j}w_iw_j(p_{\mathrm{T},i}-\lr{\pT})(p_{\mathrm{T},j}-\lr{\pT})}{\sum_{i\neq j}w_iw_j}}_{\mathrm{evt}}\\\nonumber
\lr{v_n^2} & =  \lr{ \frac{\sum_{i\neq j}w_iw_j \cos (n(\phi_i-\phi_j))}{\sum_{i\neq j}w_iw_j}}_{\mathrm{evt}}\\\label{eq:6}
\lr{v_n^2\delta \pT} &= \lr{\frac{\sum_{i\neq j\neq k}w_iw_jw_k\cos (n(\phi_i-\phi_j))(p_{\mathrm{T},k}-\lr{\pT})}{\sum_{i\neq j\neq k}w_iw_jw_k}}_{\mathrm{evt}}\;.
\end{align}}\normalsize
Averages are performed first over all multiplets in a single event and then over all events within a fixed $\nchrec$ bin. Track-wise weights $w_{i,j,k}$ account for tracking efficiency and its $\phi$-dependent variation: $w=1/[\epsilon_{\mathrm{TPC}} \times \mathrm{acc}_{\eta}(\phi)]$. The normalized azimuthal track distribution, $\mathrm{acc}_{\eta}(\phi)$, is obtained by averaging over many events in each centrality bin as a function of $\eta$. The efficiency-corrected charged particle multiplicity, $\nch$, is calculated as $\nch = \sum_{i} w_{i}$ using all reconstructed tracks within $\pT>0.2$ GeV/$c$ and $|\eta|<0.5$. 

Observables $\lr{v_n^2}$ and $\lr{(\delta \pT)^2}$ are obtained using both the standard method (particles $i$ and $j$ from $|\eta|<1$) and the two-subevent method (particles $i$ and $j$ from $-1 < \eta_i < -0.1$ and $0.1 < \eta_j < 1$, respectively). The rapidity gap (on average 1.1 unit) in the two-subevent method suppresses short-range non-flow correlations from resonance decays and jets. 

The covariance $\lr{v_n^2\delta \pT}$ is calculated by averaging over all triplets ($i$, $j$, and $k$). To assess the influence of non-flow, $\lr{v_n^2\delta \pT}$ is also calculated using two- and three-subevent methods~\cite{Jia:2017hbm,ATLAS:2017rtr,ATLAS:2018ngv,ATLAS:2022dov}. In the two-subevent method, particles $i$ and $j$ are taken from $-1 < \eta_i < -0.1$ and $0.1 < \eta_j < 1$, while particle $k$ can be taken from either subevent. In the three-subevent method, particles $i$, $j$ and $k$ are taken from distinct $\eta$ ranges: $-1 < \eta_i < -0.4$, $0.4 < \eta_j < 1$, and $|\eta_k|<0.3$. Comparing these methods quantifies the impact of non-flow correlations and longitudinal fluctuations.

The Pearson correlation coefficient $\rho_{n}^{\mathrm{pcc}}$~\cite{Bozek:2016yoj} quantifies correlations between $v_n^2$ and $[\pT]$ (Eq.~\eqref{eq:4}). The variance, var($v_n^2$), is obtained from the standard two- and four-particle cumulants: $c_n\{2\} = \lr{v_n^2}$ and $c_n\{4\} = \lr{v_n^4} - 2 \lr{v_n^2}^2$~\cite{Borghini:2000sa} as~\cite{ATLAS:2022dov},
\begin{align}\label{eq:7}
\mbox{var}(v_n^2) \equiv c_n\{2\}^2_{\mathrm{two-sub}} +c_n\{4\}_{\mathrm{standard}}\;.
\end{align}
Observable $c_n\{4\}$, a four-particle cumulant, is insensitive to non-flow but has poor statistical precision. Therefore, it is obtained from the standard method in the full event. The two-particle cumulants, $c_n\{2\}$, more susceptible to non-flow, are calculated using the two-subevent method. 

Quadrupole deformation also influences $c_2\{4\}$:
\begin{align}\label{eq:8}
c_2\{4\} = -c_1-d_1\beta_2^4\;.
\end{align}
where $c_1$ and $d_1$ are positive, centrality-dependent coefficients. The liquid-drop model calculation predicts $d_1 = 2/7 b_1^2$~\cite{Jia:2021qyu}. Previous STAR measurements suggest that $d_1\beta_2^4$ is larger than $|c_1|$ in the 0--5\% most central collisions~\cite{STAR:2015mki}. 

For higher-order flow harmonics ($n=3$,4), large statistical uncertainties in $c_3\{4\}$ and $c_4\{4\}$ preclude measuring $\rho_{3}^{\mathrm{pcc}}$ and $\rho_{4}^{\mathrm{pcc}}$. Thus, the simpler normalization of Eq.~\eqref{eq:5} is used. For $n=2$ in central collisions and generally for $n=3$ and 4, $|c_n\{4\}|\ll c_n\{2\}^2$, so $\sqrt{\mathrm{var}(v_n^2)} \approx \lr{v_n^2}$ and $\rho_{n}^{\mathrm{pcc}} \approx \rho_{n}$. However, in non-central collisions, $\sqrt{\mathrm{var}(v_2^2)} < \lr{v_2^2}$, leading to $\rho_{2}^{\mathrm{pcc}} > \rho_{2}$. In peripheral collisions, their differences are more affected by non-flow.

Previous studies.~\cite{Jia:2021qyu,STAR:2024eky} found that $\rho_{2}$ is nearly centrality-independent for spherical nuclei, making it ideal for detecting quadrupole deformation. This normalization also benefits from the simple dependence of $\lr{v_2^2}$ on $\beta_2$ (Eq.~\eqref{eq:2}). In contrast, $\mbox{var}(v_2^2)$ has a more complex form:
\begin{align}\label{eq:9}
\mbox{var}(v_2^2) = (a_1+b_1\beta_2^2)^2 -c_1-d_1\beta_2^4
\end{align}
where $a_1$ and $b_1$ are from Eq.~\eqref{eq:2}, and $c_1$ and $d_1$ are from Eq.~\eqref{eq:8}. This complicates the interpretation of $\rho_{2}^{\mathrm{pcc}}$ in terms of quadrupole deformation.

All observables are computed for four $\pT$ ranges: 0.2--2 GeV/$c$, 0.2--3 GeV/$c$, 0.5--2 GeV/$c$, and 0.5--3 GeV/$c$. Studying $\pT$ range dependence helps illuminate non-flow effects, which are generally larger at higher $\pT$. Furthermore, the hydrodynamic response may vary with $\pT$-range due to sub-leading modes of the energy density fluctuations~\cite{Mazeliauskas:2015efa,Schenke:2020uqq}. These sub-leading modes could be sensitive to the shape and radial structure of the colliding nuclei. These high-order geometric effects should be included (at least partially) by event-by-event hydrodynamic model simulations. Thus, observables measured in different $\pT$ ranges are expected to provide consistent constraints on nuclear deformation. 
 
\subsection{Influence of non-flow correlations}\label{sec:2.3}
Non-flow refers to correlations among a few particles from a common source (e.g., resonance decays, jets) that are uncorrelated with the initial geometry. Most non-flow sources are short-range (near-side), but dijet fragmentation can cause wider $\eta$ correlations (away-side). The non-flow contribution generally increases with $\pT$, decreases for higher-order cumulants, and decreases with $\eta$ gaps between particles. However, the $\eta$-gap dependence can also arise from longitudinal fluctuations.

Near-side non-flow is largely suppressed by the subevent method, which requires a large $\eta$ gap. Remaining away-side non-flow can be estimated assuming independent sources, where non-flow in $n$-particle cumulants ($\mathcal{O}$) scales with charged particle multiplicity as $1/\nch^{n-1}$:
\begin{align}\label{eq:10}
\lr{\mathcal{O}}_{\mathrm{nf}} \approx \frac{\left[\lr{\mathcal{O}}\nch^{n-1}\right]_{\mathrm{peri}}}{\nch^{n-1}}\;,
\end{align}
where ``peri'' denotes the $\nch$ range corresponding to the 90--100\% centrality bin. While widely used in small collision systems at RHIC and the LHC~\cite{CMS:2013jlh,ALICE:2013snk,ATLAS:2014qaj,PHENIX:2013ktj,STAR:2022pfn}, this method may not be appropriate for large systems where jet quenching strongly modifies away-side correlations.

Following Ref.~\cite{STAR:2024eky}, non-flow is estimated by: 1) the difference between the standard method and subevent methods, and 2) the $\nch$-scaling method (Eq.~\eqref{eq:10}). The latter is applied for $\lr{v_n^2}$ and $\lr{v_n^2\delta \pT}$, where medium effects redistribute non-flow azimuthally rather than suppress them. This approach isn't suitable for $\lr{(\delta \pT)^2}$, where medium effects should always suppress non-flow. 

The subevent requirement reduces the values of $\lr{v_n^2}$, $\lr{(\delta \pT)^2}$, and $\lr{v_n^2\delta \pT}$ in the most peripheral bin. Differences between standard and subevent methods are used to estimate residual non-flow in the subevent method. These differences, propagated to ratios between U+U and Au+Au, are 1.1\%, 3.5\%, and 11\% for $R_{v_2^2}$, $R_{(\delta \pT)^2}$, and $R_{v_2^2\delta \pT}$, respectively. 

For non-flow estimation based on the $\nch$-scaling method, we assume the entire signals in the 80--100\% centrality two-subevent results are non-flow, then estimate the non-flow fraction using Eq.~\eqref{eq:10}. This is used for $\lr{v_2^2}$ and $\lr{v_2^2\delta \pT}$. In 0--5\% most central collisions, estimated non-flow is about 6\% for $\lr{v_2^2}$ and about 1.4\% for $\lr{v_2^2\delta \pT}$. Propagated to ratios, these are 2.8\% for $R_{v_2^2}$ and 2.5\% for $R_{v_2^2\delta \pT}$.

Non-flow systematic uncertainties are taken as the larger of the two estimates for $R_{v_2^2}$ and $R_{v_2^2\delta \pT}$, while for $R_{(\delta \pT)^2}$, the difference between standard and subevent methods is used. Total non-flow uncertainties in the 0--5\% centrality are 2.8\%, 3.5\% and 11\% for $R_{v_2^2}$, $R_{(\delta \pT)^2}$ and $R_{v_2^2\delta \pT}$, respectively.  

Non-flow correlations are physical processes, not instrumental effects. Most flow measurements in large collision systems at RHIC and the LHC do not perform model-dependent non-flow subtractions. Ideally, non-flow physics should be incorporated into theoretical models used for comparison. Unfortunately, most models are unable to correctly describe the non-flow correlations (e.g., including contributions from resonance decays but not jet quenching). This should be taken into account when comparing measurements to models. 

In this paper, non-flow systematics are treated as model-dependent uncertainties and are incorporated for quantitative extraction of nuclear shape parameters in Section~\ref{sec:4.2}, but are otherwise not included in data plots.

\subsection{Systematic uncertainties}\label{sec:2.4}

Systematic uncertainties were estimated by varying track quality selections and collision $z_{\mathrm{vtx}}$ range, studying the influence of pileup, comparing results from different data-taking periods, and performing a closure test. The following summarizes uncertainties for the default $\pT$ range (0.2--3 GeV/$c$). 

{\underline{\it Track selection:}} Varying the number of fit hits on the track (16 to 19) and the DCA cut ($<3$ cm to $<2.5$ cm) yielded variations of 0.5--1\% for $\lr{\pT}$ and 1--5\% for $\lr{(\delta \pT)^2}$. Uncertainties for $\lr{v_n^2}$ reached $3\%$, $5\%$, and $10\%$ for $n=2$, 3, and 4, respectively, with larger relative uncertainties in central collisions due to smaller $\lr{v_n^2}$ values. Track selection influences on $\lr{v_n^2\delta\pT}$ lead to uncertainties up to $4\%$ ($n=2$), $10\%$ ($n=3)$, and $20\%$ ($n=4$).

{\underline{\it Collision vertex:}} Comparing results for $|z_{\mathrm{vtx}}|<12$ cm with $|z_{\mathrm{vtx}}|<30$ cm showed differences of $0.2$--0.9\% for $\lr{\pT}$ and $0.5$--3\% for other observables.

{\underline{\it Data-taking periods:}} Comparisons between normal and reverse magnetic field runs in Au+Au collisions showed consistency within statistical uncertainties. Comparisons between the 2010 and 2011 Au+Au datasets, which have different active acceptances in the TPC, were largely consistent, within the quoted uncertainties, as a function of centrality. However, significant differences are observed in several cases. We observe about 0.1\%, 1--2\%, 2--4\%, 2\%, 2--8\% variations in $\lr{\pT}$, $\lr{(\delta \pT)^2}$, $\lr{v_2^2}$, $\lr{v_3^2}$, and $\lr{v_4^2}$, respectively. In addition, a 10--15\% variation of $\lr{v_n^2\delta \pT}$ is also observed in central collisions. Half of the observed differences are included in the systematic uncertainties for each observable.

{\underline{\it Pileup and background:}} Varying the cut on the correlation between $\nchrec$ and TOF hits showed an influence of 1--3\% for $\langle v_n^2 \rangle$ and $\langle (\delta \pT)^2 \rangle$, and up to 2--10\% for $\langle v_n^2 \delta \pT \rangle$ in mid-central collisions.

{\underline{\it Acceptance correction:}} As a cross-check, a re-centering correction on flow vectors~\cite{Selyuzhenkov:2007zi} was applied instead of the $\phi$-dependent acceptance correction, $\mathrm{acc}_{\eta}(\phi)$. Results were consistent.

{\underline{\it Closure test:}} A closure test was performed. Reconstruction efficiency and its variations in $\eta$ and $\phi$ from data were used to retain a fraction of particles generated by a multi-phase transport model~\cite{Lin:2004en}. Observables were calculated with track-by-track weights  (Eq.~\eqref{eq:6}) and compared to original particles. The observables $\lr{v_n^2}$, $\lr{\pT}$, and $\lr{(\delta \pT)^2}$ were recovered within statistical uncertainties. However, a 2--4\% non-closure was observed in $\lr{v_2^2 \delta \pT}$ and $\lr{v_3^2 \delta \pT}$, and up to 15\% in $\lr{v_4^2 \delta \pT}$, which are included in the final systematic uncertainties.

{\underline{\it Additional cross-checks:}} The track reconstruction efficiency has an approximate 5\% uncertainty. Varying this efficiency resulted in variations of about 0.1\% for $\lr{\pT}$.  Variations for other observables were either less than 1\% or within statistical uncertainties.  The reconstructed $\pT$ can differ from the true value due to finite momentum resolution.  This was investigated by smearing the reconstructed $\pT$, calculating the observable, and comparing the results with the original ones. A discrepancy of approximately 0.1\% and 0.5\% was observed for $\lr{\pT}$ and $\lr{(\delta \pT)^2}$, respectively. Other observables remained consistent within their statistical uncertainties. 

Systematic uncertainties for other $\pT$ ranges are similar, generally smaller for higher $\pT$ ranges due to better track reconstruction efficiency and fewer fake tracks. However, a slightly larger difference between standard and subevent methods is observed at higher $\pT$ due to larger non-flow.

These different sources are assumed to be independent, and are hence combined in quadrature to give the total systematic uncertainties. For measurements performed in $0.2<\pT<2$ or $0.2<\pT<3$ GeV/$c$, total uncertainties for $\lr{v_2^2}$, $\lr{v_3^2}$, $\lr{v_4^2}$,  $\lr{\pT}$, $\lr{(\delta \pT)^2}$, $\lr{v_2^2\delta \pT}$,  $\lr{v_3^2\delta \pT}$, and $\lr{v_4^2\delta \pT}$ are 1--4\%, 2--8\%, 4--15\%, 0.2--1\%, 1--5\%, 3--8\%, 5--15\% and 10--20\% respectively. Uncertainties are similar for 0.2--2 GeV/$c$ and 0.3--2 GeV/$c$, but smaller for $0.5<\pT<2$ and $0.5<\pT<3$ GeV/$c$. 

Many uncertainty sources are similar in Au+Au and U+U collisions and partially cancel in the ratios between the two systems. However, detector stability, as assessed by comparing 2010 and 2011 Au+Au data, is incorporated without cancellation  into the ratio calculation. Ratios are calculated for each source of systematic uncertainty, and variations are combined to give the total uncertainty on this ratio. 
\begin{figure}[b]
\begin{center}
\includegraphics[width=1\linewidth]{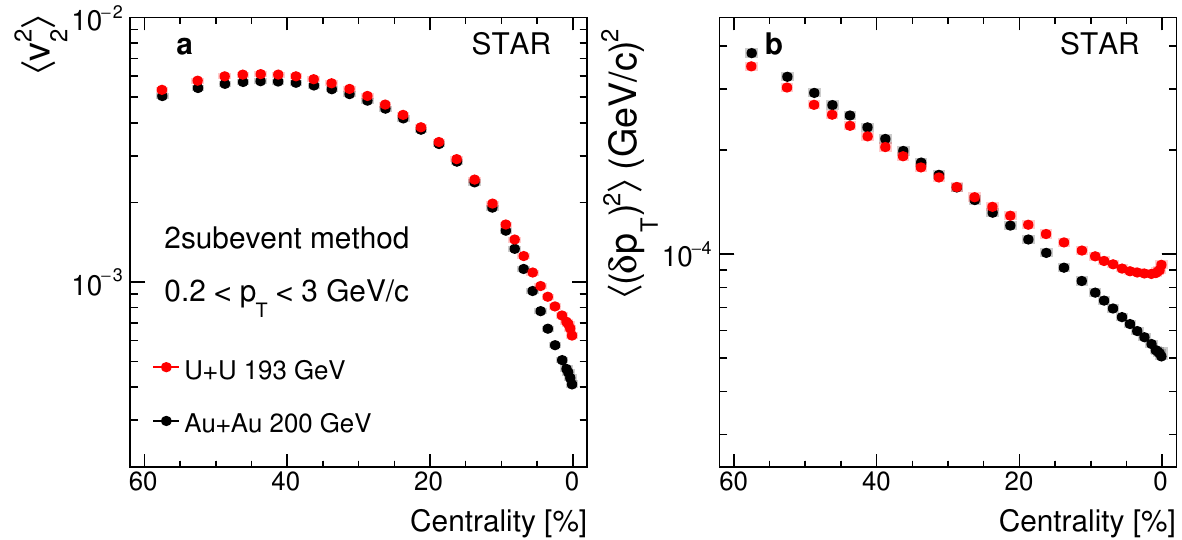}
\end{center}
\caption{\label{fig:3.1} $\lr{v_2^2}$ (left) and $\lr{(\delta\pT)^2}$ (right) as a function of centrality in U+U and Au+Au collisions for charged hadrons in $0.2<\pT<3$ GeV/$c$.  Results are obtained using the two-subevent method.}
\end{figure}

\begin{figure*}[htbp]
\begin{center}
\includegraphics[width=0.9\linewidth]{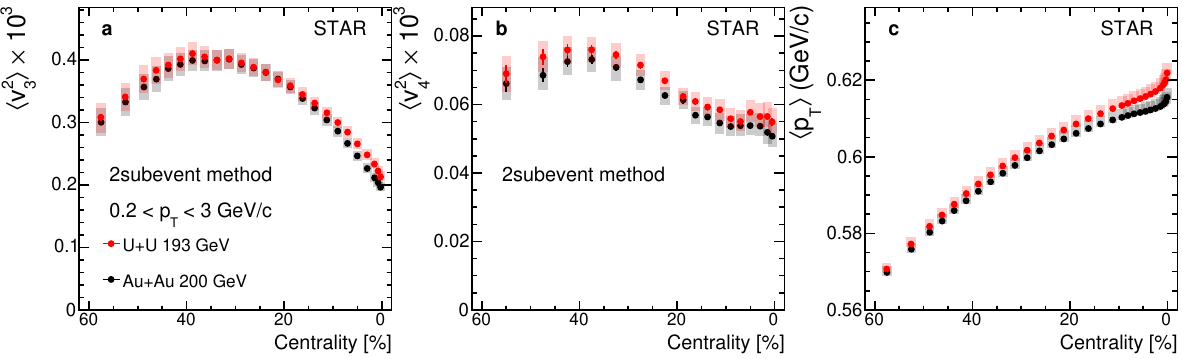}
\end{center}
\caption{\label{fig:3.2} The $\lr{v_3^2}$ (left), $\lr{v_4^2}$ (middle) and  $\lr{\pT}$ (right)  as a function of centrality in U+U and Au+Au collisions for charged hadrons in $0.2<\pT<3$ GeV/$c$.}
\end{figure*}
\begin{figure*}[htbp]
\begin{center}
\includegraphics[width=0.9\linewidth]{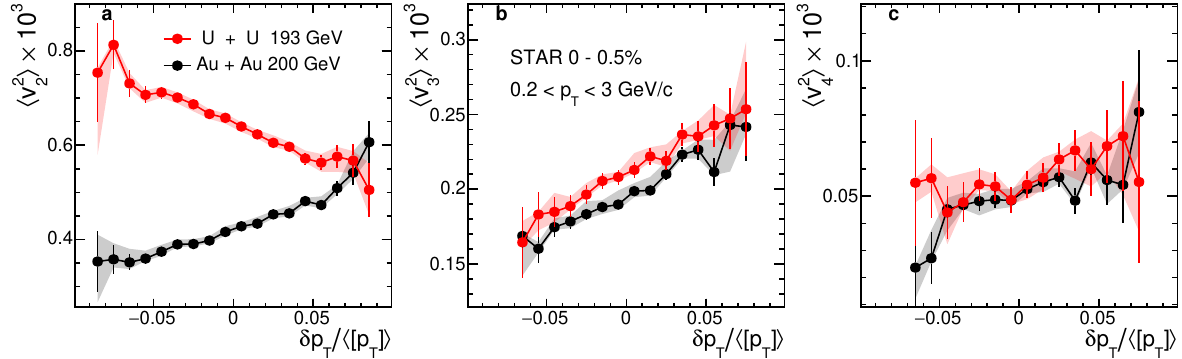}
\end{center}
\caption{\label{fig:3.3} The $\lr{v_2^2}$ (left), $\lr{v_3^2}$ (middle) and $\lr{v_4^2}$ (right) as a function of $\delta \pT/\lr{[\pT]}$ in the 0--0.5\% most central Au+Au (black) and U+U (red) collisions for charged hadrons in $0.2<\pT<3$ GeV/$c$.  Results are obtained using the two-subevent method.}
\end{figure*}

This paper presents results for $\lr{v_n^2}$, $\lr{v_n^2\delta\pT}$, $\lr{\pT}$, $\lr{(\delta\pT)^2}$, and $\rho_n$, and their derived ratios between U+U and Au+Au. Results are presented by default as a function of centrality, with corresponding results as a function of $\nch$ included in the appendix.  Statistical uncertainties are indicated by error bars, and systematic uncertainties by shaded boxes in all figures.

\section{Results}\label{sec:3}
\subsection{General trends and the signatures of deformation}\label{sec:3.1}

To isolate the effects of nuclear deformation, we compare results from collisions of highly-deformed $^{238}$U with those of nearly-spherical $^{197}$Au, as a function of centrality. The 20\% difference in mass number between these nuclei is expected to have a minor influence on the hydrodynamic response~\cite{Mantysaari:2024uwn}, ensuring that observed differences at the same centrality are primarily sensitive to nuclear deformation.

Figure~\ref{fig:3.1} shows a significant enhancement of $\lr{v_2^2}$ and $\lr{(\delta\pT)^2}$ in central U+U collisions compared to central Au+Au collisions. This enhancement is attributed to increased fluctuations in elliptic and radial flow driven by the large prolate deformation of $^{238}$U (Eq.~\eqref{eq:2}). In contrast, Fig.~\ref{fig:3.2} displays much smaller differences in $\lr{v_3^2}$, $\lr{v_4^2}$, and $\lr{\pT}$ between the two systems, which are expected since quadrupole deformation minimally impacts these observables.

However, $\lr{v_3^2}$ values in central collisions are about 7\% larger in U+U compared to Au+Au. This result is surprising. Naively, we expect the Au+Au system, being smaller, to have larger fluctuations in its initial geometry (specifically, larger $\lr{\varepsilon_3^2}$). This simple scaling, $\lr{\varepsilon_3^2} \propto 1/A$, suggests that $\lr{v_{3}^2}_{\mathrm{Au}}$ should be greater than $\lr{v_{3}^2}_{\mathrm {U}}$.

While greater viscous damping in the smaller Au+Au system would slightly reduce its $v_3$ compared to U+U, transport models still predict a slightly larger $\lr{v_3^2}$ (by about 10\%) for Au+Au, even after accounting for this effect~\cite{Jia:2021tzt}. On the other hand, this unexpected ordering of $v_3$ values can be explained by including a small amount of octupole deformation in the uranium nucleus; this octupole deformation would increase $v_{3\mathrm{U}}$, as shown in Eq.~\ref{eq:2b}. Hydrodynamic model calculations support this, confirming that a 7\% larger $\lr{v_3^2}$ in U+U is consistent with an octupole deformation parameter of $\beta_{3\mathrm{U}}\sim 0.08$--0.10~\cite{Zhang:2025hvi}. 

Similar considerations apply to the ordering of $\lr{v_4^2}$. Although the measurement uncertainties are larger for this observable, some impact from the hexadecapole deformation $\beta_{4\mathrm{U}}$ might be expected. However, transport model calculations suggest that the expected larger fluctuation in Au+Au collisions is largely canceled by the stronger viscous damping~\cite{Jia:2021tzt}. This could result in $\lr{v_4^2}$ being very similar in the two collision systems. Hence, the data suggest that the impact of $\beta_{4\mathrm{U}}$ on $\lr{v_4^2}$ is not significant compared to the measurement uncertainties.

The $\lr{\pT}$ values are slightly larger in U+U collisions, consistent with a denser QGP medium due to uranium's larger mass number. The sharp increase in $\lr{\pT}$ toward the most central collisions may be linked to genuine temperature fluctuations~\cite{Gardim:2019brr}. These fluctuations, when convoluted with a steeply falling $\nch$ distribution, lead to an increase in $\lr{\pT}$~\cite{CMS:2024sgx,ATLAS:2024jvf}. It will be interesting to investigate whether this increase could also be influenced by nuclear deformation.

To visualize the expected shape-size correlation from Fig.~\ref{fig:1}, we examine the 0--0.5\% most central collisions, where nuclear deformation has the greatest impact. These events are binned according to their event-by-event $\delta \pT$ values, and the corresponding $\lr{v_n^2}$ values are shown in Fig.~\ref{fig:3.3}. The positive correlation between $v_2$ and $\delta \pT$ in Au+Au collisions aligns with the positive correlation between $\varepsilon_2$ and $d_{\perp}$~\cite{Giacalone:2020awm,Jia:2021qyu}. Conversely, the decrease of $v_2$ with increasing $\delta \pT$ in U+U collisions supports the picture in Fig.~\ref{fig:1}~\cite{Giacalone:2019pca}: events with small (large) $[\pT]$ are enriched with body-body (tip-tip) collisions. This is a dramatic effect: $\lr{v_2^2}$ in U+U is over 60\% higher than in Au+Au at the low end of $\delta \pT$, but nearly identical at the high end.  The $v_3$ and $v_4$ values are similar in the two systems and largely unaffected by quadrupole deformation.

\begin{figure}[!htb]
\begin{center}
\includegraphics[width=1\linewidth]{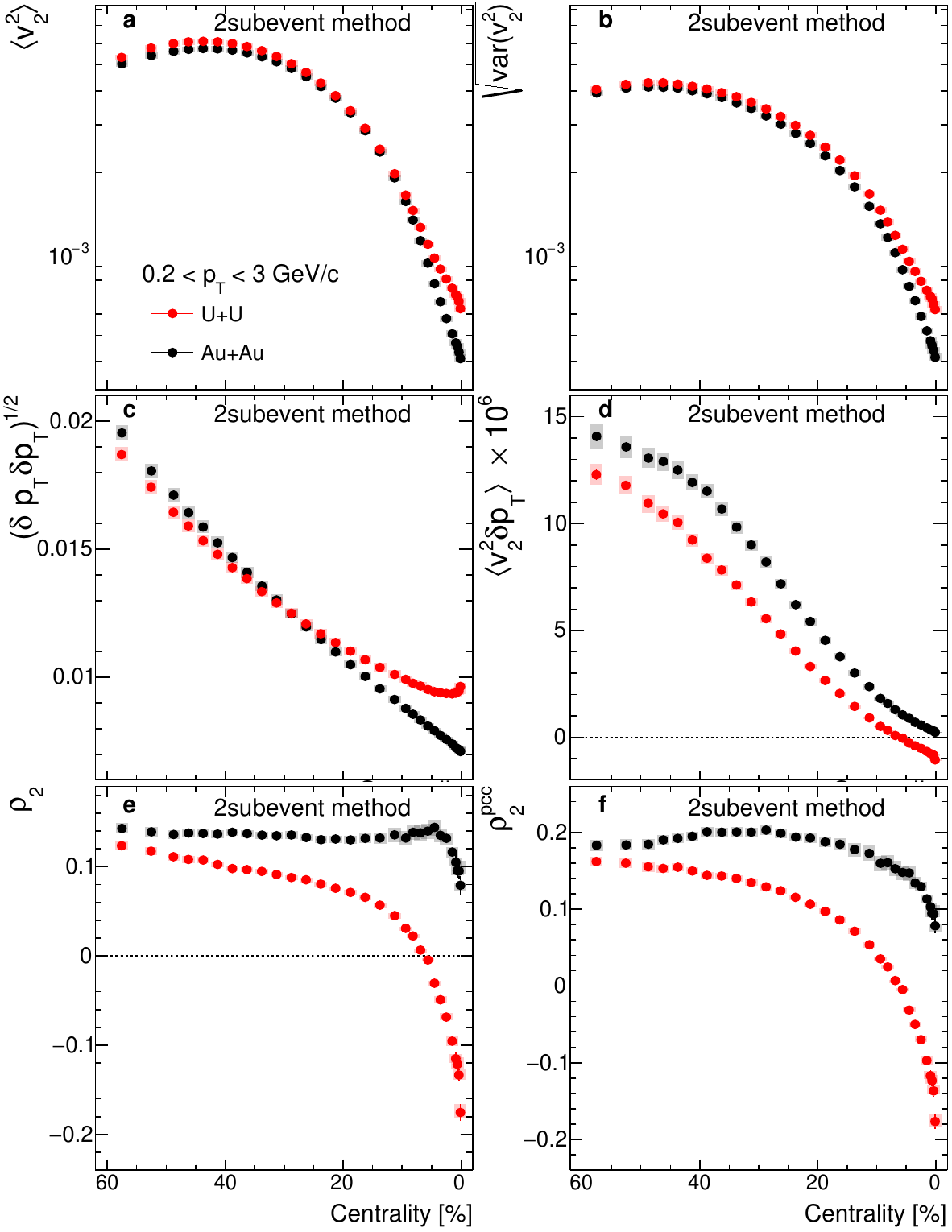}
\end{center}
\caption{\label{fig:3.4} Centrality dependence of $\lr{v_2^2}$, $\sqrt{\mathrm{var}(v_2^2)}$, $\sqrt{\lr{(\delta \pT)^2}}$, $\lr{v_2^2\delta \pT}$ and derived quantities $\rho_2$ (Eq.~\eqref{eq:5}) and $\rho_2^{\mathrm{pcc}}$ (Eq.~\eqref{eq:4}) in Au+Au and U+U collisions. All are obtained using the two-subevent method.  Note that the $\lr{v_2^2}$ and $\sqrt{\lr{(\delta \pT)^2}}$ data are taken directly from Fig~\ref{fig:3.1}.}
\end{figure}

The linear correlation between $v_n$ and $\delta\pT$ with a slope of $h$ in Fig.~\ref{fig:3.3} can be directly related to $\rho_n$. We can express the linear relationship as $v_n^2/\lr{v_n^2}-1 =  h \delta \pT/\lr{[\pT]}$. Multiplying both sides by $\delta \pT$ and averaging over the event ensemble, we get $\lr{v_n^2\delta \pT}/\lr{v_n^2} = h \lr{(\delta \pT)^2}/\lr{[\pT]}$. This leads to $\rho_n = h \sqrt{\lr{(\delta \pT)^2}}/\lr{[\pT]}$, implying that $\rho_n$ captures the linear correlation in Fig.~\ref{fig:3.3}.

The calculation of $\rho_n$ and $\rho_n^{\mathrm{pcc}}$ requires $\lr{v_n^2}$, $\sqrt{\mathrm{var}(v_n^2)}$, $\sqrt{\lr{(\delta \pT)^2}}$ and $\lr{v_n^2\delta \pT}$; these components for $n=2$ are shown in Fig.~\ref{fig:3.4} as a function of centrality. Central U+U collisions show significant enhancements in $\lr{v_2^2}$, $\sqrt{\mathrm{var}(v_2^2)}$, and $\sqrt{\lr{(\delta \pT)^2}}$ relative to Au+Au, consistent with a large $\beta_{2\rm U}$. Conversely, $\lr{v_2^2\delta\pT}$ values are suppressed in U+U collisions across all centralities, consistent with a large $\beta_{2\rm U}$ and a prolate shape (Eq.~\eqref{eq:2}). In central collisions, where the Au+Au baseline is positive, the uranium's deformation drives $\lr{v_2^2\delta \pT}$ negative in the 0--7\% centrality range. 

\begin{figure}[!htb]
\begin{center}
\includegraphics[width=1\linewidth]{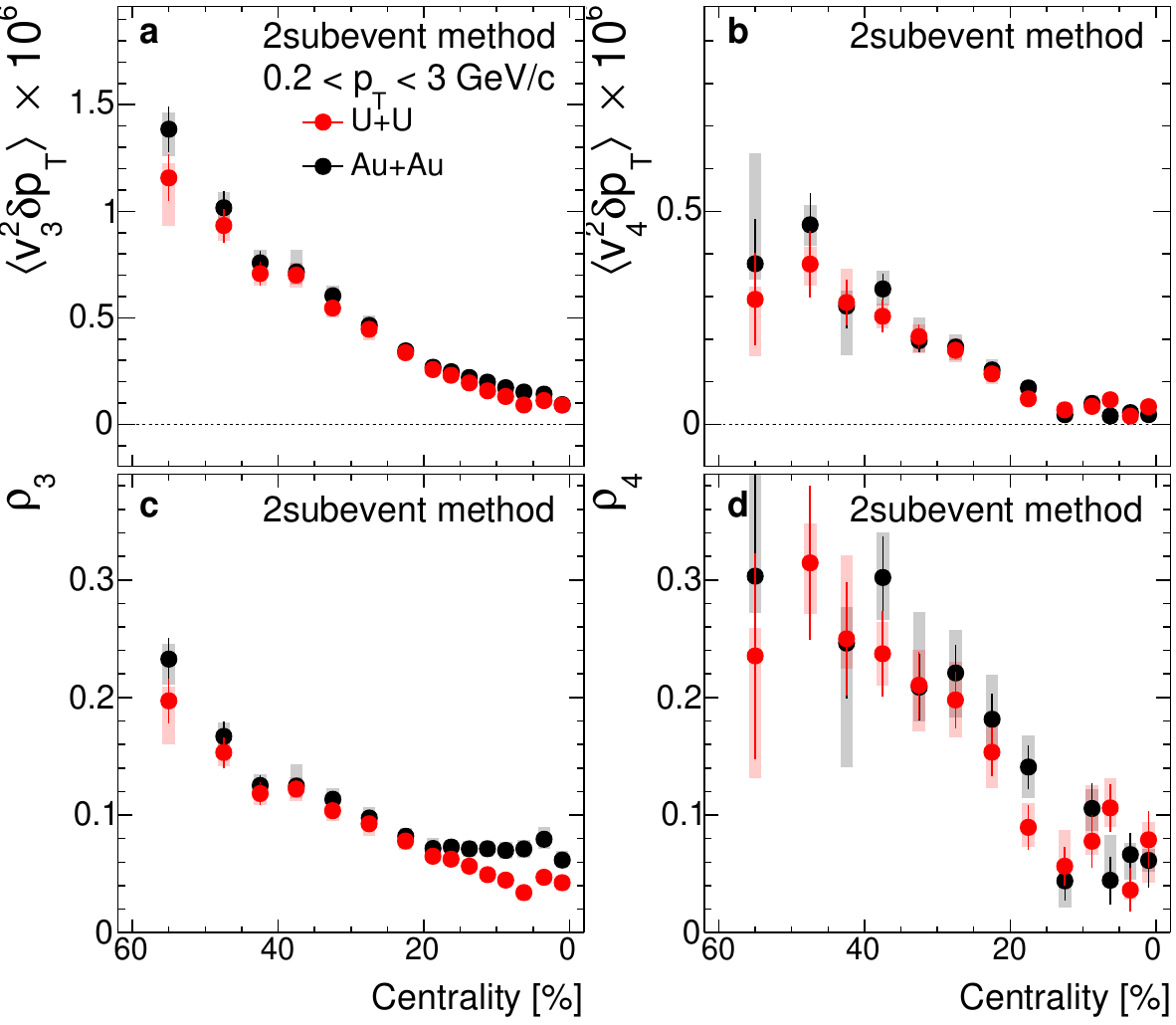}
\end{center}
\caption{\label{fig:3.5} Centrality dependence of $\lr{v_n^2\delta \pT}$ (top) and $\rho_n$ (bottom) for $n=3$ (left) and $n=4$ (right) in Au+Au and U+U collisions. All are obtained using the two-subevent method.}
\end{figure}

From these components, we calculate $\rho_2$ and $\rho_2^{\mathrm{pcc}}$ and present them in the bottom row of Fig.~\ref{fig:3.4}. The $\rho_2$ values in Au+Au collisions remain nearly constant across centrality, while in U+U they decrease gradually up to 15\% centrality, then drop sharply in more central collisions. The $\rho_2^{\mathrm{pcc}}$ values in Au+Au collisions decrease slowly toward central collisions, largely unrelated to nuclear deformation. These behaviors are defining features of $\rho_2$ and $\rho_2^{\mathrm{pcc}}$, and are robust against correlation methods and $\pT$ selection (see Section~\ref{sec:3.2}). 

A similar study for higher-order harmonic flow is presented in Fig.~\ref{fig:3.5}, which shows $\lr{v_n^2\delta\pT}$ and $\rho_n$ for $n=3$ and 4 as a function of centrality. While generally similar between systems, $\lr{v_3^2\delta\pT}$ and $\rho_3$ values are smaller in central U+U collisions, qualitatively compatible with a small octupole deformation in $^{238}$U via Eq.~\eqref{eq:2b}~\cite{Zhang:2025hvi}.

\subsection{Dependence on correlation methods and kinematic selections}\label{sec:3.2}
\begin{figure}[!t]
\begin{center}
\includegraphics[width=1\linewidth]{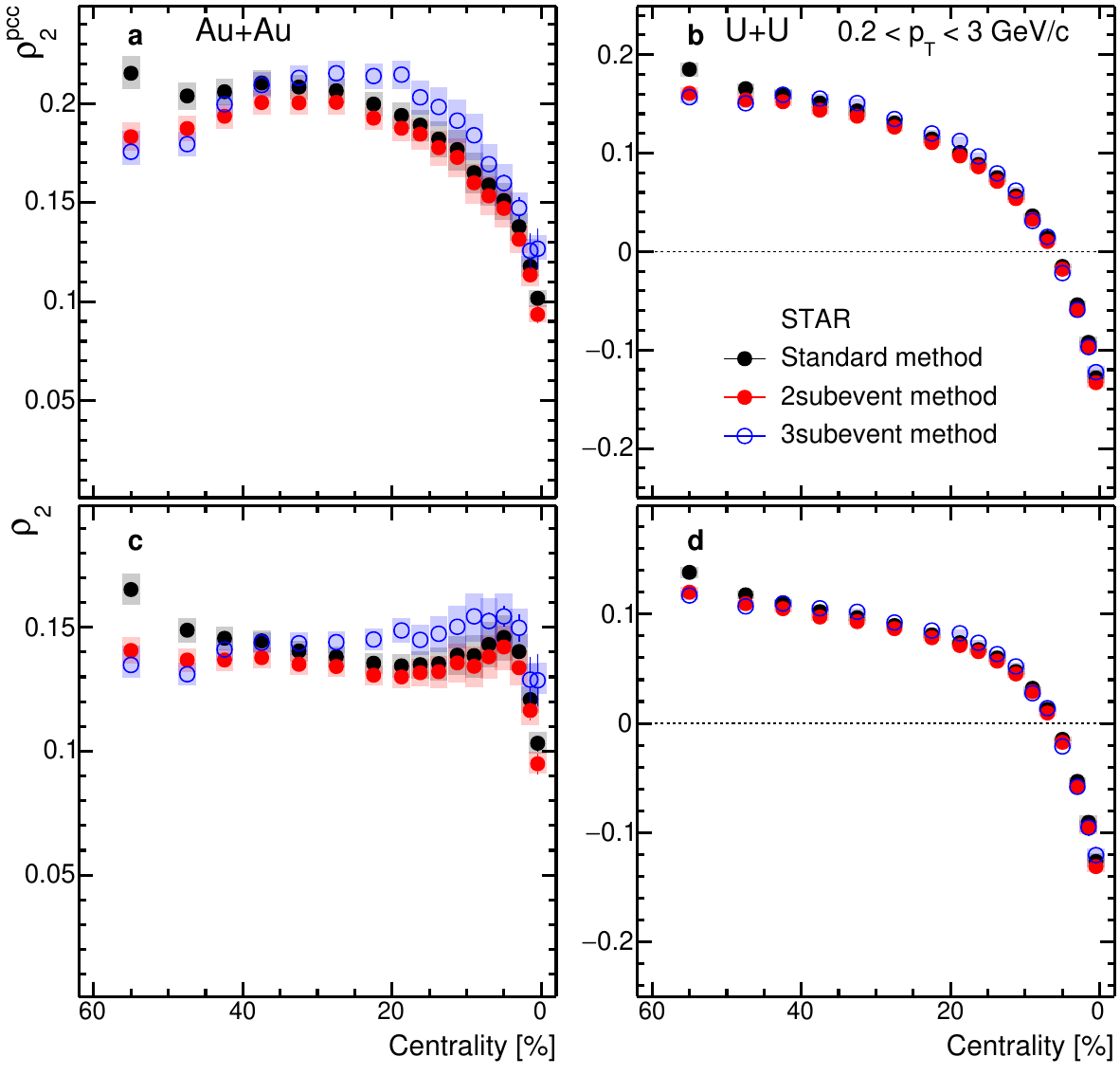}
\end{center}
\caption{\label{fig:3.6} Centrality dependence of $\rho_2^{\mathrm{pcc}}$ (top) and $\rho_2$ (bottom) in Au+Au (left) and U+U (right) collisions, compared between the standard (black), two-subevent (red) and three-subevent methods (blue).}
\end{figure}

\begin{figure}[!t]
\begin{center}
\includegraphics[width=1\linewidth]{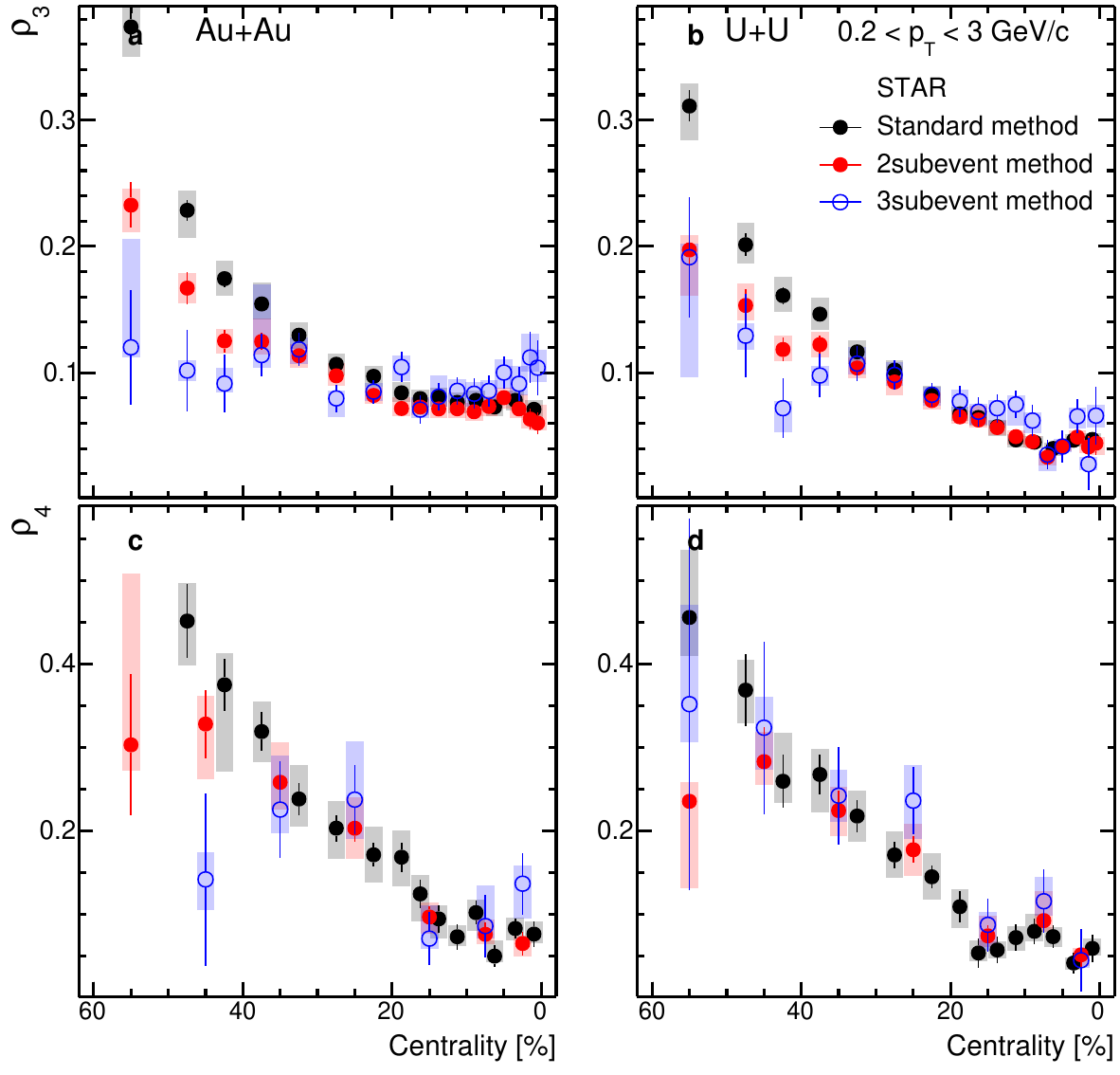}
\end{center}
\caption{\label{fig:3.7} The normalized quantities $\rho_3$ (top) and $\rho_4$ (bottom) via Eq.~\eqref{eq:5} in Au+Au (left) and U+U (right) collisions, compared between the standard (black), two-subevent (red) and three-subevent methods (blue).}
\end{figure}

All results so far utilize the two-subevent method. The impact of short-range non-flow correlations on $\lr{v_n^2\delta \pT}$ is quantified by comparing the standard, two-subevent, and three-subevent methods. These comparisons are shown in Fig.~\ref{fig:3.6} for $\rho_2^{\mathrm{pcc}}$ and $\rho_2$, and in Fig.~\ref{fig:3.7} for $\rho_3$ and $\rho_4$. Note that $\lr{v_n^2}$ and $\lr{(\delta \pT)^2}$ in these comparisons are always calculated with the two-subevent method, so method variations in $\rho_n$ originate solely from $\lr{v_n^2\delta \pT}$. 

\begin{figure}[t!]
\begin{center}
\includegraphics[width=1\linewidth]{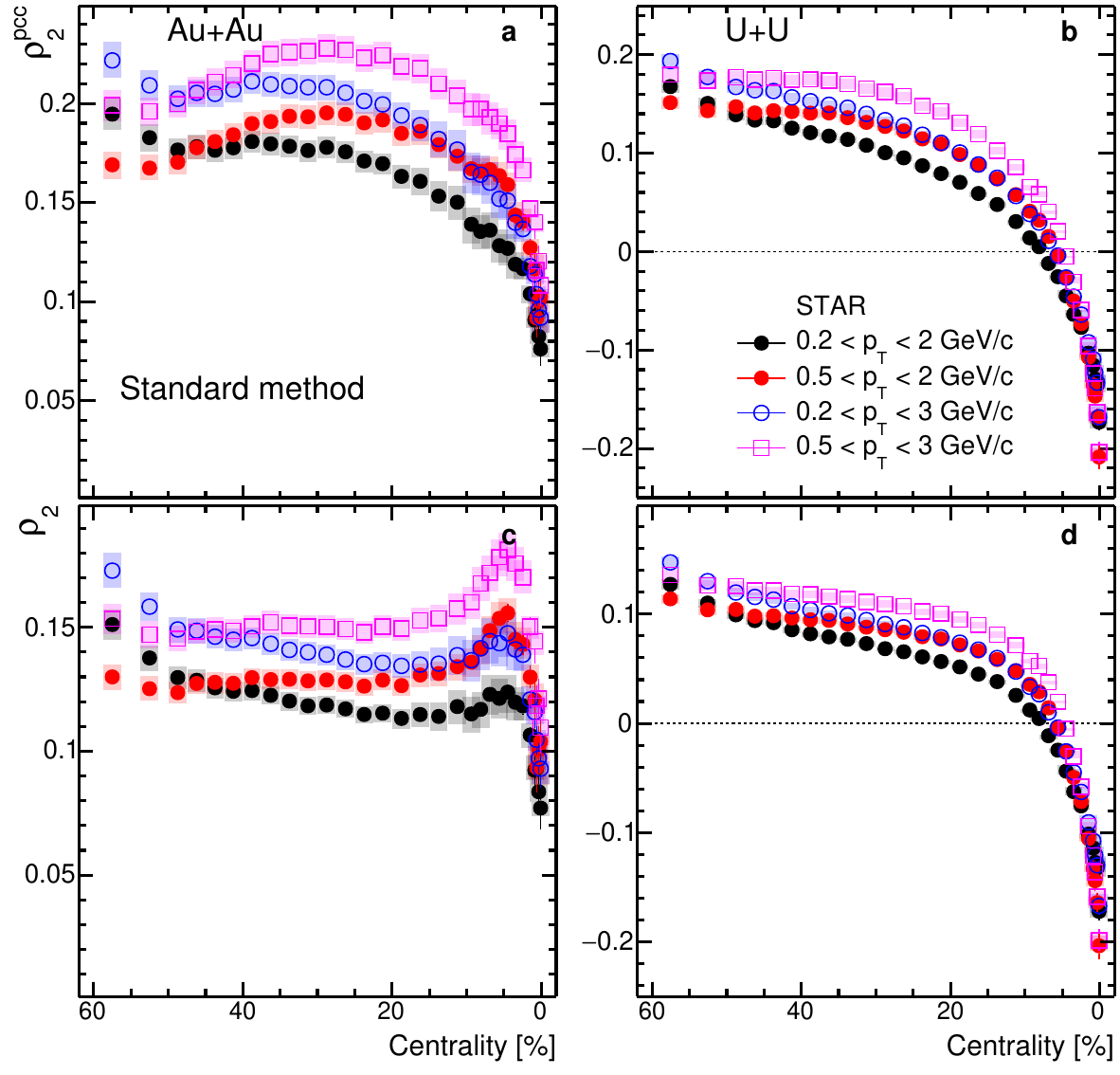}
\end{center}
\caption{\label{fig:3.8} Centrality dependence of normalized quantities $\rho_2^{\mathrm{pcc}}$ via Eq.~\eqref{eq:4} (top) and $\rho_2$ via Eq.~\eqref{eq:5} (bottom) in Au+Au (left) and U+U (right) collisions, calculated in four $\pT$ intervals.}
\end{figure}

In the 30--60\% centrality range, notable discrepancies between the standard and subevent methods reflect non-flow and longitudinal fluctuations. These differences are modest for $n=2$ but larger for $n=3$ and 4. Most variations occur between the standard and two-subevent methods, with generally smaller differences between the two- and three-subevent methods, except in peripheral collisions where statistical precision is poor.

Figure~\ref{fig:3.8} shows $\rho_2^{\mathrm{pcc}}$ and $\rho_2$ in four $\pT$ intervals, revealing significant variations (around 20\%) with the $\pT$ selection. These variations reflect incomplete cancellation of final-state effects between the numerators and denominators, resulting from differences in the expected hydrodynamic responses. Key features include: 1) $\rho_2^{\mathrm{pcc}}$ and $\rho_2$ values increase when the $\pT$ interval's lower or upper end is raised, similar to observations in the Xe+Xe and Pb+Pb collisions at the LHC~\cite{ATLAS:2022dov}. 2) Both observables decrease sharply in the UCC region in both systems.  This sharp decrease, also observed at the LHC, can be attributed to centrality smearing due to the finite centrality resolution associated with $\nchrec$~\cite{Jia:2021wbq}. 3) $\rho_2$ exhibits a peak around 5\% centrality in Au+Au, more pronounced when the lower end of the $\pT$ interval is raised. This increasing trend is potentially due to the moderately oblate deformation of Au nuclei~\cite{Jia:2021qyu}.

\begin{figure}[h!]
\begin{center}
\includegraphics[width=1\linewidth]{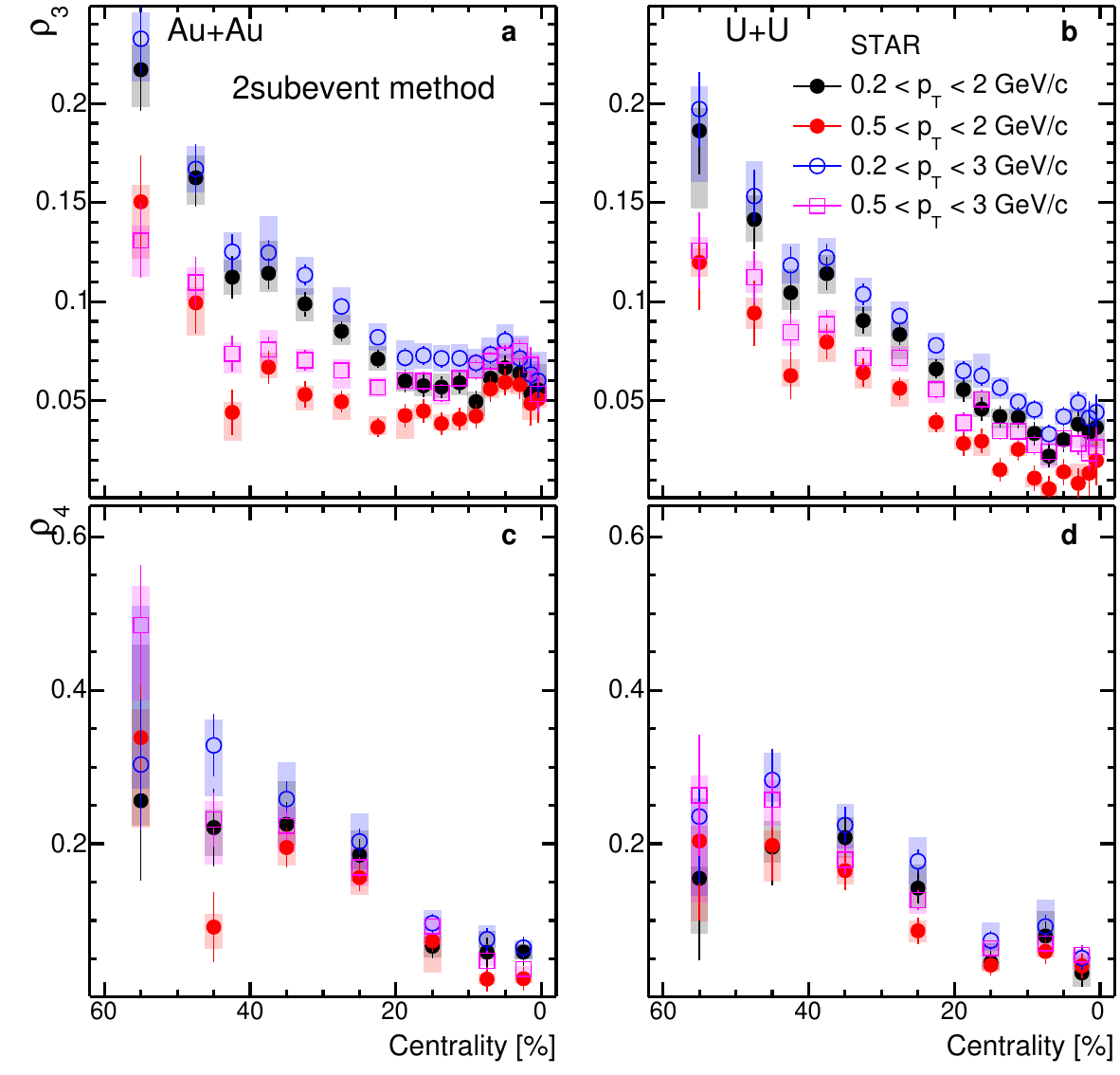}
\end{center}
\caption{\label{fig:3.9} Centrality dependence of normalized quantities $\rho_3$ (top) and $\rho_4$ (bottom) via Eq.~\eqref{eq:5} in Au+Au (left) and U+U (right) collisions, calculated in four $\pT$ intervals.}
\end{figure}

A similar study for $\rho_3$ and $\rho_4$ is presented in Fig.~\ref{fig:3.9}, based on the two-subevent method. The values of $\rho_3$ increase when raising the higher end of the $\pT$ interval, but decrease when raising the lower end, opposite to the behavior observed for $\rho_2$. In the UCC region, $\rho_3$ values in Au+Au collisions are systematically larger than in U+U collisions. The $\pT$ dependence of $\rho_4$ is similar to that for $\rho_3$, but statistical uncertainties are too large for quantitative evaluation. 

\subsection{Ratios between the two systems}\label{sec:3.3}

While normalized observables like $\rho_n$ reduce final-state effects, they do not fully cancel them, as indicated by their sensitivities to the $\pT$ selection. A more robust approach involves direct ratios of the same observable between U+U and Au+Au (Eq.~\eqref{eq:3}) at matching centrality. These ratios also reduce the influence of initial-state properties common to both U and Au, such as nucleon sizes, nucleon distances, and nucleon substructure fluctuations.

\begin{figure}[h!]
\begin{center}
\includegraphics[width=1\linewidth]{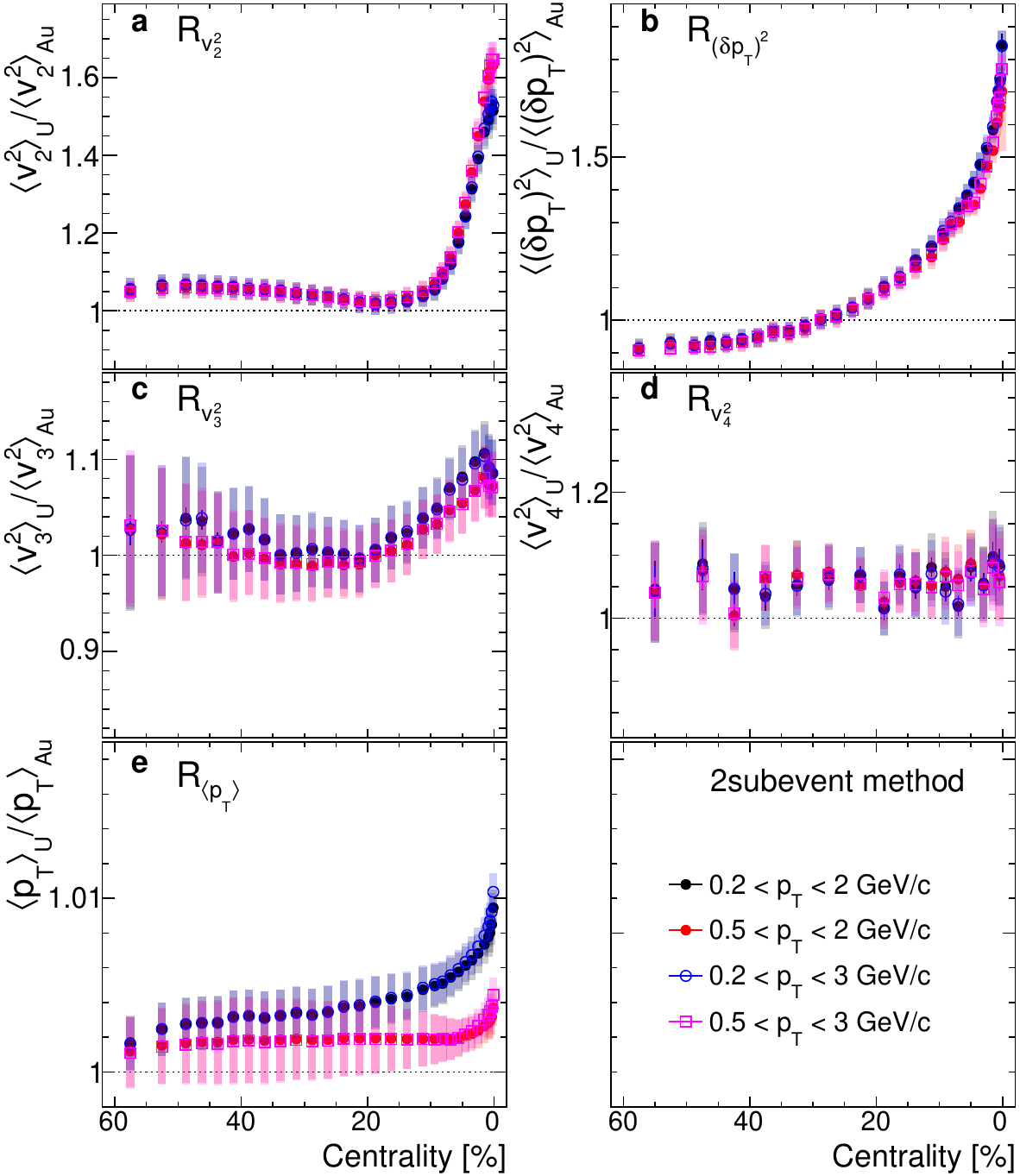}
\end{center}
\caption{\label{fig:3.10a} Centrality dependence of ratios of $\lr{v_2^2}$, $\lr{(\delta \pT)^2}$, $\lr{v_3^2}$, $\lr{v_4^2}$ and $\lr{\pT}$ between U+U and Au+Au collisions in four $\pT$ ranges.}
\end{figure}

Figure~\ref{fig:3.10a} presents ratios of one- and two-particle observables. Although individual observables, such as $\lr{v_n^2}$ and $\lr{(\delta \pT)^2}$, can change by over a factor of two across the four $\pT$ intervals, their ratios are remarkably $\pT$-independent, suggesting significant cancellation of final-state effects. In central collisions, $R_{v_2^2}$ and $R_{(\delta \pT)^2}$ show a dramatic increase due to uranium's larger deformation (Eq.~\eqref{eq:2}). The centrality range for this increase is narrow for $R_{v_2^2}$ (0--10\%) but much wider for $R_{(\delta \pT)^2}$ (0--30\%). This difference arises because $\lr{v_2^2}$ in non-central collisions is dominated by the elliptic shape of the overlap region, thereby diluting the impact of nuclear deformation. In contrast, $\lr{(\delta \pT)^2}$ is not affected by this average geometry~\cite{Jia:2022qgl}. The ratio $R_{v_3^2}$ shows a slight increase in central collisions, possibly from octupole deformation in Uranium, while $R_{v_4^2}$ shows no centrality-dependent modification.

\begin{figure}[h!]
\begin{center}
\includegraphics[width=1\linewidth]{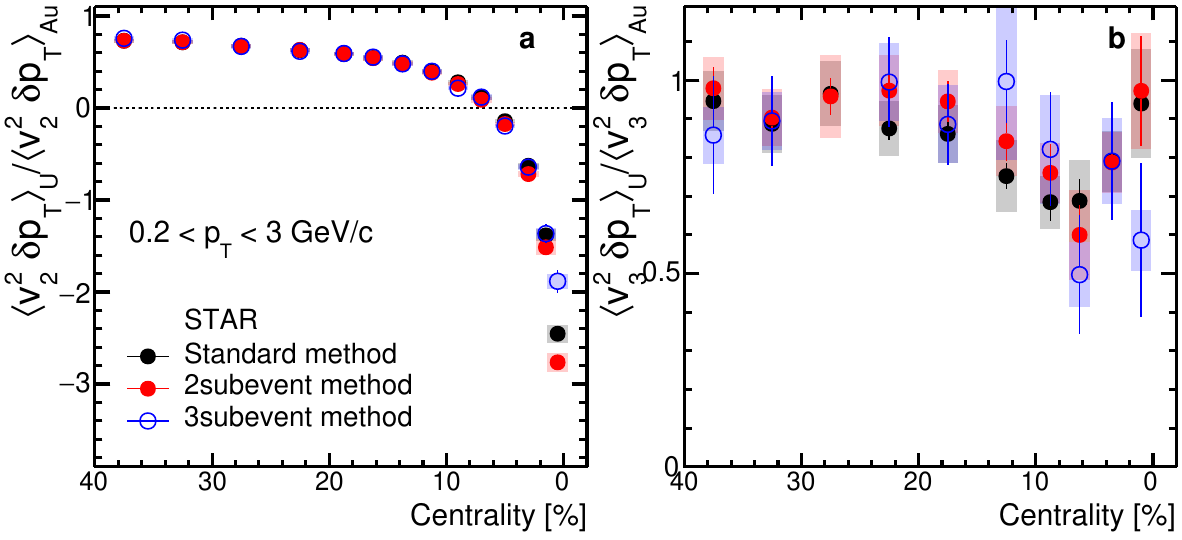}
\end{center}
\caption{\label{fig:3.10} Centrality dependence of ratios of $\lr{v_2^2\delta \pT}$ (left) and $\lr{v_3^2\delta \pT}$ (right) between U+U and Au+Au collisions, compared between different methods in $0.2<\pT<3$ GeV/$c$.}
\end{figure}

\begin{figure}[h!]
\begin{center}
\includegraphics[width=1\linewidth]{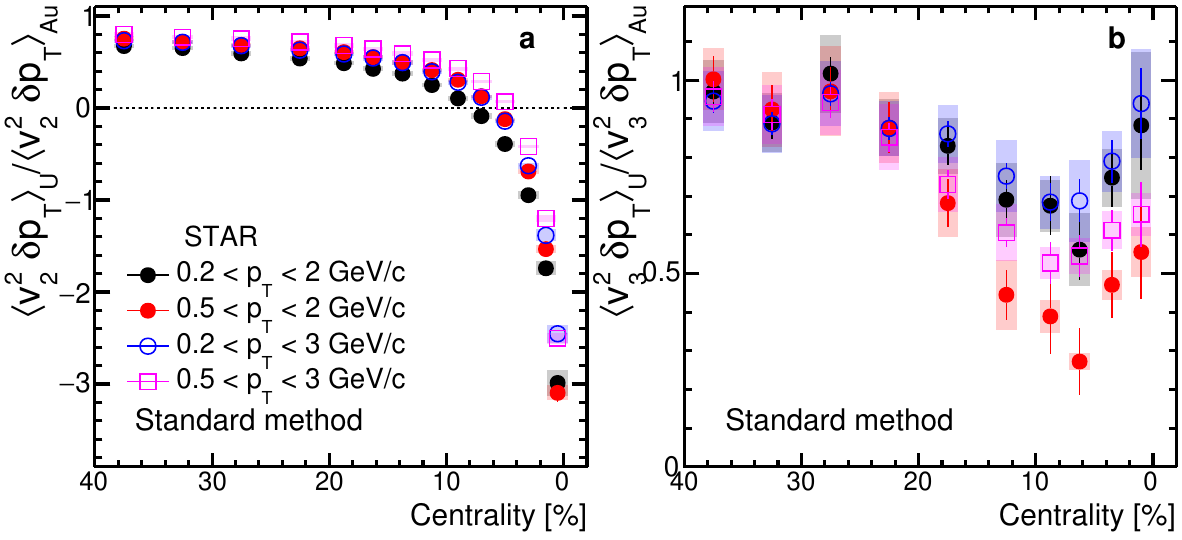}
\end{center}
\caption{\label{fig:3.11} Centrality dependence of ratios of $\lr{v_2^2\delta \pT}$ (left) and $\lr{v_3^2\delta \pT}$ (right) between U+U and Au+Au collisions, obtained for four $\pT$ intervals.}
\end{figure}

The $\lr{\pT}$ ratio increases slightly (about 1\%) toward central collisions, possibly due to a denser medium in U+U collisions and a bias towards tip-tip collisions in UCC~\cite{STAR:2015mki}. Note that $\pT$ interval choice affects $\lr{\pT}$, with increasing lower $\pT$ limits naturally lowering $\lr{\pT}$. This kinematic effect dilutes the increase in the UCC region, accounting for the split of the ratios into two groups based on the lower $\pT$ limit. Changing the upper limit has minimal impact due to the relatively fewer particles at high $\pT$. 

Figure~\ref{fig:3.10} shows the ratio of $\lr{v_n^2\delta \pT}$ between U+U and Au+Au for standard and subevent methods. Figure~\ref{fig:3.11} shows the same ratio for four $\pT$ intervals. In mid-central and peripheral collisions, $R_{v_n^2\delta \pT}$ values are independent of the method and $\pT$ interval, but modest dependencies appear in the UCC region. 

Many ratios in Fig.~\ref{fig:3.10a}--\ref{fig:3.11} deviate significantly from unity in the UCC region. As mentioned earlier and discussed in Sec.~\ref{sec:4}, these deviations can be used to constrain the quadrupole and higher-order deformation of $^{238}$U.

A challenge in this analysis is that observables are dominated by baseline contributions from spherical nuclei ($a_n$ terms in Eqs.~\eqref{eq:2}), which are strong functions of impact parameter and increase rapidly towards more peripheral collisions. In contrast, deformation-related terms ($b_n$ terms) are largest in central collision and decrease slowly towards peripheral collisions~\cite{Jia:2021tzt,Jia:2021qyu}. Therefore, the ratios and their centrality dependencies discussed above largely reflect impact parameter dependencies.

To remove the $a_n$ terms, which have similar values in U+U and Au+Au, we calculate baseline-subtracted ratios~\cite{Jia:2021qyu}:
\small{\begin{align}\nonumber
\rho_{2}^{\mathrm {sub}} &= \frac{\lr{v_2^2\delta \pT}_{\mathrm {U}}-\lr{v_2^2\delta \pT}_{\mathrm {Au}}}{(\lr{v_2^2}_{\mathrm {U}}-\lr{v_2^2}_{\mathrm {Au}})\sqrt{\lr{(\delta \pT)^2}_{\mathrm {U}}-\lr{(\delta \pT)^2}_{\mathrm {Au}}}}\;,\\\label{eq:12}
\rho_{2}^{\mathrm {pcc,sub}} &= \frac{\lr{v_2^2\delta \pT}_{\mathrm {U}}-\lr{v_2^2\delta \pT}_{\mathrm {Au}}}{\sqrt{\mathrm{var}(v_2^2)_{\mathrm {U}}-\mathrm{var}(v_2^2)_{\mathrm {Au}}}\sqrt{\lr{(\delta \pT)^2}_{\mathrm {U}}-\lr{(\delta \pT)^2}_{\mathrm {Au}}}}.
\end{align}}\normalsize

This difference $\Delta\mathcal{O} = \mathcal{O}_{\mathrm U}- \mathcal{O}_{\mathrm{Au}}$ cancels most spherical baseline contributions while isolating the deformation component:
\begin{align}\nonumber
\rho_{2}^{\mathrm {sub}} &\approx- \frac{b_3 \left(\cos(3\gamma_{\mathrm U}) \beta_{2\mathrm U}^3 - \cos(3\gamma_{\mathrm{Au}})\beta_{2\mathrm{Au}}^3\right)}{b_1\sqrt{b_2}(\beta_{2{\mathrm U}}^2-\beta_{2{\mathrm {Au}}}^2)^{3/2}}\\\label{eq:13}
&\hspace*{-0.4cm}\overset{\beta_{2\mathrm{Au}}\rightarrow0}{=\joinrel=\joinrel=\joinrel=} -\frac{b_3}{b_1\sqrt{b_2}}\cos(3\gamma_{\mathrm U})\;.
\end{align}
A liquid-drop model estimates the coefficients $-\frac{b_3}{b_1\sqrt{b_2}}=-\sqrt{10}/7$~\cite{Jia:2021qyu}. In this limit, $\rho_{2}^{\mathrm {sub}}=-0.76$ assuming $(\beta_{2\mathrm U}, \gamma_{\mathrm U})=(0.28,0^{\circ})$ and $(\beta_{2\mathrm{Au}}, \gamma_{\mathrm {Au}})=(0.14,45^{\circ})$. The second line of Eq.~\eqref{eq:13} shows that in the large $\beta_{2\mathrm U} $ limit and ignoring $\beta_{2\mathrm{Au}}$, $\rho_{2}^{\mathrm {sub}}$ depends only on uranium's triaxiality.

The subtraction method is valid when the baseline difference is much smaller than the deformation contribution. This holds for $\lr{v_2^2\delta \pT}$ over a wide centrality range, but only a limited range for $\lr{v_2^2}$ (0--5\%), $\sqrt{\mathrm{var}(v_2^2)}$ (0--10\%), and $\lr{(\delta \pT)^2}$ (0--20\%), see Fig.~\ref{fig:3.4}.
\begin{figure}[h!]
\begin{center}
\includegraphics[width=1\linewidth]{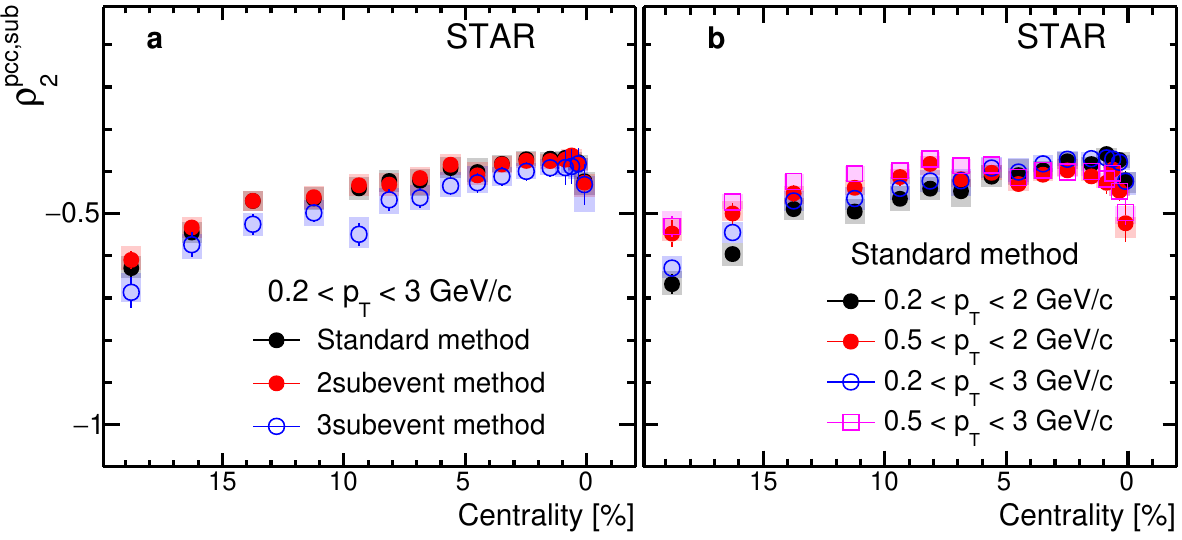}
\includegraphics[width=1\linewidth]{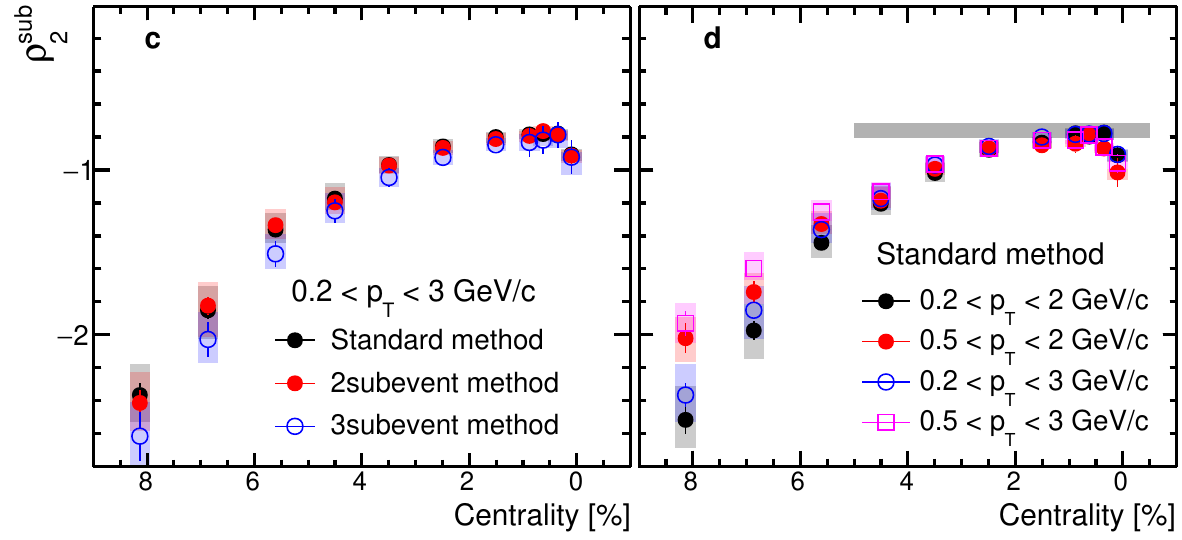}
\end{center}
\caption{\label{fig:3.12} Centrality dependence of the baseline-subtracted and normalized quantities $\rho_{2}^{\mathrm {pcc,sub}}$ (top) and  $\rho_{2}^{\mathrm {sub}}$ (bottom) defined via Eq.~\eqref{eq:12}. They are compared between different methods in $0.2<\pT<3$ GeV/$c$ (left) or four $\pT$ intervals (right). The thick grey line in the bottom right panel indicates the expectation from the liquid drop model estimation at -0.76 (see text).}
\end{figure}

The interpretation of $\rho_{2}^{\mathrm {pcc,sub}}$ is more complex due to a more involved form for $\mathrm{var}(v_2^2)$ derived from Eq.~\eqref{eq:9}. Assuming identical $a_1$, $b_1$, $c_1$ and $d_1=2/7b_1^2$ for the two collision systems:
\begin{align}\label{eq:14}
\Delta \mathrm{var}(v_2^2)  = b_1^2(\beta_{2\mathrm U}^2-\beta_{2\mathrm{Au}}^2)(2a_1/b_1+5/7(\beta_{2\mathrm U}^2+\beta_{2\mathrm{Au}}^2))\;,
\end{align}
we obtain: 
\begin{align}\nonumber
\rho_{2}^{\mathrm {pcc,sub}} &\approx  -\frac{b_3 \left(\cos(3\gamma_{\mathrm U}) \beta_{2\mathrm U}^3 - \cos(3\gamma_{\mathrm{Au}})\beta_{2\mathrm{Au}}^3\right)}{b_1\sqrt{b_2(2a_1/b_1+5/7(\beta_{2\mathrm U}^2+\beta_{2\mathrm{Au}}^2))}(\beta_{2{\mathrm U}}^2-\beta_{2{\mathrm{Au}}}^2)}\\\label{eq:15}
& \hspace*{-0.4cm}\overset{\beta_{2\mathrm{Au}}\rightarrow0}{=\joinrel=\joinrel=\joinrel=}-\frac{b_3\beta_{2{\mathrm U}}}{b_1\sqrt{b_2(2a_1/b_1+5/7\beta_{2{\mathrm U}}^2)}}\cos(3\gamma_{\mathrm U})\;.
\end{align}

The $a_1/b_1$ term, affected by the spherical baseline, reduces the magnitude of $\rho_{2}^{\mathrm {pcc,sub}}$ relative to $\rho_{2}^{\mathrm {sub}}$. Also, $\rho_{2}^{\mathrm {pcc,sub}}$ can be obtained over a wider centrality range because the centrality interval over which $\Delta\mathrm{var}(v_2^2)>0$ is wider than that for $\Delta\lr{v_2^2}>0$ (see Fig.~\ref{fig:3.4}). 

Figure~\ref{fig:3.12} shows $\rho_{2}^{\mathrm {pcc,sub}}$ and $\rho_{2}^{\mathrm {sub}}$ in central collisions. These observables are robust against variations in correlation methods and $\pT$ intervals. They are negative over a significantly wider centrality range than the original observables $\rho_{2}^{\mathrm {pcc}}$ and $\rho_{2}$. Furthermore, the values of $\rho_{2}^{\mathrm {sub}}$ in the UCC region are remarkably close to the analytical liquid-drop limit of -0.76. 

\section{Model comparisons}\label{sec:4}
\subsection{Comparisons with initial-state and hydrodynamic models for $v_n$--$\pT$ correlations}\label{sec:4.1}
The correlation between $v_n$ and $[\pT]$ arises from hydrodynamic responses to correlated fluctuations between elliptic eccentricity, $\varepsilon_2$, and the inverse of the transverse area, $d_{\perp}$. Most flow observables are influenced by both initial-state geometry and final-state QGP evolution.

{\bf Initial state model.} Models that consider only the initial state, such as the Glauber model, rely on linear response relations: $v_n\propto \varepsilon_n$ and $\delta \pT/\pT \propto \delta d_{\perp}/d_{\perp}$. These relations yield estimators for $\rho_2$ and $\rho_2^{\mathrm{pcc}}$ in terms of eccentricity and $d_{\perp}$~\cite{Schenke:2020uqq,Bozek:2020drh}:
\begin{align}\nonumber
\hat{\rho}_{2}&= \frac{\lr{\varepsilon_2^2\delta d_{\perp}/d_{\perp}}}{\lr{\varepsilon_2^2}\sqrt{\lr{(\delta  d_{\perp}/d_{\perp})^2}}}\\\label{eq:16}
\hat{\rho}_{2}^{\mathrm{pcc}}&= \frac{\lr{\varepsilon_2^2\delta d_{\perp}/d_{\perp}}}{\sqrt{\mathrm{var}(\varepsilon_2^2)}\sqrt{\lr{(\delta  d_{\perp}/d_{\perp})^2}}}\;.
\end{align}

Similarly, ratios between U+U and Au+Au, mimicking $R_{v_2^2\delta\pT}$ or $R_{\rho_2}$ (Eq.~\eqref{eq:3}), as well as the subtracted forms to mimic $\rho_{2}^{\mathrm {sub}}$ and $\rho_{2}^{\mathrm {pcc,sub}}$, can be formed.
\begin{align}\nonumber
\hat{\rho}_{2}^{\mathrm {sub}} &= \frac{\Delta\lr{\varepsilon_2^2\delta d_{\perp}/d_{\perp}}}{\Delta\lr{\varepsilon_2^2}\sqrt{\Delta\lr{(\delta  d_{\perp}/d_{\perp})^2}}}\\\label{eq:18}
\hat{\rho}_{2}^{\mathrm {pcc,sub}} &=\frac{\Delta\lr{\varepsilon_2^2\delta d_{\perp}/d_{\perp}}}{\sqrt{\Delta\mathrm{var}(\varepsilon_2^2)}\sqrt{\Delta\lr{(\delta  d_{\perp}/d_{\perp})^2}}}\;.
\end{align}
Here, $\Delta \mathcal{O}$ represents the difference of an observable $\mathcal{O}$ between U+U and Au+Au collisions. For this study, we use the standard quark Glauber model~\cite{Loizides:2016djv} with deformation parameters from Ref.~\cite{Jia:2021qyu}. These normalized quantities serve as estimators for the corresponding ratios of final-state observables and do not capture final-state effects in the data (Fig.~\ref{fig:3.8}).

\begin{figure*}[!htbp]
\begin{center}
\includegraphics[width=1\linewidth]{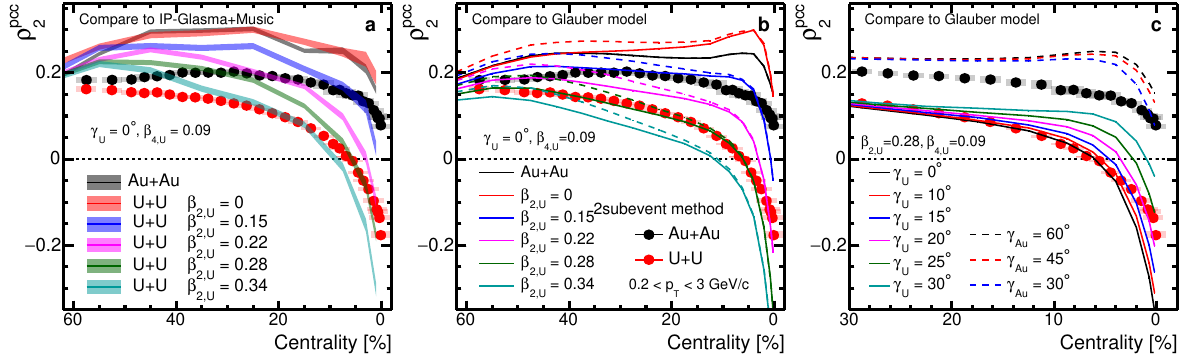}
\end{center}
\caption{\label{fig:4.1} $\rho_2^{\mathrm{pcc}}$ in Au+Au (black symbols) and U+U (red symbols) collisions from $0.2<\pT<3$ GeV/$c$ range, compared with IP-Glasma+MUSIC hydrodynamic model (left) and Glauber model calculation (middle) with $\gamma_{\mathrm{U}}=0^{\circ}$ and different $\beta_{2\mathrm{U}}$ values, and Glauber model calculation with $\beta_{2\mathrm{U}}=0.28$ and different $\gamma_{\mathrm{U}}$ and $\gamma_{\mathrm{Au}}$ (right). In all cases, $\beta_{2\mathrm{Au}}$ is fixed at 0.14, and $\beta_{4\mathrm{U}}=0.09$, except in the middle panel where $\beta_{4\mathrm{U}}=0$ (dashed lines) are also shown.}
\end{figure*}
\begin{figure*}[!htbp]
\begin{center}
\includegraphics[width=1\linewidth]{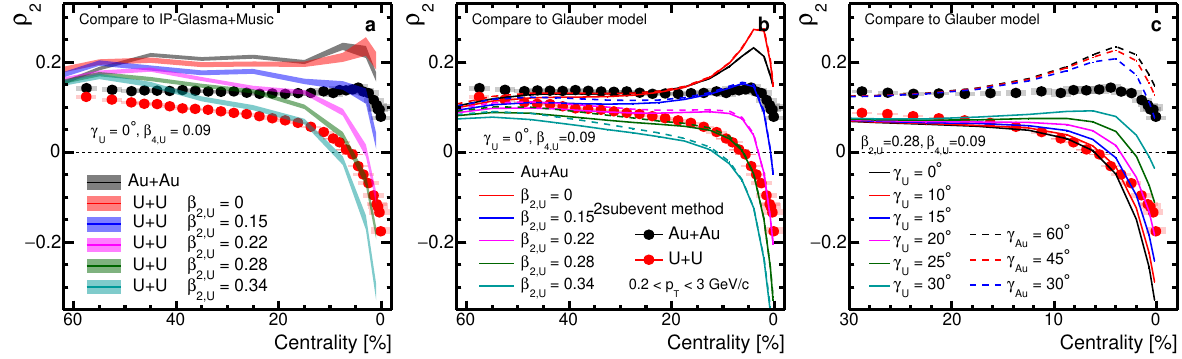}
\end{center}
\caption{\label{fig:4.2} Same as Fig.~\ref{fig:4.1} but for $\rho_2$.}
\end{figure*}

{\bf Full hydrodynamic simulations.} Final-state effects can be quantified using full event-by-event hydrodynamic model simulations with realistic initial conditions. We utilize the state-of-the-art IP-Glasma+MUSIC model~\cite{Schenke:2020uqq,Schenke:2020mbo}, which dynamically generates initial conditions from the gluon saturation model (IP-Glasma) and simulates full 3D viscous hydrodynamics via the MUSIC code. Final-state particles are generated using the Cooper-Fry prescription for hadronization, followed by UrQMD hadronic transport. This model is computationally expensive, so it can only be studied with limited statistics. 

{\bf Nuclear structure setup.} Table~\ref{tab:1} details the Woods-Saxon parameters, including nuclear deformation for Au and U, used in the Glauber and IP-Glasma+MUSIC models. Nucleon-nucleon inelastic cross-sections are 40.6 mb for U+U at 193 GeV and 42 mb for Au+Au at 200 GeV. Following Refs.~\cite{Bally:2023dxi,STAR:2024eky}, we set default $^{197}$Au parameters as $\beta_{2\mathrm{Au}}=0.14$ and $\gamma_{\mathrm{Au}}=45^{\circ}$. For $^{238}$U, known for its prolate deformation and moderate hexadecapole deformation from low-energy experiments and calculations, we can $\beta_{2\mathrm{U}}$ values from 0 to 0.34 in both models. We also vary $\beta_{4\mathrm{U}}$ (0 to 0.09), $\gamma_{\mathrm U}$ (0$^{\circ}$ to 30$^{\circ}$), and $\gamma_{\mathrm{Au}}$ (30$^{\circ}$ to 60$^{\circ}$) to test sensitivity to hexadecapole deformation and triaxiality. Variations of other structure parameters are only used in Section~\ref{sec:4.2} for quantitative estimation of deformation parameters. Approximately 50 million Glauber events are generated following Ref.~\cite{Jia:2021qyu} and 100,000--400,000 IP-Glasma+MUSIC events are generated for each parameter set, using the code provided by the authors of Refs.~\cite{Schenke:2020uqq,Schenke:2020mbo}. About 5\% of these generated events ends up in the UCC region. Each IP-Glasma+MUSIC event is oversampled at least 100 times to minimize non-flow and statistical fluctuations in the hadronic transport.

\begin{table}[!h]
\centering
\caption{\label{tab:1} Choices of Woods-Saxon parameters, including nuclear deformations, in the Glauber and IP-Glasma+MUSIC models. Default values are in bold; the rest are variations to constrain $(\beta_{2\mathrm{U}},\gamma_{\mathrm U})$ and their theoretical uncertainties, especially for the IP-Glasma+MUSIC model discussed in Section~\ref{sec:4.2}.}
\begin{tabular}{c|l|c|c|c|c|c}\hline 
           model        &species        & $R_0$(fm)           & a (fm)         & $\beta_2$        & $\beta_4$    & $\gamma$ ($^{\circ}$) \\\hline
\multirow{3}{*}{Glauber}&$^{197}$Au      & {\bf 6.62}       & {\bf 0.52}      &  {\bf 0.14}       & 0            & 30,{\bf 45},60 \\\cline{2-7}
                              &$^{238}$U & {\bf 6.81}       & {\bf 0.55}      &  0,0.15,0.22      & 0            & {\bf 0},10,15  \\
                              &         &                  &                 &{\bf 0.28},0.34    & {\bf 0.09}   & 20,25          \\\hline
\multirow{4}{*}{Hydro}  &$^{197}$Au      & {\bf 6.62},6.38  & {\bf 0.52}      & {\bf 0.14},0.12   & 0            & 37,{\bf 45},53  \\\cline{2-7}
                              &$^{238}$U & {\bf 6.81},7.07  & {\bf 0.55}      &  0,0.15           & 0            & {\bf 0}, 10    \\
                              &         &                  &                 &  0.22,0.25        & {\bf 0.09}   & 15, 20 \\
                              &         &                  &                 & {\bf 0.28},0.34   &              &        \\\hline
\end{tabular}
\end{table}

{\bf Comparison of $\rho_2^{\mathrm{pcc}}$ and $\rho_2$.} Figure~\ref{fig:4.1}a compares $\rho_2^{\mathrm{pcc}}$ in the $0.2<\pT<3$ GeV/$c$ range with IP-Glasma+MUSIC for varying $\beta_{2\mathrm{U}}$. While the model over-predicts data in peripheral collisions, prolate deformation significantly reduces $\rho_2$ in more central collisions. For $\beta_{2\mathrm{U}}=0.28$, the model predicts a zero-crossing around 7\% centrality, consistent with data. This zero-crossing point serves as an important anchor, as it is (by construction) independent of normalization and correlation methods, and weakly dependent on $\pT$ selections (Fig.~\ref{fig:3.8}).

Figure~\ref{fig:4.1}b shows the comparison with Glauber model estimators $\hat{\rho}_2^{\mathrm{pcc}}$. The magnitude and trend of the Glauber model are similar to the IP-Glasma+MUSIC model, indicating a dominance of initial conditions on the behavior of $\rho_2^{\mathrm{pcc}}$. The Glauber model also correctly predicts the zero-crossing location. Both models predict a steeper decrease in the UCC region than observed in the data, likely due to weaker centrality smearing in the models. Hexadecapole deformation mildly reduces $ \hat{\rho}_2^{\mathrm{pcc}} $ in non-central collisions.

Figure~\ref{fig:4.1}c investigates the impact of triaxiality. Data are compared to Glauber calculations for $\beta_{2\mathrm {U}}=0.28$ and $\beta_{2\mathrm {Au}}=0.14$ with varying triaxiality. Increasing $\gamma$ raises $\rho_2^{\mathrm{pcc}}$ in the 0--20\% centrality range in U+U, shifting the zero-crossing point towards more central collisions. Au+Au results are less sensitive to $\gamma$ due to the smaller $\beta_{2\mathrm{Au}}$. 

Figure~\ref{fig:4.2} presents the same comparison for $\rho_2$. The IP-Glasma+MUSIC model predicts a flat centrality dependence for Au+Au, consistent with data, though with a larger magnitude. For small $\beta_2$ values, both models (especially Glauber) predict a slight peak around 3\% centrality, more pronounced than in the data. 

We attribute the modest discrepancies between models and data for $\rho_2^{\mathrm{pcc}}$ and $\rho_2$ to residual final-state effects, not included in the Glauber model~\cite{Schenke:2020uqq,Schenke:2020mbo} and not fully captured by the IP-Glasma+MUSIC model~\cite{Schenke:2020uqq}. Example of such effects could be subleading flow~\cite{Mazeliauskas:2015efa} or non-linear effects between $\varepsilon_n$ and $v_n$ and between $\delta d_{\perp}$ and $\delta [\pT]$~\cite{Noronha-Hostler:2015dbi}.

{\bf Ratios Between U+U and Au+Au.} Incomplete cancellation of final-state effects is expected as hydrodynamic responses are observable-dependent and thus differ between numerators and denominators.  Ratios of the same quantity between U+U and Au+Au are expected to cancel most residual sensitivities and model dependencies in initial conditions, making them more sensitive to genuine nuclear structure effects.

\begin{figure}[t!]
\begin{center}
\includegraphics[width=1\linewidth]{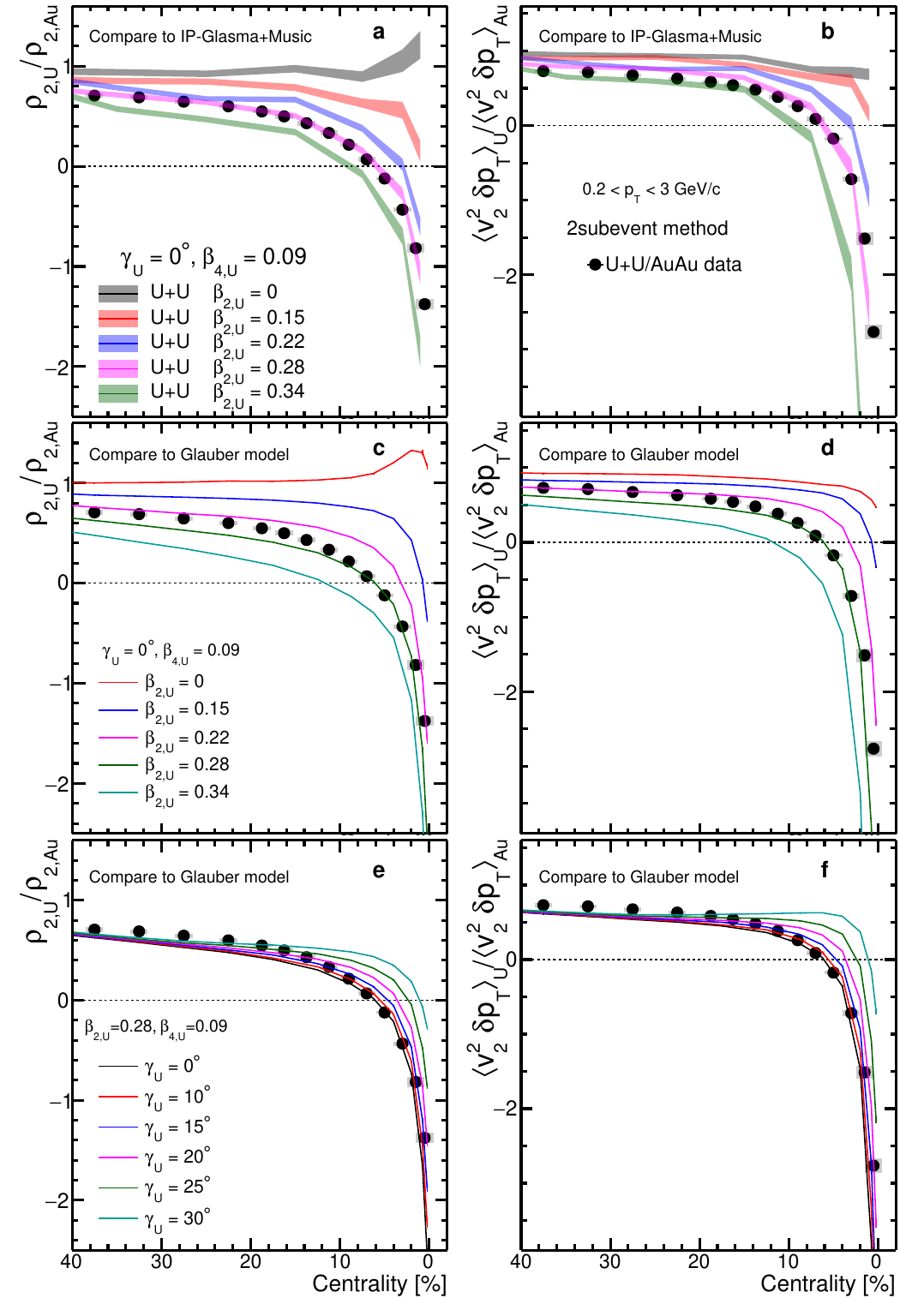}
\end{center}
\caption{\label{fig:4.3} Ratios of $\rho_2$ (left panels) and $\lr{v_2^2\delta \pT}$ (right panels) between U+U and Au+Au collisions, compared with the IP-Glasma+MUSIC (top row) and the Glauber model (middle row) with $\gamma_{\mathrm{U}}=0^{\circ}$ and different $\beta_{2\mathrm{U}}$ values, and Glauber model with $\beta_{2\mathrm{U}}=0.28$ and different $\gamma_{\mathrm{U}}$ and $\gamma_{\mathrm{Au}}$ (bottom row). In all cases, Au deformation parameters are fixed at $(\beta_{2\mathrm{Au}},\gamma_{\mathrm{Au}})= (0.14,45^{\circ})$, and $\beta_{4\mathrm{U}}=0.09$, except in middle row where $\beta_{4\mathrm{U}}=0$ (dashed lines) are also shown.}
\end{figure}

Figure~\ref{fig:4.3} compares ratios for $\rho_2$ and $\lr{v_2^2\delta \pT}$ with model calculations. IP-Glasma+MUSIC with $\beta_{2\mathrm{U}}=0.28$ (top row) agrees with data over a much wider centrality range than that in Figs.~\ref{fig:4.1} and \ref{fig:4.2}. The Glauber model (middle row) also shows much better agreement. Centrality smearing effects in the UCC region are expected to be similar between Au+Au and U+U collisions~\cite{Jia:2021wbq}, reducing their impact in ratios and leading to better model agreement compared to Figs.~\ref{fig:4.1} and \ref{fig:4.2}. The bottom row demonstrates the sensitivity of these ratios to $\gamma_{\mathrm{U}}$ in the Glauber model. A triaxiality of more than $15^{\circ}$ would visibly shift the zero-crossing point and cause strong deviations from data, also supporting a prolate shape for the $^{238}$U nucleus.

\begin{figure}[!htb!]
\begin{center}
\includegraphics[width=1\linewidth]{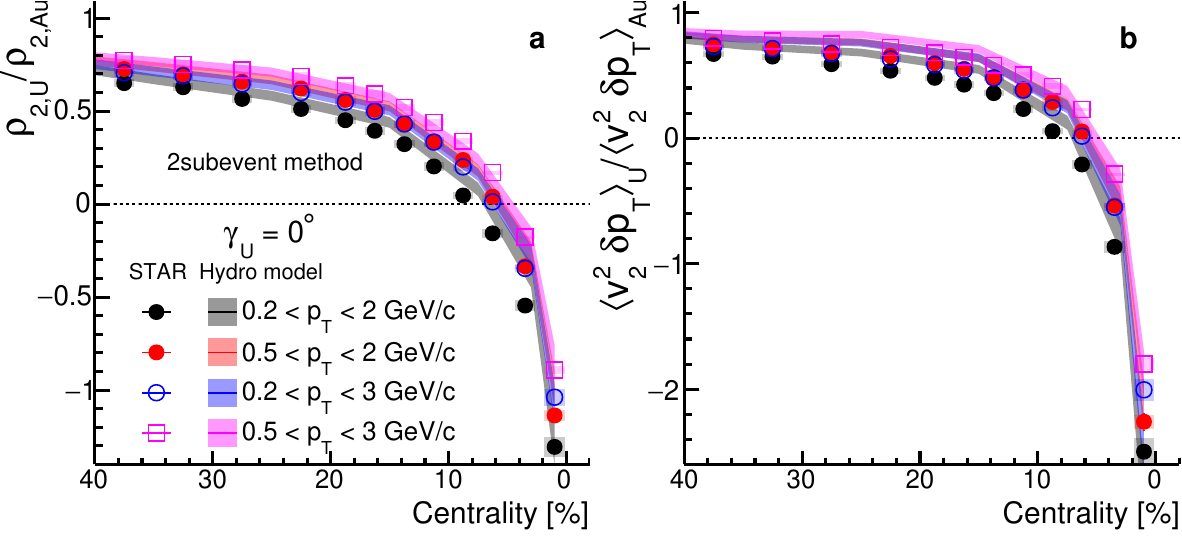}
\end{center}
\caption{\label{fig:4.4} Ratio of $\rho_2$ (left) and $\lr{v_2^2\delta \pT}$ (right) between U+U and Au+Au collisions in four $\pT$ intervals, compared with hydrodynamic model predictions. Data in the UCC region have been rebined for better comparison with the model.}
\end{figure}

We have observed that $R_{\rho_2}$ and $R_{v_2^2\delta \pT}$ have small residual dependencies on the choice of $\pT$ intervals. These $\pT$ dependencies, especially the relative ordering between different $\pT$ intervals, are well-reproduced by the IP-Glasma+MUSIC (Figure~\ref{fig:4.4}). This further validates the use of these ratios for constraining uranium deformation.  

\begin{figure}[h!]
\begin{center}
\includegraphics[width=1\linewidth]{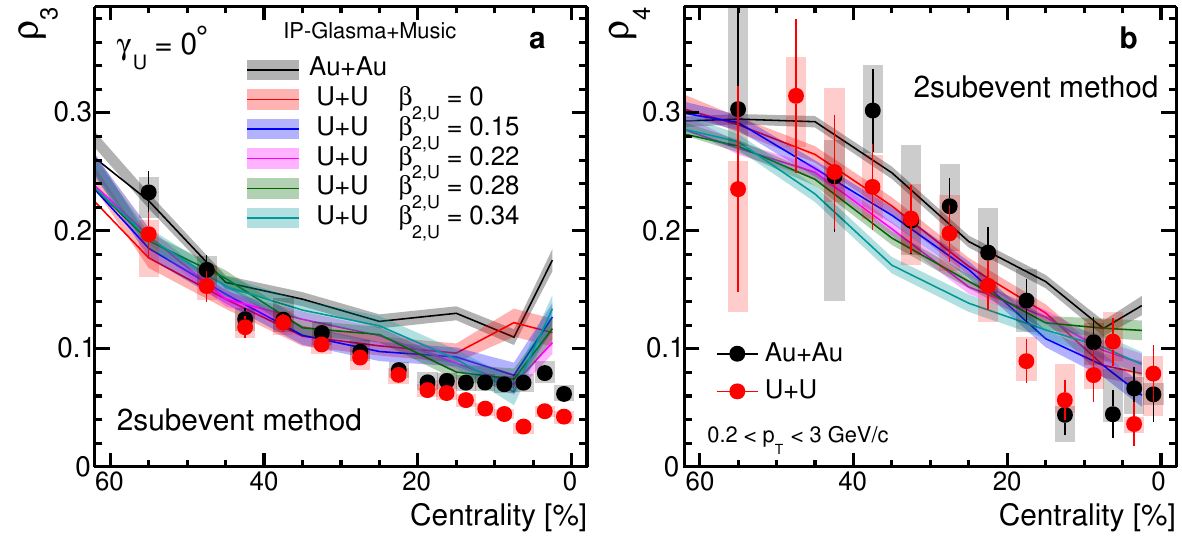}
\end{center}
\caption{\label{fig:4.5} Centrality dependence of $\rho_3$ (left) and $\rho_4$ (right) from U+U and Au+Au collisions in $0.2<\pT<3$ GeV/$c$, compared with hydrodynamic model predictions.}
\end{figure}

Figure~\ref{fig:4.5} show IP-Glasma+MUSIC predictions for $\rho_3$ and $\rho_4$. The model provides a qualitative description of the trends in the data.

\begin{figure}[h!]
\begin{center}
\includegraphics[width=1\linewidth]{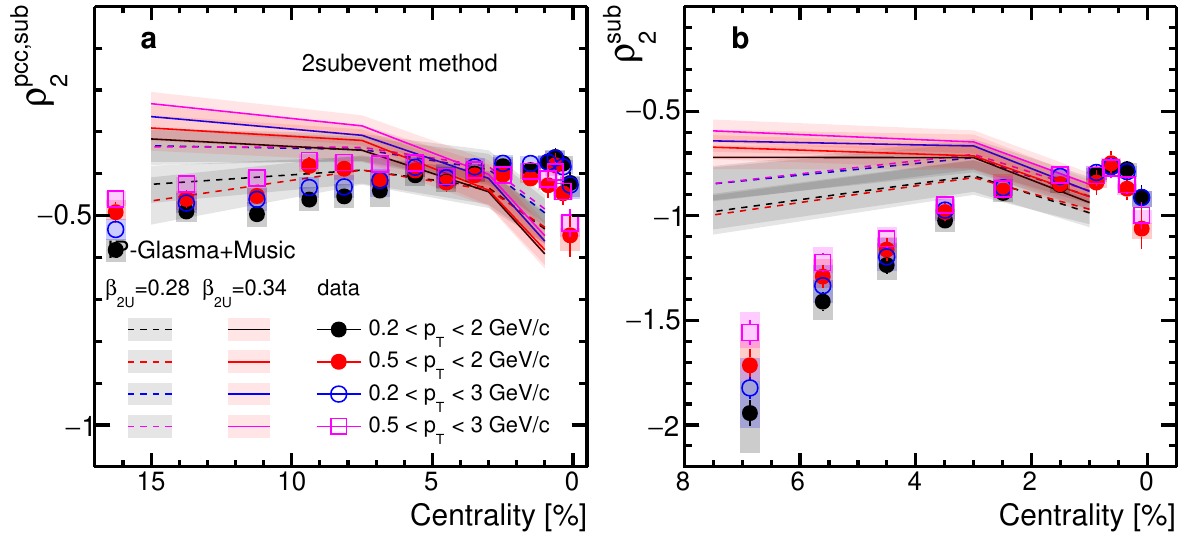}
\end{center}
\caption{\label{fig:4.6} Centrality dependence of $\rho_{2}^{\mathrm {pcc,sub}}$ (left) and $\rho_{2}^{\mathrm {sub}}$ (right) defined via Eq.~\eqref{eq:10} in four $\pT$ ranges, compared with IP-Glasma+MUSIC calculations with $\gamma_{\mathrm{U}}=0^{\circ}$ and $\beta_{2\mathrm{U}}=0.28$ (dashed lines) and $\beta_{2\mathrm{U}}=0.34$ (solid lines).}
\end{figure}

\begin{figure}[h!]
\begin{center}
\includegraphics[width=1\linewidth]{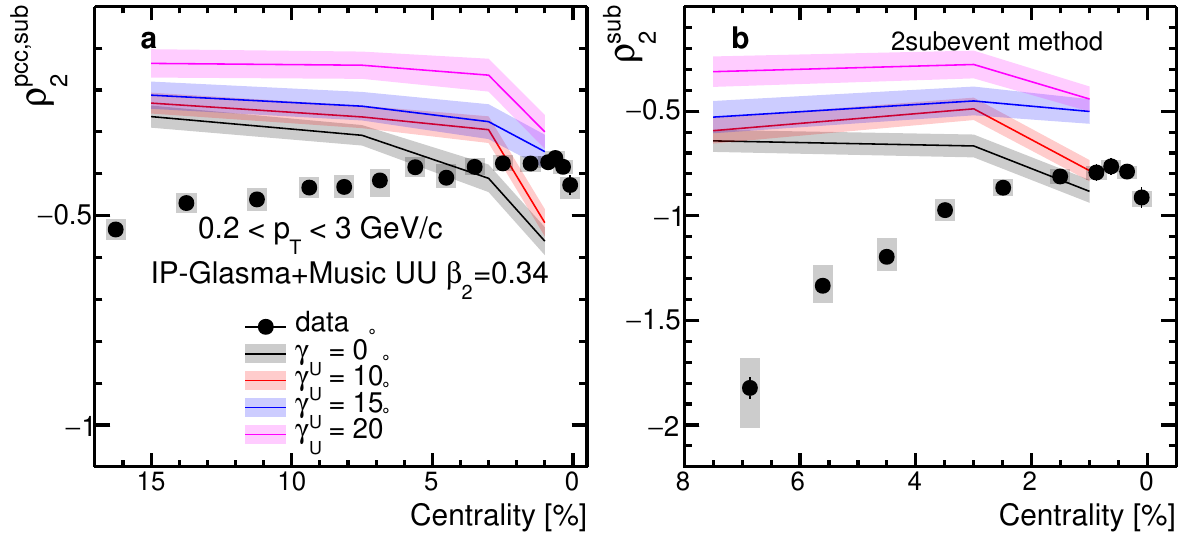}
\end{center}
\caption{\label{fig:4.6b} Centrality dependence of $\rho_{2}^{\mathrm {pcc,sub}}$ (left) and $\rho_{2}^{\mathrm {sub}}$ (right) defined via Eq.~\eqref{eq:10}. Compared with IP-Glasma+MUSIC calculations for $\beta_{2\mathrm{U}}=0.34$ and four $\gamma_{\mathrm{U}}$ values.}
\end{figure}

\begin{figure}[h!]
\begin{center}
\includegraphics[width=1\linewidth]{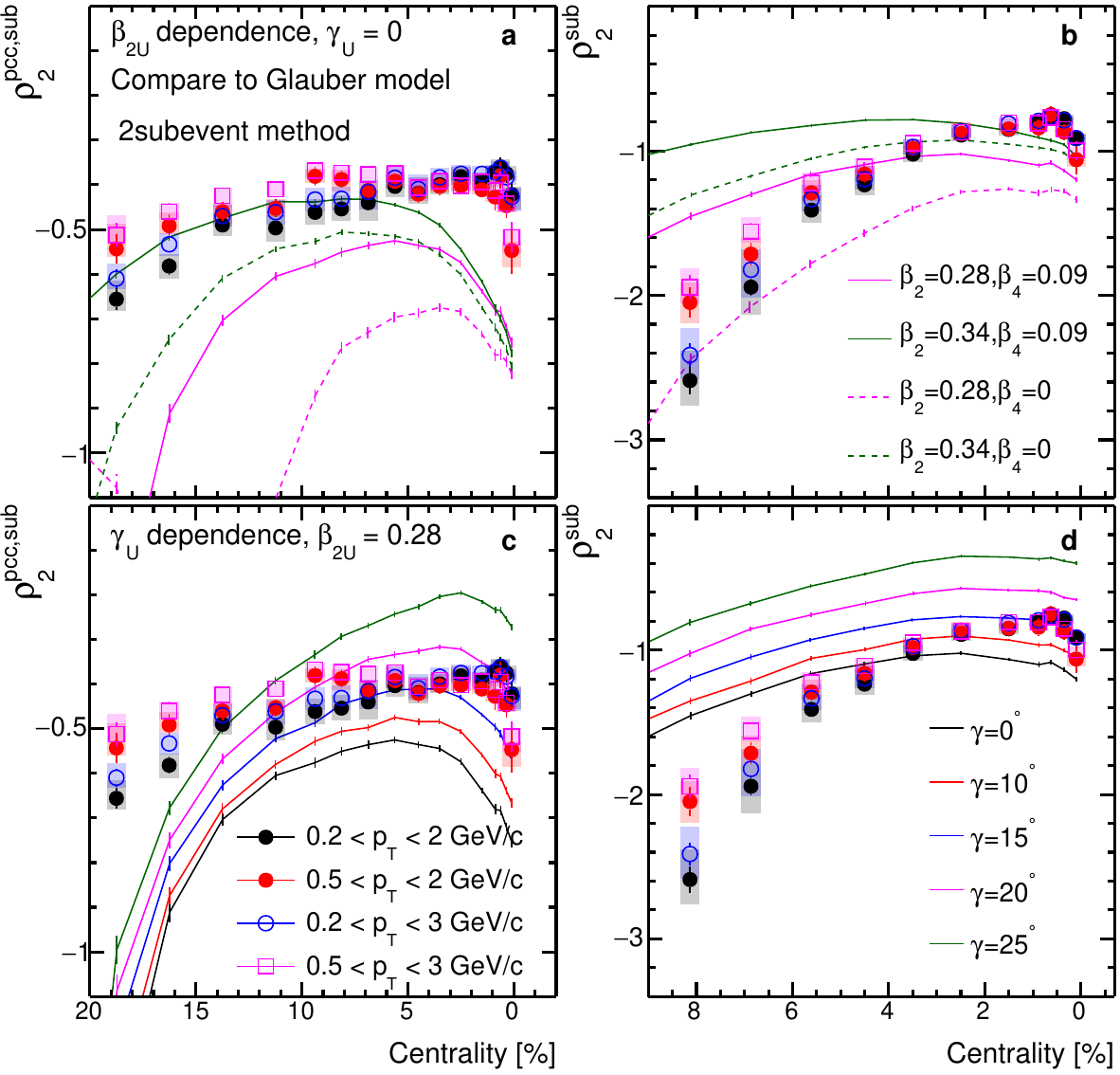}
\end{center}
\caption{\label{fig:4.7} Centrality dependence of $\rho_{2}^{\mathrm {pcc,sub}}$ (left column) and $\rho_{2}^{\mathrm {sub}}$ (right column) via Eq.~\eqref{eq:10} in four $\pT$ intervals. Compared with the Glauber model calculations for $\beta_{2\mathrm{U}}=0.28$ and 0.34 and $\beta_{4\mathrm{U}}=0$ and 0.09 (top panels) and $\beta_{2\mathrm{U}}=0.28$ and $\beta_{4\mathrm{U}}=0.09$ with different $\gamma_{\mathrm{U}}$ values (bottom panels). Glauber model calculations have no $\pT$ dependence by construction.}
\end{figure}

{\bf Comparison for observables obtained via the subtraction method.} Section\ref{sec:3.3} introduced subtracted observables ($\rho_2^{\mathrm{sub}}$, $\rho_2^{\mathrm{pcc,sub}}$) to remove spherical baseline contributions and expose the impact of triaxiality. Figures~\ref{fig:4.6} and \ref{fig:4.6b} compare these subtracted quantities with IP-Glasma+MUSIC, and Fig.~\ref{fig:4.7} shows the comparison with Glauber model.


For sufficiently large $\beta_{2\mathrm{U}}$ ($\gtrsim0.28$), $\rho_{2}^{\mathrm {pcc,sub}}$ and $\rho_{2}^{\mathrm {sub}}$ in the UCC region should approach a limiting value dependent only on $\gamma_{\mathrm{U}}$ (Eqs.~\eqref{eq:13} and \eqref{eq:15}). IP-Glasma+MUSIC predictions for $\beta_{2\mathrm{U}}=0.28$ and 0.34 indeed converge in the 0--5\% centrality range (Fig.~\ref{fig:4.6}). Deviations in less central collisions are expected due to incomplete subtraction of spherical baseline, unrelated to deformation. Increasing $\gamma_{\mathrm{U}}$ also increase these observables (Fig.~\ref{fig:4.6b}); a relatively small triaxiality ($\gamma_{\mathrm{U}}\lesssim 15^{\circ}$) is preferred by the data.

This limiting behavior can be studied with better statistical precision in the Glauber model (Fig.~\ref{fig:4.7}). The top panels compare data with calculation for different $\beta_{2\mathrm{U}}$ and $\beta_{4\mathrm{U}}$ with fixed $\gamma_{\mathrm{U}}=0^{\circ}$. Glauber predictions for $\beta_{2\mathrm{U}}=0.28$ and 0.34 are very close, but slightly lower than data for $\rho_{2}^{\mathrm {pcc,sub}}$. Furthermore, the model values slightly increase with $\beta_{4\mathrm{U}}$.

Bottom panels of Fig.~\ref{fig:4.7} show the effect of varying $\gamma_{\mathrm{U}}$ with $\beta_{2\mathrm{U}}$ fixed at 0.28. Larger $\gamma_{\mathrm{U}}$ increases $\rho_{2}^{\mathrm {pcc,sub}}$ and $\rho_{2}^{\mathrm {sub}}$, with $\gamma_{\mathrm{U}}=15^{\circ}$ achieving the best agreement with data in the ultra-central region.

\begin{figure}[htbp]
\begin{center}
\includegraphics[width=1\linewidth]{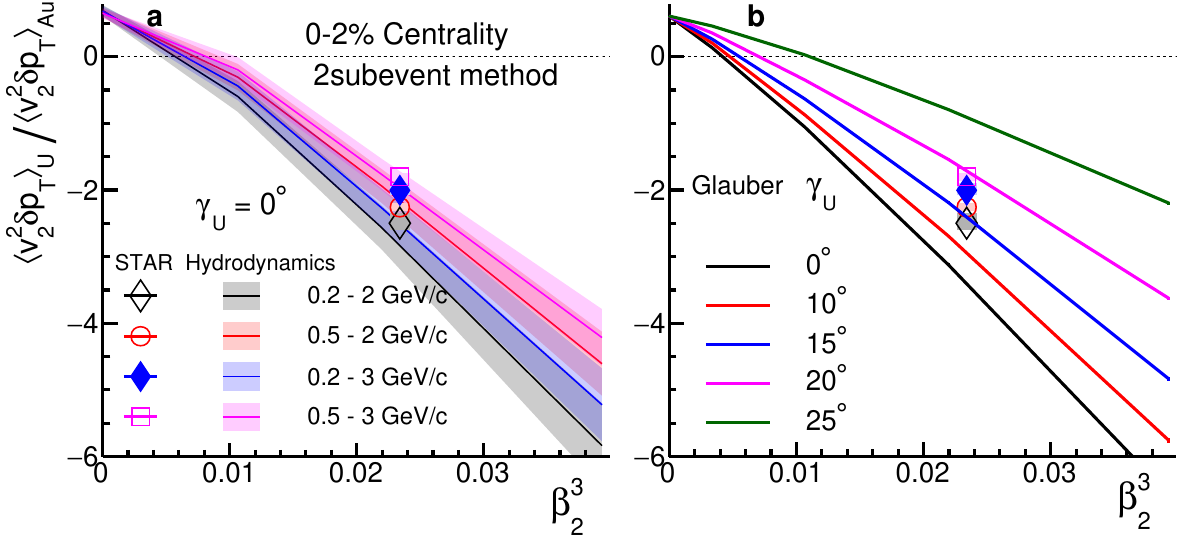}
\includegraphics[width=1\linewidth]{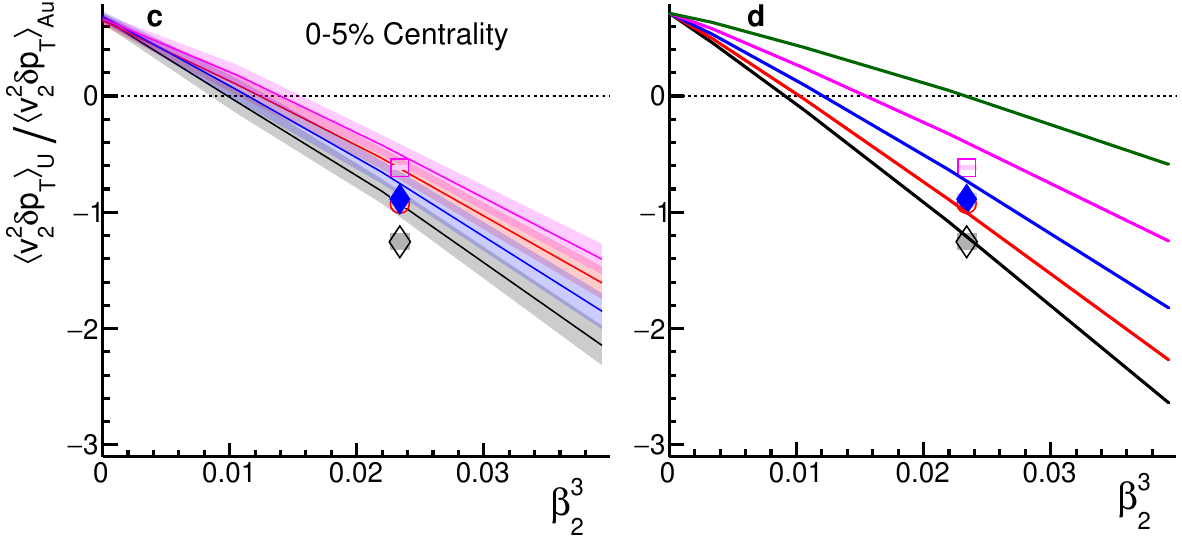}
\end{center}
\caption{\label{fig:4.8}  Ratio of covariance $\lr{v_2^2\delta\pT}_{\rm UU}/\lr{v_2^2\delta\pT}_{\rm AuAu}$ measured in 0--2\% (top row) and 0--5\% (bottom row) most central collisions for four $\pT$ intervals, placed at $\beta_{2,\mathrm{U}}^3=0.28^3=0.022$. Data are compared with the hydrodynamic model assuming $\gamma_{\mathrm{U}}=0$ (left column) and the Glauber model with different $\gamma_{\mathrm{U}}$ (right column). Uncertainties in the hydrodynamic model are correlated between different $\pT$ intervals.}
\end{figure}

{\bf Role of $\beta_{2\mathrm{U}}$ and $\gamma_{\mathrm{U}}$ in ultra-central collisions.} Data in the UCC region demonstrate the strongest sensitivity to nuclear deformation. Figure~\ref{fig:4.8} show the $\beta_{2\mathrm{U}}^3$ dependence of $R_{v_2^2\delta p_T}$ in 0--2\% and 0--5\% centrality bins across four $\pT$ intervals. Note that the influence of deformation is stronger in 0--2\% centrality, while the wider 0--5\% interval reduces centrality smearing effects. Experimental data are plotted at $\beta_{2,\mathrm{U}}^3=0.28^3=0.022$ and compared with model calculations. IP-Glasma+MUSIC predictions confirm a nearly linear dependence on $\beta_2^3$, consistent with Eq.~\eqref{eq:2}. Matching data to the model yields average values of $\beta_{2\mathrm{U}}\approx 0.272$--0.289 (0--2\% centrality) and $\beta_{2\mathrm{U}}\approx 0.291$--0.306 (0--5\% centrality) for $\gamma_{\mathrm{U}}=0^{\circ}$.
\begin{figure}[htbp]
\begin{center}
\includegraphics[width=1\linewidth]{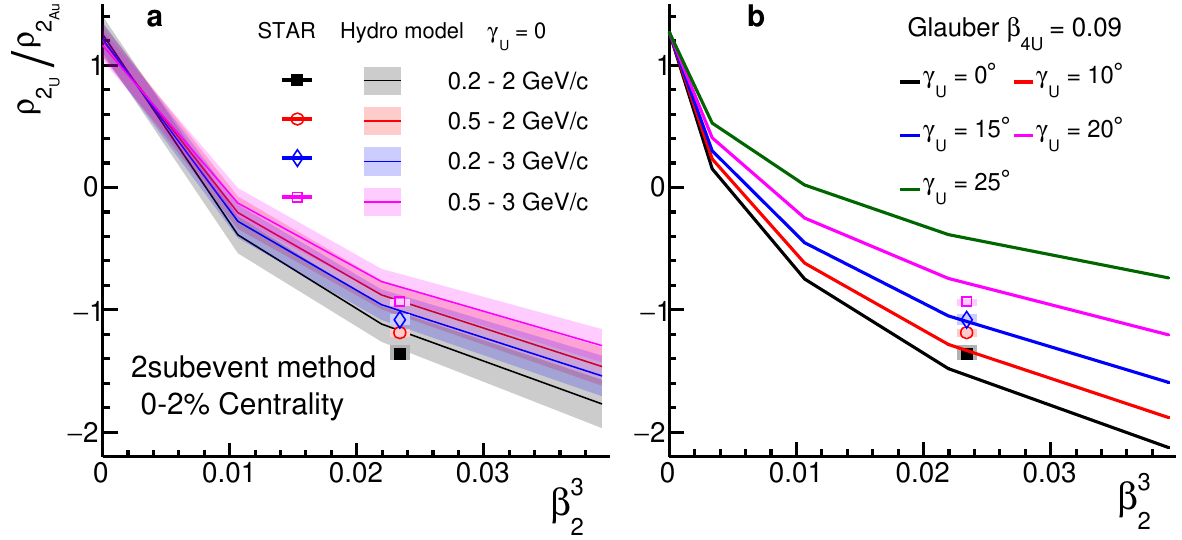}
\includegraphics[width=1\linewidth]{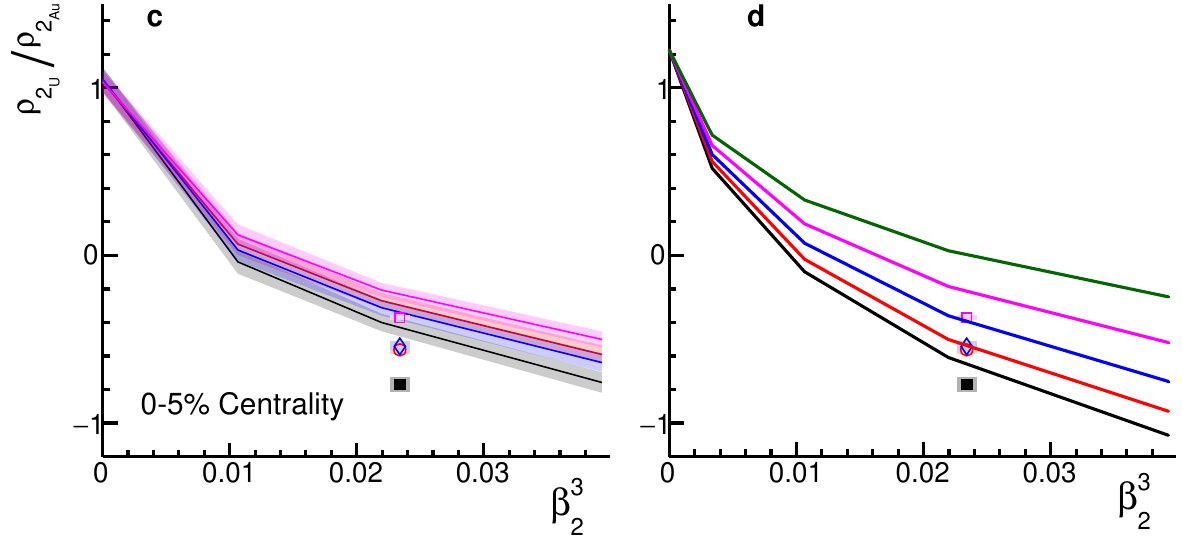}
\end{center}
\caption{\label{fig:4.9}  Ratio of covariance $\rho_{2,\rm UU}/\rho_{2,\rm AuAu}$ measured in 0--2\% (top row) and 0--5\% (bottom row) most central collisions for four $\pT$ intervals, placed at $\beta_{2,\mathrm{U}}^3=0.28^3=0.022$. Data are compared with the hydrodynamics model assuming $\gamma_{\mathrm{U}}=0$ (left column) and the Glauber model with different $\gamma_{\mathrm{U}}$ (right column).  Due to oversampling, statistical uncertainties in the model are dominated by fluctuations in initial configurations and are therefore correlated across different $\pT$ intervals.}
\end{figure}

The negative data values in Fig.~\ref{fig:4.8} require a positive $\cos(3\gamma)$, implying $\gamma_{\mathrm {U}}<30^{\circ}$. Since the Glauber model shows similar agreement to the full hydrodynamic model for $R_{v_2^2\delta \pT}$, it can provide high-precision estimates of $\beta_{2\mathrm U}$ and $\gamma_{\mathrm U}$. Right panels of Fig.~\ref{fig:4.8} compared the $\beta_2^{3}$ dependence of $R_{v_2^2\delta \pT}$ data with Glauber model's prediction for different $\gamma_{\rm U}$ values. The calculations exhibit a linear dependence on $\beta_2^3$, independent of the $\gamma$.  As the Glauber model does not simulate $\pT$ dependence, we can only extract a broad constraint on $\gamma_{\rm U}$. The extracted values are $\gamma_{\mathrm{U}}\lesssim 17^{\circ}$ (0--2\% centrality) and $\gamma_{\mathrm{U}}\lesssim 13^{\circ}$ (0--5\% centrality) assuming $\beta_{2\mathrm{U}}=0.28$.

The ratio of $\rho_2$ can also constrain nuclear deformation. However, this observable does not have a simple linear dependence on $\cos(3\gamma)\beta_2^3$ (Fig.~\ref{fig:4.9}), due to additional $\beta_2^2$ dependence introduced by $\lr{v_2^2}$ and $\lr{v_2^2\delta \pT}$ in the denominator of $\rho_2$ (see Eq.~\eqref{eq:13}). Therefore, the ratio of $\rho_2$ is not as robust for extracting nuclear deformations.

\begin{figure*}[!htb!]
\begin{center}
\includegraphics[width=0.75\linewidth]{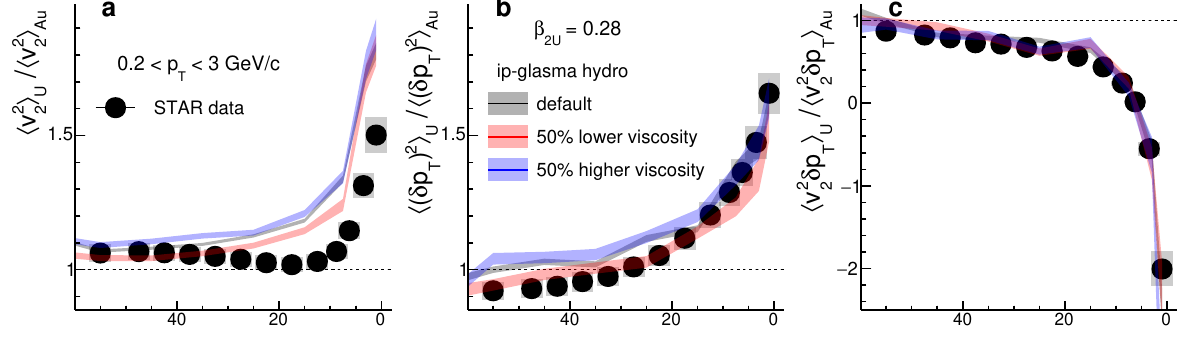}
\includegraphics[width=0.75\linewidth]{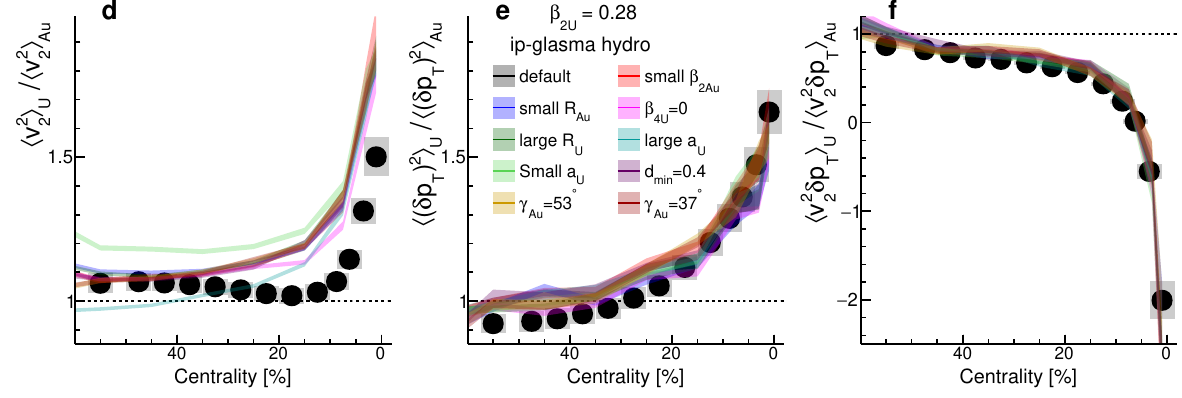}
\end{center}
\caption{\label{fig:4.10} IP-Glasma+MUSIC model prediction of the ratios of $\lr{v_2^2}$ (left column), $\lr{(\delta \pT)^2}$ (middle column), and $\lr{v_2^2\delta \pT}$ (right column) between U+U and Au+Au collisions in $0.2<\pT<3$ GeV/$c$, calculated for varying shear and bulk viscosities (top row) and varying nuclear structure parameters (bottom row), compared with the data. Checks consistent with the default calculation within their statistical uncertainties are not included in the model uncertainties in Fig.~\ref{fig:4.12}.}
\end{figure*}

\subsection{Quantitative constraints on Uranium deformation parameters}\label{sec:4.2}
Having explored nuclear deformation's influence on flow observables and established that U+U to Au+Au ratios exhibit the clearest signatures of deformation, we now perform a more quantitative, simultaneous extraction of $\beta_{2\mathrm{U}}$ and  $\gamma_{\mathrm{U}}$. This is accomplished by using three observables $R_{v_2^2}$, $R_{(\delta\pT)^2}$, and $R_{v_2^2\delta\pT}$ together. If we ignore the modest deformation in the Au nucleus, Eq.~\eqref{eq:2} suggests:
\begin{align}\nonumber
R_{v_2^2}         &\approx \frac{a_{\rm 1,U}}{a_{\rm 1,Au}} \left(1+\frac{b_{1}}{a_{1}}\beta_{\rm 2}^2 \right)_{\rm U}\;,\\\nonumber
R_{(\delta \pT)^2} &\approx \frac{a_{\rm 2,U}}{a_{\rm 2,Au}}\left(1+\frac{b_{2}}{a_{2}}\beta_{\rm 2}^2 \right)_{\rm U}\;,\\\label{eq:19}
R_{v_2^2\delta \pT}&\approx \frac{a_{\rm 3,U}}{a_{\rm 3,Au}}\left(1-\frac{b_{3}}{a_{3}}\beta_{\rm 2}^3\cos(3\gamma) \right)_{\rm U}\;
\end{align}

In the absence of deformation, these ratios approach $a_{n,\rm U}/a_{n,\rm Au}$, which are less than unity in UCC where $a_{n}$ are dominated by random fluctuations and are larger in Au+Au collisions. Following Ref.~\cite{STAR:2024eky}, we consider additional uncertainties for these ratios:
\begin{itemize}
\item {\bf Non-flow estimate:} Non-flow estimates, representing physical correlations, are included in the measured data ratios as model-dependent contributions (see Section~\ref{sec:2.3}).

\item {\bf Impact of viscosity:} Varying QGP shear and bulk viscosities by $\pm50$\% in IP-Glasma+MUSIC shows that while individual flow observables can change significantly, their ratios in U+U and Au+Au remain relatively stable (top panels of Fig.~\ref{fig:4.10}). Half the maximum variation is included in the model uncertainty.

\item {\bf Impact of other structure parameters:} To derive uncertainties on $\beta_{2\mathrm{U}}$ and $\gamma_{\mathrm{U}}$, other Woods–Saxon inputs in Table~\ref{tab:1} are varied: 1) Au deformation parameters are varied within $\beta_{2\mathrm{Au}}\approx 0.12$--0.14 and $\gamma_{\mathrm{Au}}\approx 37$--$53^{\circ}$ according to Refs.~\cite{Ryssens:2023fkv,Bally:2023dxi}. 2) The role of high-order deformation is checked by varying $\beta_{4\mathrm{U}}$ from 0 to 0.09. 3) The assumed nuclear radius $R_0$ and nuclear skin $a$ are varied. 4) Parameters common to both collision systems, like the minimum inter-nucleon distance in nuclei $d_{\mathrm {min}}$ are also varied. The impact of these variations on ratios of flow observables is shown in the bottom panels of Fig.~\ref{fig:4.10}. Dependencies on these structural parameters are small in the most central collisions.
\end{itemize}

\begin{figure*}[!htb!]
\begin{center}
\includegraphics[width=0.75\linewidth]{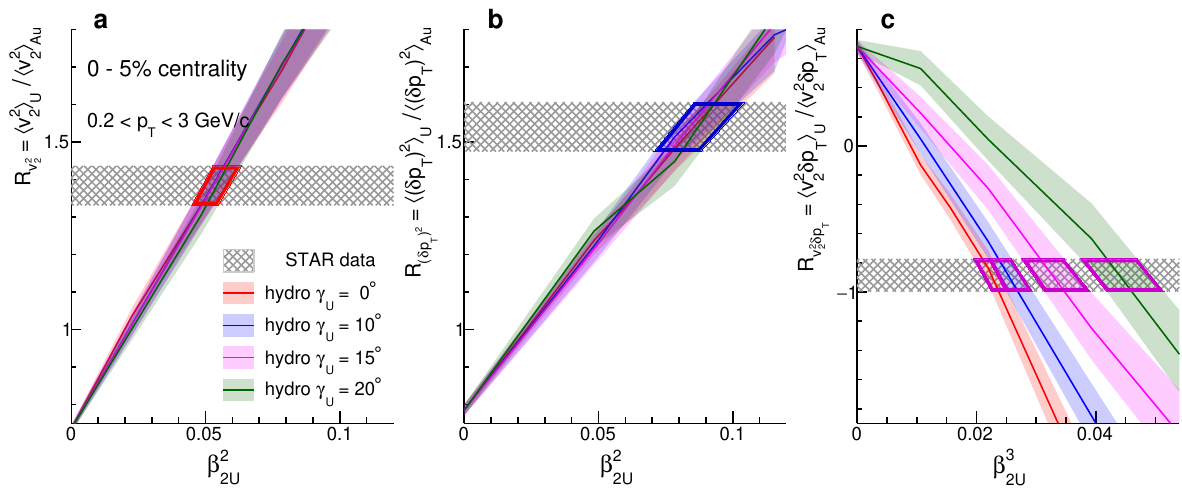}
\end{center}
\caption{\label{fig:4.12} Ratio values in 0--5\% most central collisions for $0.2<\pT<3$ GeV/$c$ (hatch bands) for $\lr{v_2^2}$ (\textbf{a}), $\lr{(\delta\pT)^2}$ (\textbf{b}), and $\lr{v_2^2\delta\pT}$ (\textbf{c}) are compared with model calculations as a function of $\beta_{2\mathrm{U}}^2$ or $\beta_{2\mathrm{U}}^3$ for four $\gamma_{\mathrm{U}}$ values. Colored quadrilaterals delineate the allowed ranges of $\beta_{2\mathrm{U}}^2$ or $\beta_{2\mathrm{U}}^3$ from this data-model comparison.}
\end{figure*}

Figure~\ref{fig:4.10} shows that IP-Glasma+MUSIC generally matches central data for $R_{(\delta \pT)^2}$ and $R_{v_2^2\delta \pT}$ but consistently overestimates $R_{v_2^2}$ in 0--30\% centrality. This overestimation for $R_{v_2^2}$ may stem from a stronger sensitivity to centrality smearing, other structural parameters (such as nuclear radius and skin depth)~\cite{Jia:2022qgl}, and possible longitudinal flow decorrelations~\cite{Jia:2024xvl}. Thus, $R_{v_2^2}$ comparison is considered to set a lower bound for $\beta_{2\mathrm{U}}$ in this model.

To minimize various confounding geometry effects, we focus on 0--5\% ultra-central collisions, where deformation signals dominate. Figure~\ref{fig:4.12} compares the 0.2--3 GeV/$c$ data with IP-Glasma+MUSIC curves, generated by varying $\beta_{2\mathrm{U}}$ (0 to 0.34) and $\gamma_{\mathrm{U}}$ (0$^\circ$ to 20$^\circ$). Model uncertainties incorporate both final-state viscosities and initial-parameter scans are shown in Fig.~\ref{fig:4.10}. Calculated $R_{v_2^2}$ and $R_{(\delta \pT)^2}$ change linearly with $\beta_{2\mathrm{U}}^2$, while $R_{v_2^2\delta \pT}$ follows a $\beta_{2\mathrm{U}}^3\cos(3\gamma_{\mathrm{U}})$ trend, remarkably consistent with Eq.~\eqref{eq:19}~\footnote{However, the predicted $R_{(\delta \pT)^2}$ values display a small dependence on $\gamma_{\mathrm{U}}$ when $\gamma_{\mathrm{U}}>$30$^\circ$.} Intersections between data and model delineate the preferred ranges of $\beta_{2\mathrm{U}}$. For $R_{v_2^2\delta \pT}$, the favored $\beta_{2\mathrm{U}}$ range varies with $\gamma_{\mathrm{U}}$. 

\begin{figure}[!htb!]
\begin{center}
\includegraphics[width=1.0\linewidth]{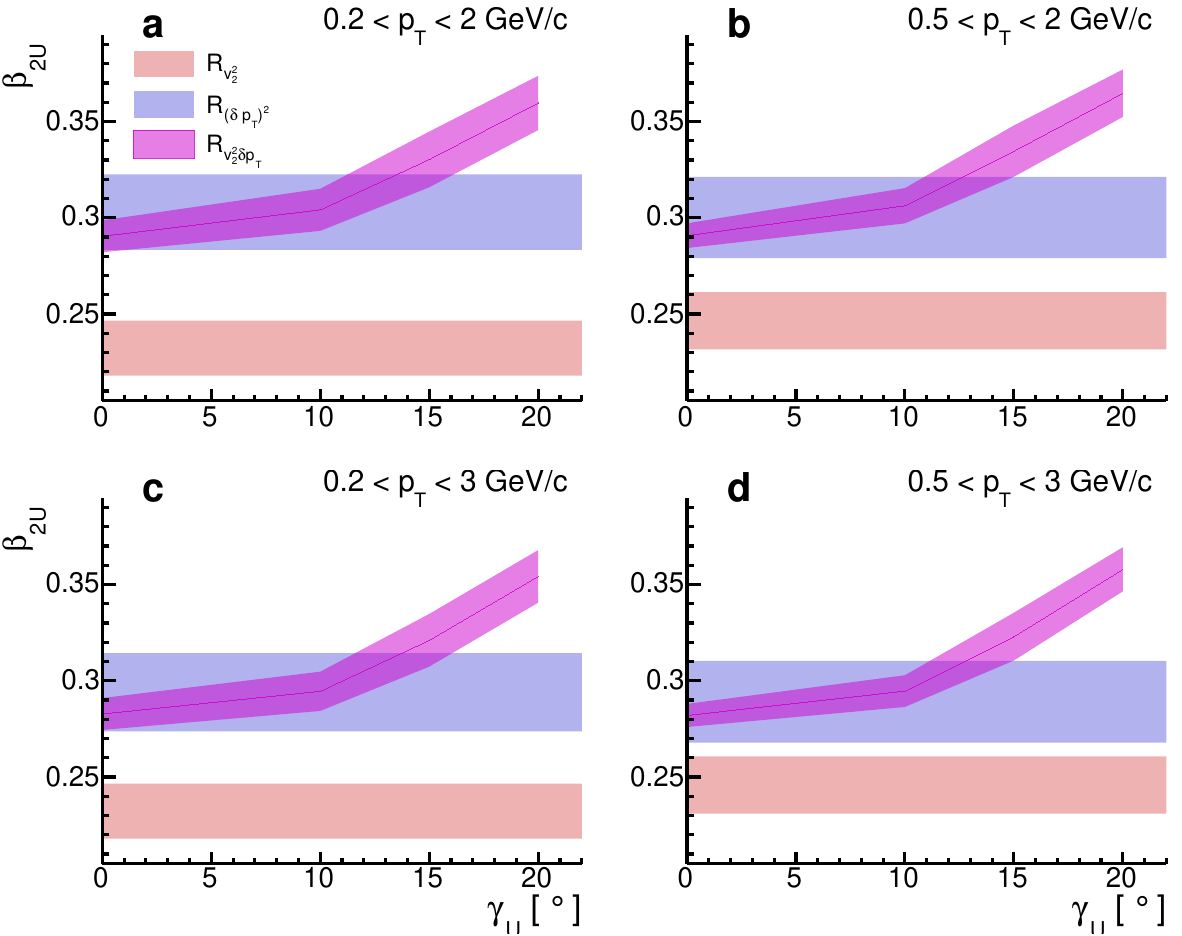}
\end{center}
\caption{\label{fig:4.13} Constraints on $\beta_{2\mathrm{U}}$ and $\gamma_{\mathrm{U}}$ from the three ratios, obtained for four $\pT$ ranges.} 
\end{figure}

\begin{table}[!h]
\centering
\caption{\label{tab:2} One standard deviation constraint on the uranium shape parameters from the three ratios in 0--5\% centrality for four $\pT$ intervals. Results averaged over the four $\pT$ ranges are also displayed.}
\begin{tabular}{c|c|c}\hline 
$\pT$ range           &   $R_{(\delta \pT)^2}$ and $R_{v_2^2\delta \pT}$ combined &  $R_{v_2^2}$   \\
 (GeV/$c$)            &   $\beta_{2\mathrm{U}}$, $\gamma_{\mathrm{U}} (^{\circ})$   &  $\beta_{2\mathrm{U}}$  \\\hline
0.2--2                &  $0.305 \pm 0.014$, $8.6 \pm 4.8$  & $0.232 \pm 0.014$ \\
0.5--2                &  $0.305 \pm 0.014$, $8.2 \pm 4.6$  & $0.247 \pm 0.017$ \\
0.2--3                &  $0.297 \pm 0.014$, $8.5 \pm 4.8$  & $0.234 \pm 0.014$ \\
0.5--3                &  $0.296 \pm 0.013$, $8.1 \pm 4.6$  & $0.246 \pm 0.017$ \\\hline
combined              &  $0.300 \pm 0.016$, $8.3 \pm 4.7$  & $0.240 \pm 0.018$ \\\hline
\end{tabular}
\end{table}

We repeat this comparison for all four $\pT$ ranges. Figure~\ref{fig:4.13} shows the extracted best-fit regions of ($\beta_{2\mathrm{U}},\gamma_{\mathrm{U}}$). $R_{v_2^2}$ sets the lower bound of $\beta_{2\mathrm{U}}$, while $R_{(\delta \pT)^2}$ and $R_{v_2^2\delta \pT}$ further constrain the upper range and triaxiality. A pseudo-experiment procedure combines uncertainties from $R_{(\delta \pT)^2}$ and $R_{v_2^2\delta \pT}$, assuming independent Gaussian probability density functions:
\begin{align}\label{eq:20}
P(\beta_{2\mathrm U},\gamma_{\mathrm U}) \propto \exp\left(-\frac{(\beta_{2\mathrm U}-\bar{\beta}_a)^2}{2 \sigma_a^2}-\frac{(\beta_{2\mathrm U}-\bar{\beta}_b(\gamma_{\mathrm U}))^2}{2 \sigma_b^2(\gamma_{\mathrm U})}\right)\;.
\end{align}
Here, $\bar{\beta}_a$ and $\sigma_a$ are the mean and uncertainty of $\beta_{2\mathrm U}$ extracted from $R_{(\delta \pT)^2}$. Similarly, $\bar{\beta}_b$ and $\sigma_b$ are the mean and uncertainty of $\beta_{2\mathrm U}$ from $R_{v_2^2\delta \pT}$, and they depend on $\gamma_{\mathrm U}$. We sample a uniform prior distribution in $\beta_{2\mathrm U}$ and $\gamma_{\mathrm U}$ to obtain the posterior distribution, from which we obtained the mean and one standard deviation uncertainty of $\beta_{2\mathrm U}$ and $\gamma_{\mathrm U}$. 

The constraints are listed in Table~\ref{tab:2} for the four $\pT$ intervals. Averaging across these $\pT$ windows, we find $\beta_{2\mathrm{U}}\approx0.300 \pm 0.016$ and $\gamma_{\mathrm{U}}\approx 8.3 \pm 4.7^{\circ}$ from $R_{(\delta \pT)^2}$ and $R_{v_2^2\delta \pT}$. In contrast, $R_{v_2^2}$ alone provides a lower-limit estimate of $\beta_{2\mathrm{U}}\approx0.24\pm0.02$. This finding is consistent with recent model studies~\cite{Ryssens:2023fkv,Xu:2024bdh}, where $\beta_{2\mathrm{U}}\sim 0.25$ could describe $R_{v_2^2}$ data. Similar nuclear structure constraints have also reported using a different hydrodynamic model framework~\cite{Fortier:2023xxy,Fortier:2024yxs}. 

We caution that shape parameters are extracted within the Woods-Saxon parametrization, where deformation values are defined at the nuclear surface. In the low-energy community, $\beta_2$ from the spectroscopic method corresponds to a $\beta_2$ defined using the body distribution of the nucleons~\cite{Ryssens:2023fkv}. In the presence of a positive $\beta_{4}$, the $\beta_2$ value from Woods-Saxon parametrization can be slightly reduced compared to the actual $\beta_2$ for the body deformation~\cite{LOBNER1970495}. Therefore, it is valuable to derive an independent constraint on the $\beta_4$ from high-energy collisions, which should be explored in the future.

Another point is worth mentioning. In Ref.~\cite{STAR:2024eky}, an independent estimate of $\beta_{2\mathrm{U}}$ and $\gamma_{\mathrm{U}}$ was obtained from another hydrodynamic model, Trajectum~\cite{Giacalone:2023cet}. Their reported values, $(\beta_{2\mathrm{U}},\gamma_{\mathrm{U}})=(0.286 \pm 0.025,8.7^{\circ} \pm 4.5^{\circ})$, are obtained for one $\pT$ bin, by combining the two constraints and including the model differences as additional uncertainties, and hence are not the same as what is reported in this analysis. 

Overall, the extracted deformation parameters align with low-energy nuclear measurements under the rigid-rotor assumption~\cite{Pritychenko:2013gwa}. These constraints are achieved without referencing transitions to excited nuclear states, demonstrating how ultra-central high-energy collisions probe ground-state nuclear shapes.

\section{Summary and outlook} 
The correlation between elliptic flow ($v_2$) and event-by-event average transverse momentum ($[\pT]$) in high-energy heavy-ion collisions offers a unique probe to the shape of the colliding nuclei, notably their quadrupole deformation ($\beta_2$) and triaxiality ($\gamma$). This sensitivity is robustly demonstrated by comparing collisions of strong prolate-deformed $^{238}$U nuclei with weakly oblate-deformed $^{197}$Au nuclei, focusing on three observables: $\lr{v_2^2}$, $\lr{(\delta\pT)^2}$, and $\lr{v_2^2\delta\pT}$. 

We construct a normalized quantity, $\rho_2 = \frac{\lr{v_2^2\delta \pT}}{\lr{v_2^2} \sqrt{(\delta \pT)^2}}$, which reduces sensitivity to final-state expansion dynamics. In 0--7\% ultra-central U+U collisions, $\rho_2$ exhibits a steep transition from positive to negative values, whereas it stays nearly constant in Au+Au collisions. Further suppression of residual final-state effects is achieved by using the ratio $R_{v_2^2\delta\pT} = \lr{v_2^2\delta \pT}_{\mathrm{U}}/\lr{v_2^2\delta\pT}_{\mathrm{Au}}$. This ratio shows a sharp decrease in central collisions, nearly independent of the correlation methods or $\pT$ selections. This provides direct evidence from the high-energy regime that $^{238}$U possesses strong prolate deformation (large $\beta_{2\mathrm{U}}$) and small triaxiality ($\gamma_{\mathrm{U}}\sim 0^{\circ}$). Independent evidence for large $\beta_{2\mathrm{U}}$ also comes from the strong rise of $R_{v_2^2} = \lr{v_2^2}_{\mathrm{U}}/\lr{v_2^2}_{\mathrm{Au}}$ and $R_{(\delta \pT)^2} = \lr{((\delta \pT)^2)}_{\mathrm{U}}/\lr{((\delta \pT)^2)}_{\mathrm{Au}}$ in ultra-central collisions. In contrast, observables associated with higher-order harmonics ($v_3$ and $v_4$), which are less affected by quadruple deformation, show much smaller differences between U+U and Au+Au systems.

The $\lr{v_2^2\delta\pT}$ data were compared with both a Glauber model, which accounts for the influence of nuclear deformation on initial collision conditions, and a state-of-the-art hydrodynamic model simulation (IP-Glasma+MUSIC), which incorporates final-state effects. Both models employ the Woods-Saxon parameterization for nuclear deformation. These models quantitatively reproduce the centrality dependence of $R_{v_2^2\delta\pT}$ when $\beta_{2\mathrm{U}}= 0.28$ and $\gamma_{\mathrm{U}}=0^{\circ}$. In ultra-central collisions, where deformation effects are dominant, the IP-Glasma+MUSIC model yields $\beta_{2\mathrm{U}}\approx  0.27$--0.30, while the Glauber model supports a triaxiality value of $\gamma_{\mathrm{U}}\lesssim15^{\circ}$.

For a more precise extraction of the uranium shape, we performed a joint fit of the hydrodynamic model to both $R_{(\delta \pT)^2}$ data (mainly constraining $\beta_{2\mathrm{U}}$) and $R_{v_2^2\delta\pT}$ data (sensitive to both $\beta_{2\mathrm{U}}$ and $\gamma_{\mathrm{U}}$). By accounting for variations in medium evolution and initial-state parameters, and incorporating non-flow effects in the data, we determine $\beta_{2\mathrm{U}}=0.300 \pm 0.016$ and $\gamma_{\mathrm{U}}=8.3 \pm 4.7^{\circ}$. These values are broadly consistent with low-energy estimates under the rigid-rotor assumption. This marks a valuable extraction of the nuclear ground-state shape purely from high-energy collisions. 

Observables involving higher-order flow harmonics, such as $\lr{v_3^2}$, $\lr{v_4^2}$, $\lr{v_3^2\delta\pT}$ and $\lr{v_4^2\delta\pT}$, are also sensitive to octupole $\beta_3$ and hexadecapole $\beta_4$ deformations. Our findings that $\lr{v_3^2}_{\mathrm{U}}/\lr{v_3^2}_{\mathrm{Au}}>1$ and $\lr{v_3^2\delta \pT}_{\mathrm{U}}/\lr{v_3^2\delta\pT}_{\mathrm{Au}}<1$ in ultra-central collisions are consistent with the presence of a modest $\beta_{3\mathrm {U}}\sim 0.08$--0.10~\cite{Zhang:2025hvi}. This represents the first experimental evidence from the high-energy regime for octupole collectivity of $^{238}$U. While a similar analysis was performed for $\beta_{4\mathrm {U}}$, the uncertainties in analogous ratios involving $\lr{v_4^2}$ and $\lr{v_4^2\delta\pT}$ were too large for a definite statement.

The successful application of imaging nuclear shapes via high-energy heavy-ion collisions raises questions about the relationship between these short-timescale snapshot measurements (where the collision crossing time is $\lesssim 0.1$ fm/$c$) and traditional low-energy methods, which probe much longer timescales. This approach also raises issues about possible partonic modifications within highly boosted nuclei. Addressing these questions is crucial for advancing our understanding of the elusive initial state of heavy-ion collisions. Further studies, incorporating various hydrodynamic modeling frameworks and an explicit treatment of longitudinal decorrelations, are necessary to refine this imaging technique. This is particularly important in small collision systems, where uncertainties in the properties and initial conditions of the QGP could reduce the sensitivity to nuclear structure differences between isobar-like species.

To establish the ``imaging-by-smashing'' method as a potential discovery tool, it must first be calibrated using isobar nuclei with known shapes. This calibration allows for the tuning of key response coefficients (Eqs.~\eqref{eq:2} and \eqref{eq:2b}) within the theoretical models used for analysis. The method must then be validated with additional species with known nuclear shapes to cross-check these coefficients and expose any limitations or inconsistencies~\cite{Jia:2025wey}. Furthermore, uncertainty quantification must account for model dependencies inherent in shape parameters derived from low-energy experiments~\cite{Dobaczewski:2025rdi}. These calibration and validation steps are also crucial for placing stricter constraints on the initial conditions, dynamics, and properties of the QGP. With proper validation, this imaging method can become a complementary tool for studying the collective, many-body structure of atomic nuclei across the nuclide chart and at various energy scales. This goal can be achieved by leveraging collider and fixed-target facilities to collide carefully selected isobar species~\cite{Bally:2022vgo}.

\subsection{Data availability statement} 
All raw data for this study were collected using the STAR detector at Brookhaven National Laboratory and are not available to the public. Derived data supporting the findings of this study are publicly available in the HEPData repository (https://www.hepdata.net/record/159930) or from the corresponding author on request.

\subsection{Acknowledgments}
We thank Chun Shen for providing the IP-Glasma+MUSIC code. We thank the RHIC Operations Group and SDCC at BNL, the NERSC Center at LBNL, and the Open Science Grid consortium for providing resources and support.  This work was supported in part by the Office of Nuclear Physics within the U.S. DOE Office of Science, the U.S. National Science Foundation, National Natural Science Foundation of China, Chinese Academy of Science, the Ministry of Science and Technology of China and the Chinese Ministry of Education, NSTC Taipei, the National Research Foundation of Korea, Czech Science Foundation and Ministry of Education, Youth and Sports of the Czech Republic, Hungarian National Research, Development and Innovation Office, New National Excellency Programme of the Hungarian Ministry of Human Capacities, Department of Atomic Energy and Department of Science and Technology of the Government of India, the National Science Centre and WUT ID-UB of Poland, the Ministry of Science, Education and Sports of the Republic of Croatia, German Bundesministerium f\"ur Bildung, Wissenschaft, Forschung and Technologie (BMBF), Helmholtz Association, Ministry of Education, Culture, Sports, Science, and Technology (MEXT), and Japan Society for the Promotion of Science (JSPS). 

\bibliography{../vnpTlong}{} \bibliographystyle{apsrev4-1}
\appendix
\section*{Comprehensive data plots}
\begin{figure*}[!htbp]
\begin{center}
\includegraphics[width=0.85\linewidth]{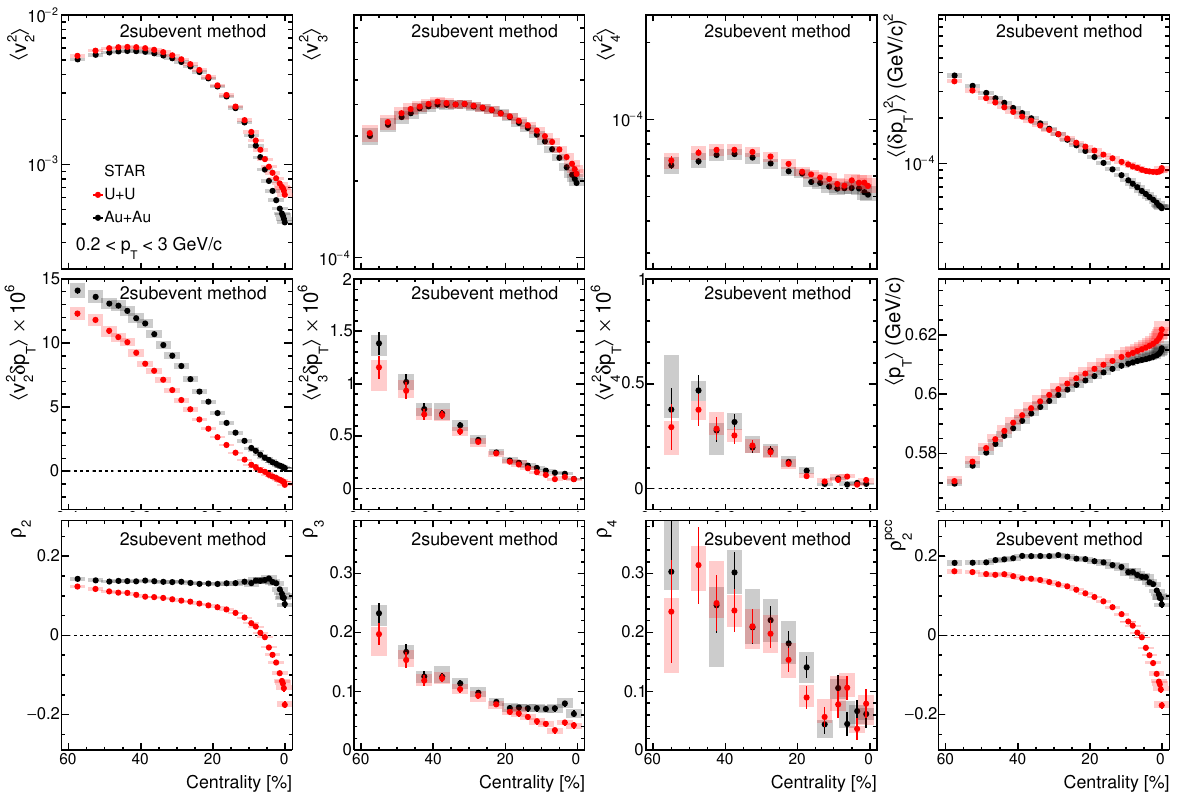}
\end{center}
\caption{\label{fig:a1} Observables as a function of centrality in U+U and Au+Au collisions for charged hadrons in $0.2<\pT<3$ GeV/$c$.}
\end{figure*}

\begin{figure*}[!htbp]
\begin{center}
\includegraphics[width=0.85\linewidth]{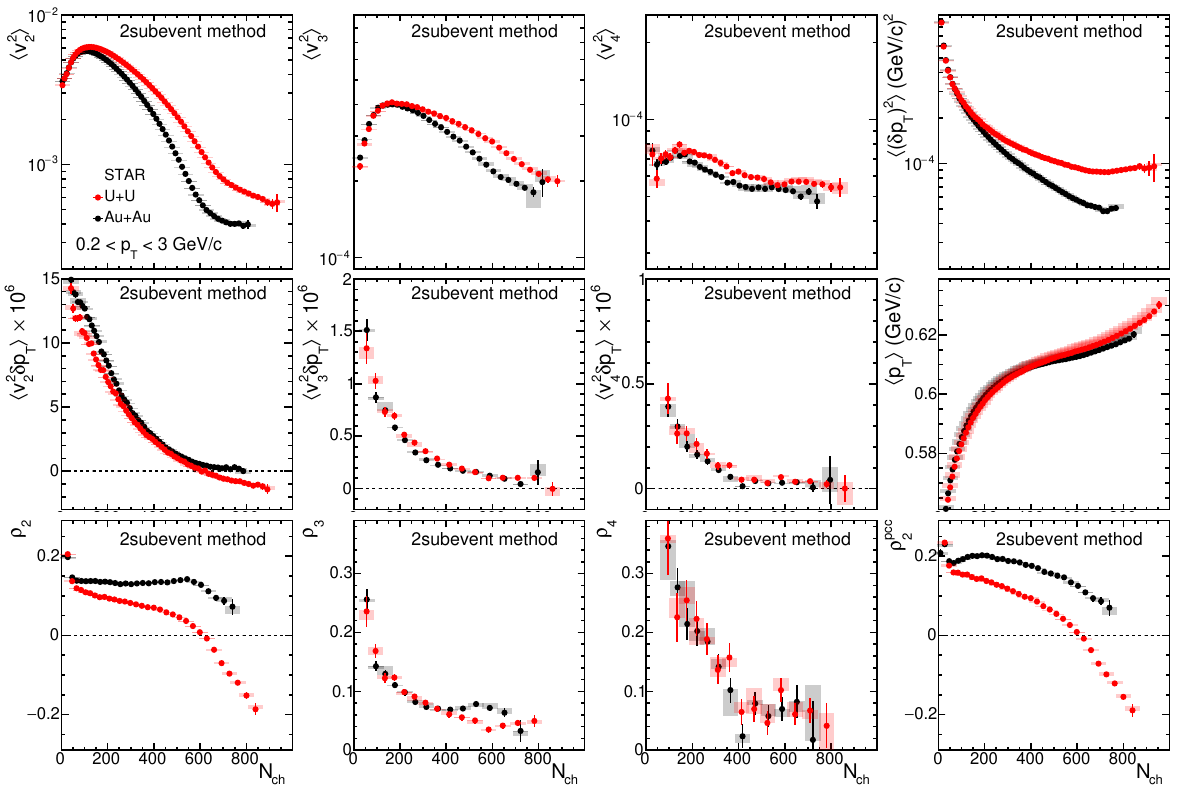}
\end{center}
\caption{\label{fig:a2} Observables as a function of $\nch$ in U+U and Au+Au collisions for charged hadrons in $0.2<\pT<3$ GeV/$c$.}
\end{figure*}

\begin{figure*}[h!]
\begin{center}
\includegraphics[width=0.85\linewidth]{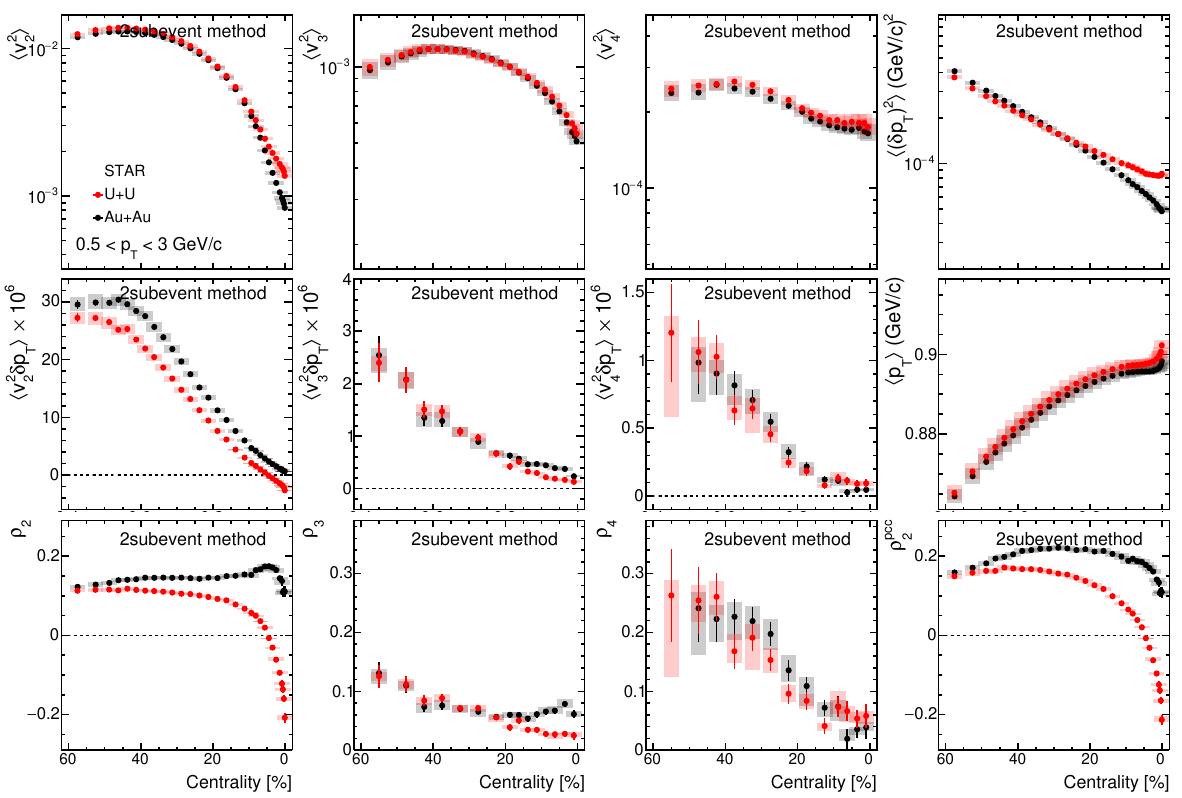}
\end{center}
\caption{\label{fig:a3} Observables as a function of centrality in U+U and Au+Au collisions for charged hadrons in $0.5<\pT<3$ GeV/$c$.}
\end{figure*}

\begin{figure*}[h!]
\begin{center}
\includegraphics[width=0.85\linewidth]{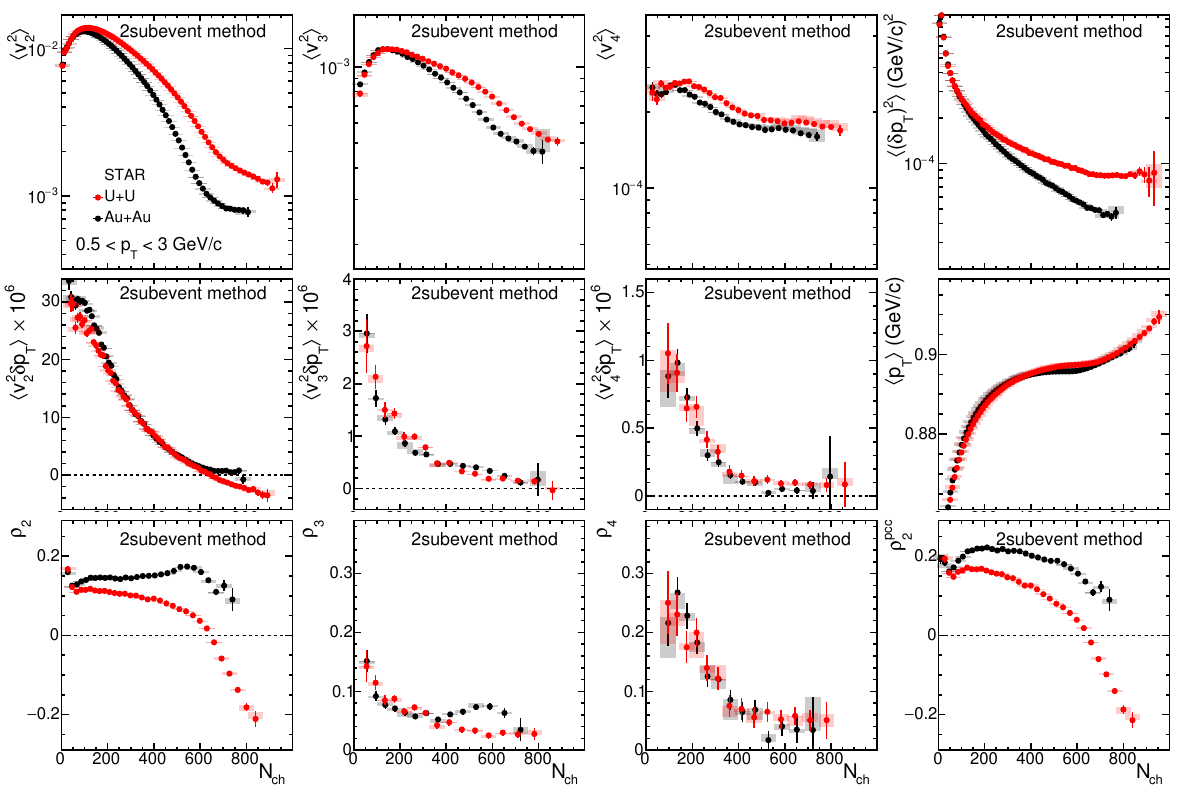}
\end{center}
\caption{\label{fig:a4} Observables as a function of $\nch$ in U+U and Au+Au collisions for charged hadrons in $0.5<\pT<3$ GeV/$c$.}
\end{figure*}

\begin{figure*}[!htbp]
\begin{center}
\includegraphics[width=0.85\linewidth]{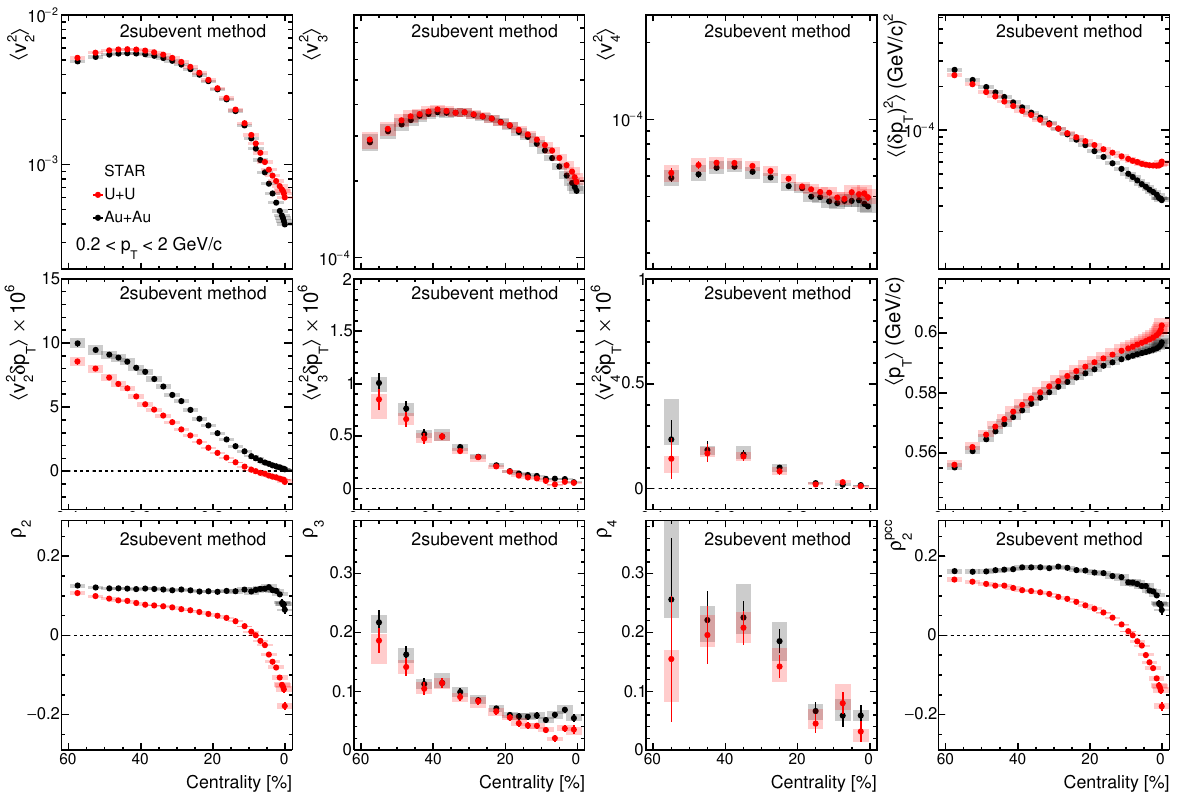}
\end{center}
\caption{\label{fig:a5} Observables as a function of centrality in U+U and Au+Au collisions for charged hadrons in $0.2<\pT<2$ GeV/$c$.}
\end{figure*}

\begin{figure*}[!htbp]
\begin{center}
\includegraphics[width=0.85\linewidth]{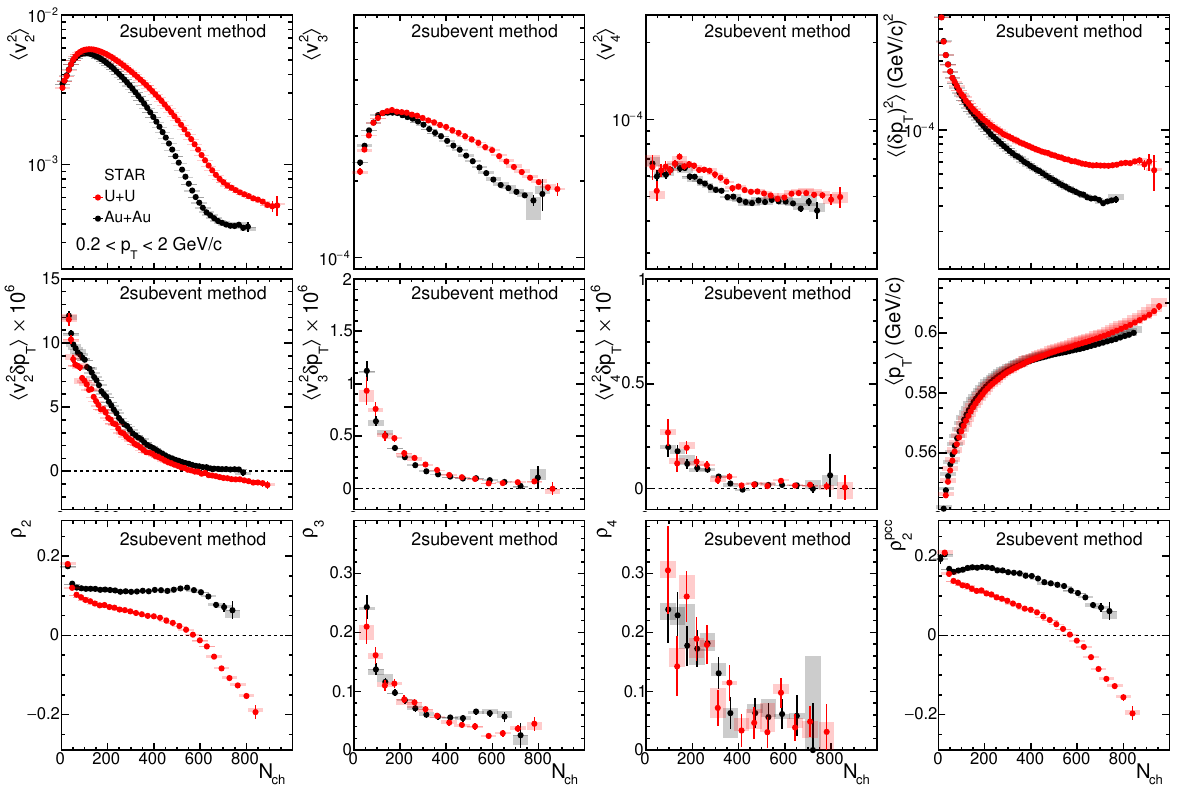}
\end{center}
\caption{\label{fig:a6} Observables as a function of $\nch$ in U+U and Au+Au collisions for charged hadrons in $0.2<\pT<2$ GeV/$c$.}
\end{figure*}

\begin{figure*}[!htbp]
\begin{center}
\includegraphics[width=0.85\linewidth]{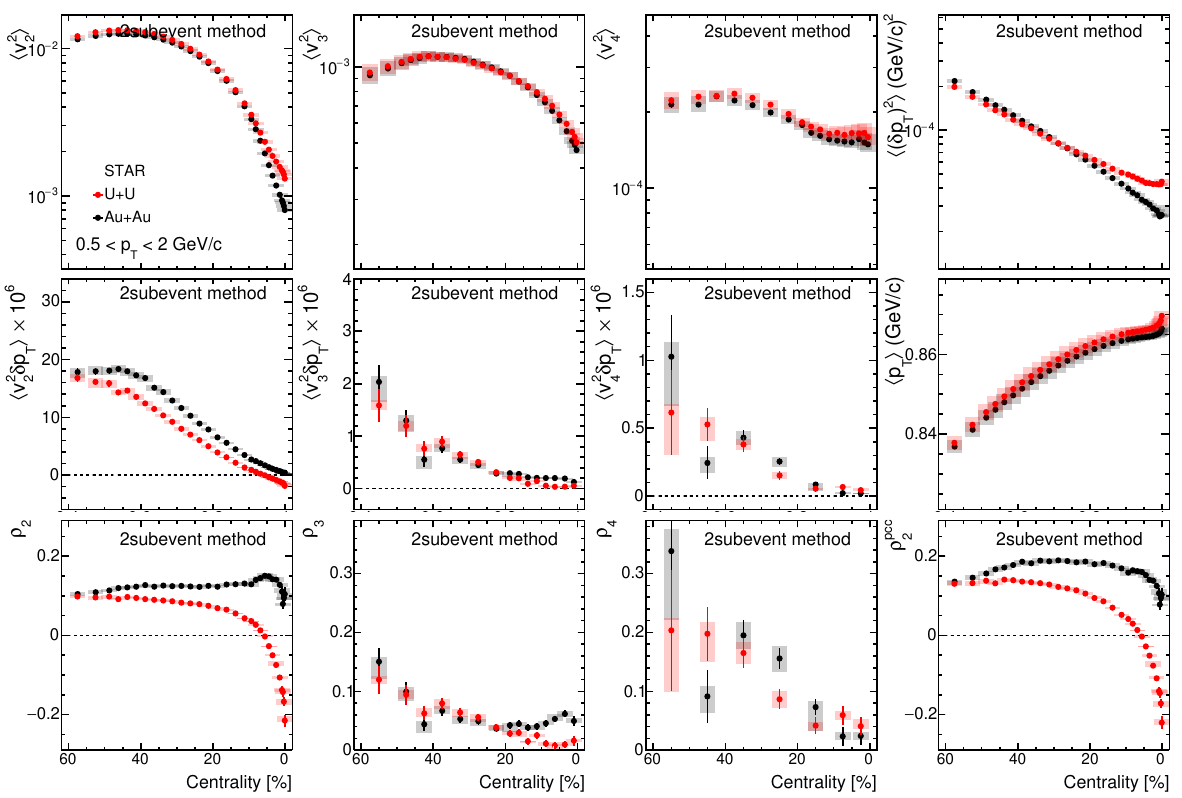}
\end{center}
\caption{\label{fig:a7} Observables as a function of centrality in U+U and Au+Au collisions for charged hadrons in $0.5<\pT<2$ GeV/$c$.}
\end{figure*}

\begin{figure*}[!htbp]
\begin{center}
\includegraphics[width=0.85\linewidth]{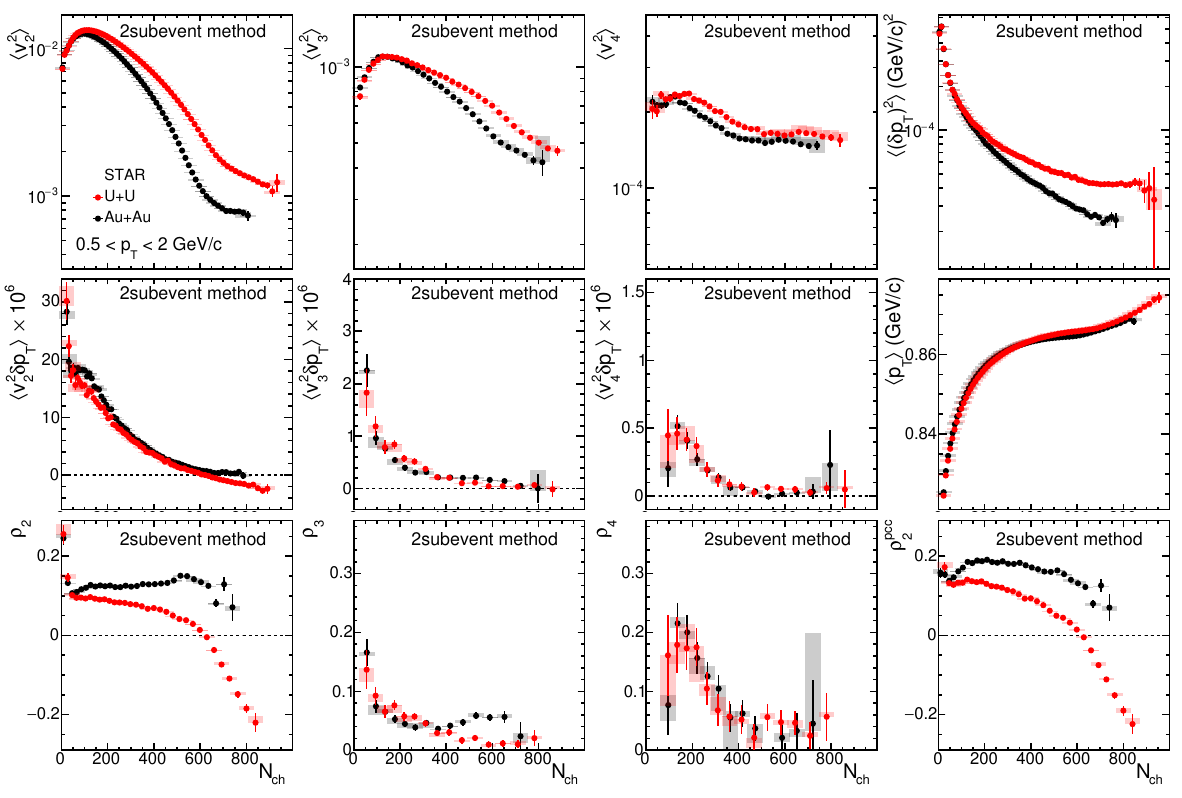}
\end{center}
\caption{\label{fig:a8} Observables as a function of $\nch$ in U+U and Au+Au collisions for charged hadrons in $0.5<\pT<2$ GeV/$c$.}
\end{figure*}

\begin{figure*}[!htbp]
\begin{center}
\includegraphics[width=0.85\linewidth]{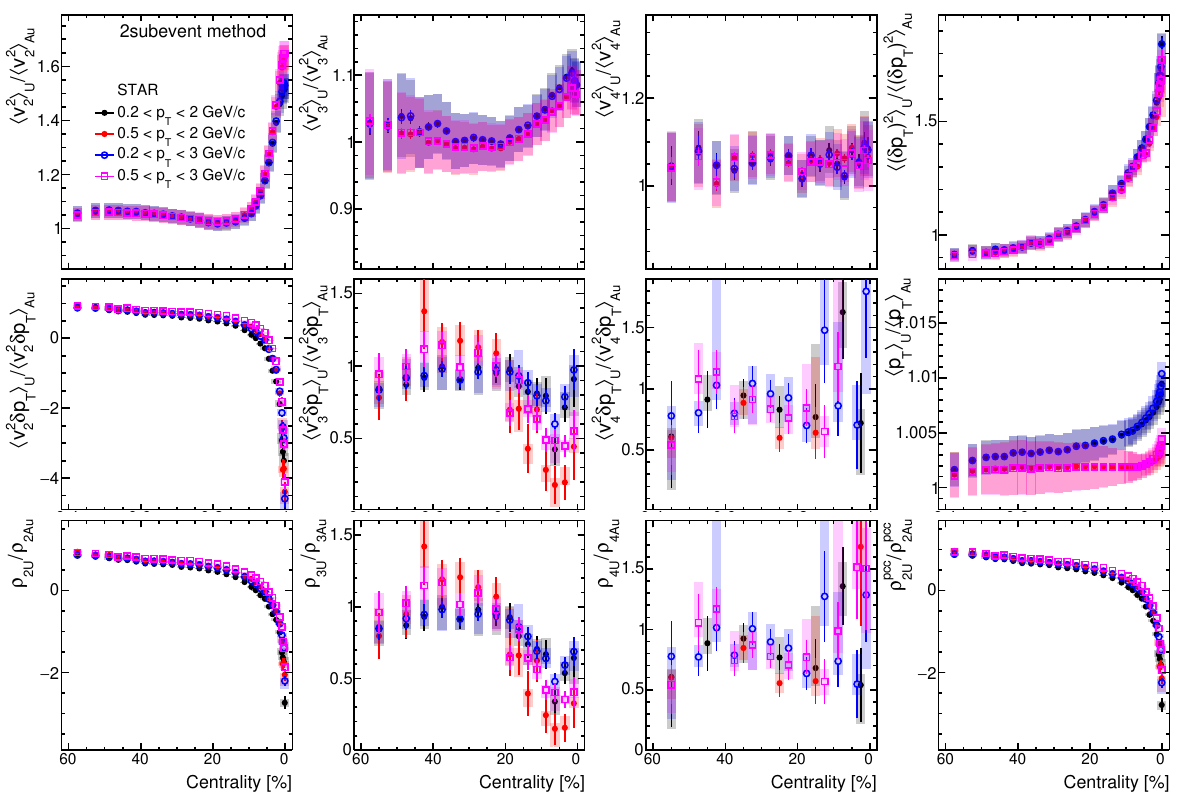}
\end{center}
\caption{\label{fig:a9} Ratios of observables as a function of centrality for charged hadrons in four $\pT$ ranges.}
\end{figure*}

The results in the paper are mainly presented as a function of centrality. However, it is also helpful to present them as a function of $\nch$ for direct comparison with models.  We collect the data for all observables used in this paper in U+U and Au+Au collisions, $\lr{v_n^2}$, $\lr{\pT}$, $\lr{(\delta\pT)^2}$, and $\lr{v_n^2\delta\pT}$ for $n=2,3,4$, and present them in Figs.~\ref{fig:a1}--\ref{fig:a8} for four $\pT$ ranges as a function of either centrality or $\nch$. The ratios between the two collision systems as a function of centrality are shown in Fig.~\ref{fig:a9}.





\end{document}